\documentclass[final]{fcs}
\usepackage[none]{hyphenat}

\usepackage{graphicx}%
\usepackage{multirow}%
\usepackage{amsmath,amssymb,amsfonts}%
\usepackage{amsthm}%
\usepackage{mathrsfs}%
\usepackage[title]{appendix}%
\usepackage{xcolor}%
\usepackage{textcomp}%
\usepackage{manyfoot}%
\usepackage{booktabs}%
\usepackage{algorithm}%
\usepackage{algorithmicx}%
\usepackage{algpseudocode}%
\usepackage{listings}%

\usepackage{array}
\usepackage{textcomp}
\usepackage{stfloats}
\usepackage{xurl}
\usepackage{verbatim}
\usepackage{geometry}
\usepackage{threeparttable}
\usepackage{subcaption} 
\usepackage{color, xcolor}
\usepackage{tabularx}
\usepackage{xltabular}
\usepackage{longtable}
\usepackage{tikz}
\usepackage{lipsum}
\usepackage{colortbl}
\usepackage{xtab}
\usepackage[percent]{overpic}  
\usepackage{forest}
\definecolor{hidden-draw}{RGB}{20,68,106}
\usepackage{adjustbox}
\usepackage{diagbox}
\usepackage{rotating}  
\usepackage{lscape}    
\usepackage{afterpage}
\usepackage{enumitem}
\usepackage{varwidth}
\usepackage{hyperref}

\usepackage{microtype}
\hypersetup{
    colorlinks=true, 
    linkcolor=blue,   
    citecolor=blue    
}

\title{A Comprehensive Exploration of Personalized Learning in Smart Education: From Student Modeling to Personalized Recommendations}
\author{Siyu Wu}

\author{Yang Cao}

\author{Runze Li}

\author{Jiajun Cui}

\author{Hong Qian}

\author{Bo Jiang}
\author*{Wei Zhang}

\address{East China Normal University, Shanghai 200062, China}

\fcssetup{
  received       = {month dd, yyyy},
  accepted       = {month dd, yyyy},
  corr-email     = {zhangwei.thu2011@gmail.com},
}

\begin{abstract}
With the development of artificial intelligence, personalized learning has attracted much attention as an integral part of intelligent education.
In recent years, countries and regions such as China, the United States, and the European Union have increasingly recognized the importance of personalized learning, emphasizing its potential to integrate large-scale education with individualized instruction effectively.
This survey provides a comprehensive analysis of personalized learning by reviewing relevant studies published in major conferences and journals between January 2017 and April 2025.
We examine its definition, objectives, and underlying educational theories, highlighting its pedagogical significance. Furthermore, we explore personalized learning from two key dimensions: student modeling and personalized recommendations. 
Student modeling is analyzed from both cognitive and non-cognitive perspectives, while recommendation approaches are categorized based on their specific objectives. 
Additionally, we investigate the interplay between these components and their role in enhancing personalized learning.
Beyond theoretical and algorithmic insights, this survey reviews real-world applications, demonstrating personalized learning’s effectiveness in educational practice. 
Finally, we discuss key challenges and future directions, offering a multidimensional perspective that bridges theory and practice.
\end{abstract}

\begin{document}
\section{Introduction}
\label{sec1}
Throughout history, the landscape of education has continually evolved, undergoing significant and transformative changes.
Personalized learning, positioned as a pivotal focus on education, has emerged as a crucial strategy to address the distinct needs and objectives of individual learners.

\begin{figure*}[htbp]
    \centering
    \includegraphics[width=0.85\textwidth]{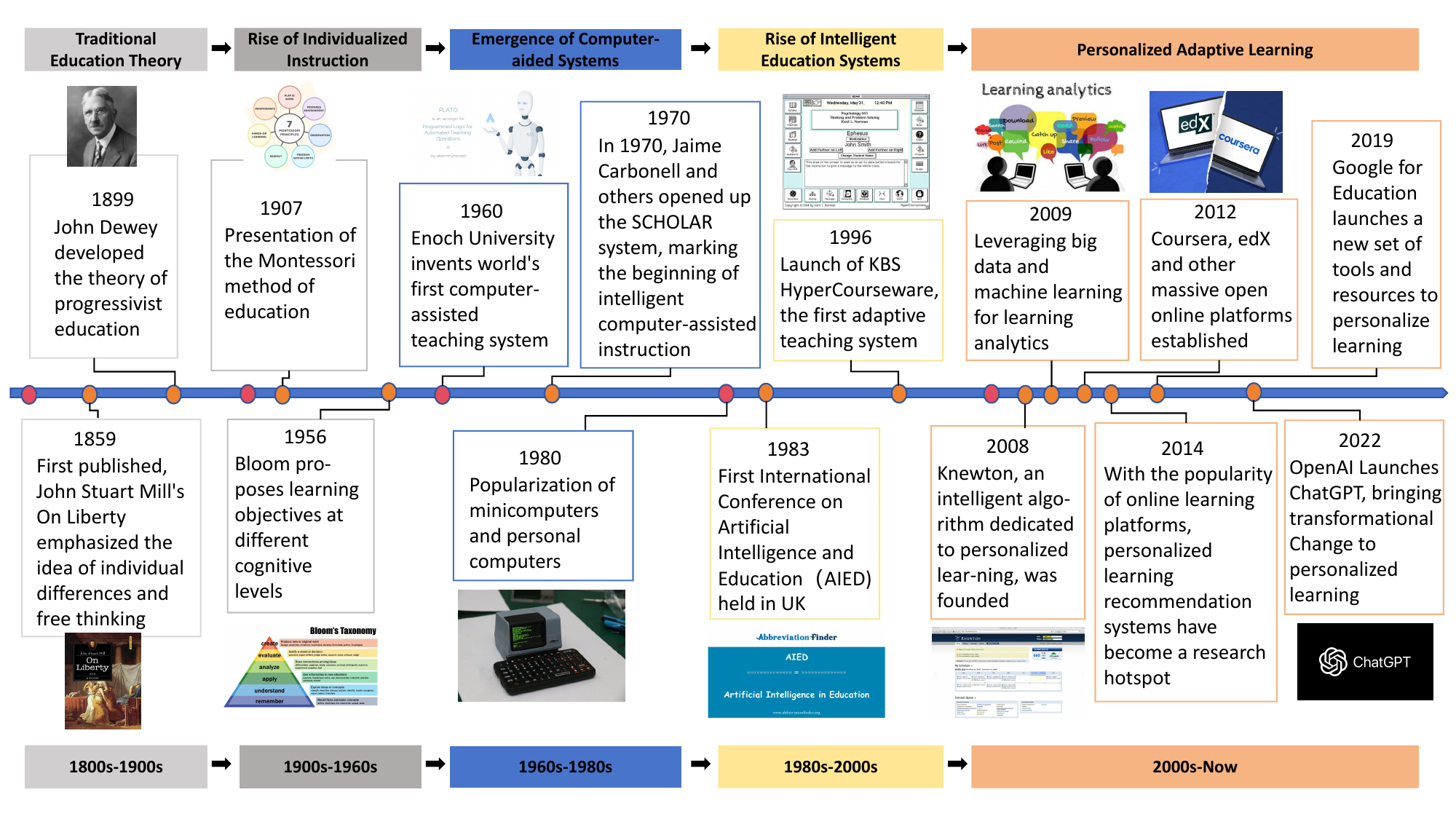}
    \caption{The evolution of personalized learning}
    \label{fig:develop}
\end{figure*}

Personalized learning has evolved from early educational research on individual differences \cite{bloom2020taxonomy} to computer-assisted instruction \cite{skinner1965review}, eventually becoming an integral part of intelligent educational systems through recommender systems and student modeling \cite{brusilovsky1998methods,siemens2011penetrating}.
With the advent of big data technologies, personalized learning has entered a data-driven era, enabling more precise adaptation to individual learner needs \cite{berland2014educational}.

This evolution has led to personalized adaptive learning, which dynamically adjusts instructional content and strategies based on students' abilities and progress.
Figure~\ref{fig:develop} outlines this developmental trajectory, illustrating key milestones from the early 1800s to the present.
The trajectory of personalized learning has traversed an evolutionary continuum, progressing from simplicity to complexity, from mechanization to intelligence, and from theoretical conceptualization to practical implementation. 
By tailoring content and learning paths to individual characteristics, personalized learning enhances engagement, improves learning outcomes, and accommodates diverse educational needs \cite{LinYHC13, Zhang23a, Xu22}.
It has demonstrated effectiveness across traditional classrooms, online education \cite{HuangLY23}, workforce training \cite{ChaipidechSKC22}, and career development \cite{WangZZZCX20}.

As education increasingly shifts towards intelligent and adaptive models, research in personalized learning has become essential for addressing student diversity, improving motivation, and reducing dropout rates \cite{GomezZSF14, NabilSE21}.
The growing prominence of personalized learning in academia, policy discussions, and media further highlights its significance \cite{FitzGeraldKJCFH18}.

Although personalized learning has made significant progress, existing review studies often focus on specific aspects and lack a comprehensive and systematic overview of the field.
To fill this gap, this study systematically analyzes personalized learning, exploring its definitions, goals, educational theories, and key technologies.
Special emphasis is placed on student modeling and personalized recommendations, as these components form the foundation of personalized learning implementation.
By synthesizing recent advances and identifying future research directions, this review aims to provide valuable insights for researchers and practitioners in the field.

\subsection{Comparison to Relevant Surveys}
Recent studies have explored personalized learning from various perspectives, yielding valuable insights. 
Chen et al. \cite{ChenW21} examined the impact of individual differences, including learning styles and abilities, on personalized learning. 
Shemshack et al. \cite{ShemshackS20} analyzed personalized learning terminology, complementing our definition in Chapter~\ref{sec2}.
Bernacki et al. \cite{bernacki2021systematic} reviewed empirical research using Preferred Reporting Items for Systematic Reviews and Meta-Analyses (PRISMA) guidelines, focusing on specific contexts and learner characteristics.
While sharing a focus on educational theories, their study does not delve as deeply into the intricate relationship between educational theories and personalized learning.
Essa et al. \cite{essa2023personalised} extensively studied learning styles, primarily emphasizing machine learning methods for learner classification.
While our review includes learning style identification, we adopt a broader scope, incorporating sentiment and behavioral analysis within learning analytics.
Many reviews focus on personalized learning recommendations, emphasizing algorithms such as deep learning, machine learning, and meta-learning \cite{ZhongWYDWT20,TianYYS20,ren2020survey}.
Some studies specialize in resource or learning path recommendations \cite{yunyue21}, often limiting their scope to specific techniques.
In contrast, our review provides a more comprehensive analysis covering not only personalized recommendations but also other important aspects of personalized learning, such as cognitive diagnosis and learning styles.

\subsection{Contributions}
This paper presents a systematic and comprehensive review of personalized learning, encompassing its definitions, goals, methodologies, educational theories, practical applications, and the interplay between student modeling and personalized recommendations.
Through detailed classification and in-depth analysis, it provides readers with a holistic understanding of the field.
The main contributions of this paper are as follows:
\begin{itemize}
    \item \textbf{New taxonomy.}
    Although comprehensive personalized learning systems typically consist of four key components: student modeling, content modeling, personalized recommendations, and instructional interventions \cite{nkambou2010advances}, this paper focuses on the two core components, student modeling and personalized recommendations, as they are sufficient to capture the complete process of personalized learning implementation. 
    To refine the classification framework, student modeling is divided into cognitive and non-cognitive categories.
    Cognitive modeling encompasses cognitive diagnosis and student performance prediction (cognitive), whereas non-cognitive modeling includes learning style analysis, sentiment analysis, behavior analysis, and student performance prediction (non-cognitive). 
    Similarly, personalized recommendations are categorized into path recommendations, course recommendations, and exercise recommendations. 
    Each classification is further delineated based on technical methodologies, research focus, and application domains.
    For example, in student behavior analysis, existing studies are categorized according to technical approaches such as clustering algorithms, association rules, machine learning, and hybrid methods, offering a systematic organization of prior research.
    \item \textbf{A comprehensive and systematic review.}
    This paper offers a comprehensive, multidimensional review of personalized learning, covering its definition, theoretical foundations, technical pipeline from student modeling to personalized recommendations, and real-world applications.
    Beyond technical methodologies, we examine its scientific basis and practical implications from an educational theory perspective. 
    \item \textbf{Latest information.}
    This paper systematically collects and analyzes the latest research findings from January 2017 to January 2025, providing a comprehensive reflection on the latest advances and trends in the field of personalized learning.
    \item \textbf{Application notes.}
    This paper will systematically present real-world use cases of personalized learning that demonstrate the effects of personalized learning in real educational settings, providing a link between theory and practice.
    \item \textbf{Future directions.}
    We critically analyze the limitations of current state-of-the-art approaches and identify key challenges in the field across four critical areas: data quality, assessment systems, technological limitations, and ethical concerns.
    Based on these insights, we propose promising future research directions to address these challenges and advance the field of personalized learning.
\end{itemize}

\begin{figure*}[!htp]
    \setlength{\abovecaptionskip}{0.cm}
    \setlength{\belowcaptionskip}{-0.cm}
    \centering
    \subfigure[Article filtering process funnel]{
        \begin{minipage}[t]{0.5\textwidth}
            \centering
            \includegraphics[width=0.9\textwidth]{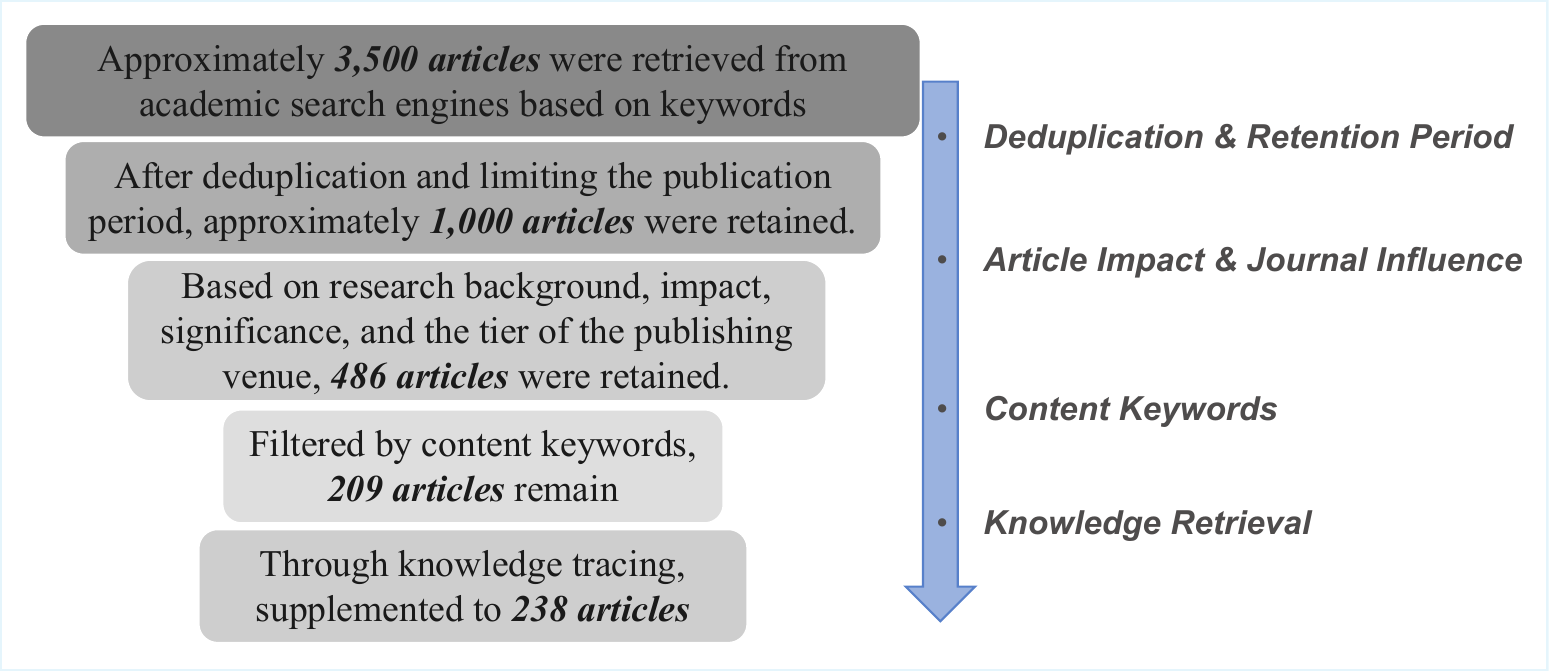}
            \label{fig:funnel}
        \end{minipage}
    }
    \subfigure[Proportion of core article themes]{
        \begin{minipage}[t]{0.46\textwidth}
            \centering
            \includegraphics[width=0.9\textwidth]{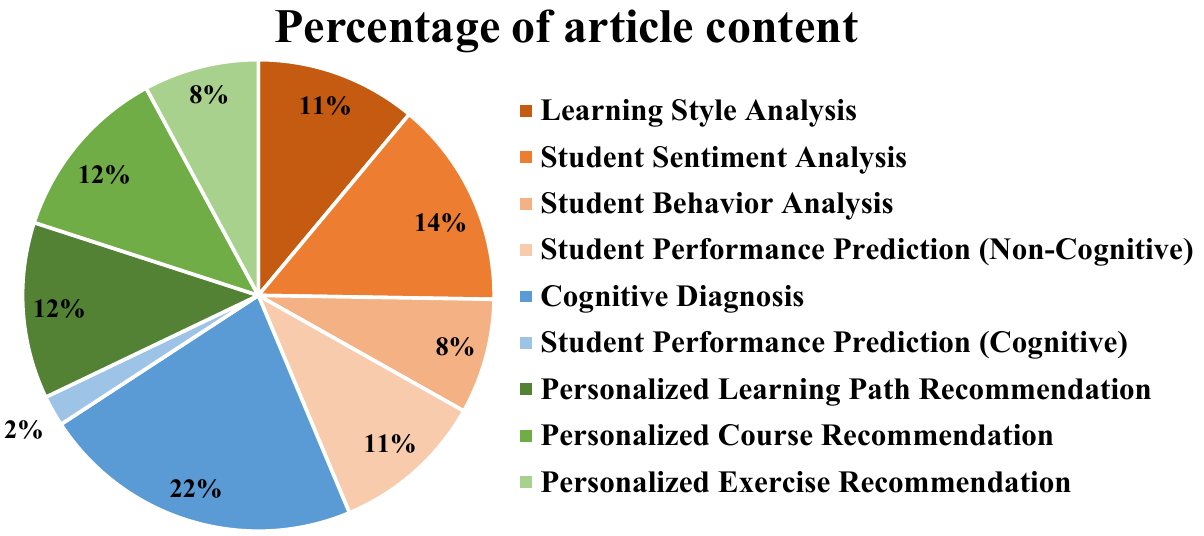}   
            \label{fig:bing}
        \end{minipage}    
    }
    \caption{Survey methodology, article filtering, and reference proportion}
    \label{fig:envs}
\end{figure*}

\subsection{Survey Methodology}
We systematically selected articles from the field of personalized learning through a four-step procedure: retrieval, initial screening, fine-grained selection, and supplementation. 
The detailed workflow and the number of the corresponding articles are illustrated in Figure~\ref{fig:funnel}. 
In the first step of retrieval, we conducted keyword-based queries on major academic search engines, using terms such as \textit{(“[intitle: personalized AND intitle: learning]”} and \textit{(“[intitle: adaptive AND intitle: learning]”)}. This process yielded a comprehensive initial set of publications.
During the initial screening, we removed duplicates and retained articles published between 2017 and 2025.
To ensure both comprehensiveness and academic rigor, we prioritized publications from high-impact journals and premier conferences (e.g., EDM, Computers \& Education, IEEE TLT, AAAI, KDD, SIGIR), while also incorporating other influential or contextually significant works. 
The selection process was guided by multiple criteria, including research novelty, analytical rigor, and overall scholarly impact.

The subsequent fine-grained selection involved a thorough examination of the remaining corpus to identify works whose primary contribution lies in student modeling or personalized recommendation, thereby ensuring the thematic coherence of this survey. 
Given that student modeling encompasses both cognitive and non-cognitive aspects, we employed refined keywords for filtering, such as:
\textit{(“[intitle: student* AND (intitle: modeling OR intitle: diagnosis OR intitle: analytics)]")}
OR \textit{(“[intitle: learner* AND (intitle: modeling OR intitle: analysis)]")}
OR \textit{(“[intitle: sentiment AND intitle: analysis]")}.
Similarly, for personalized recommendation, we used:
\textit{(“[intitle: personalized AND (intitle: recommendation OR intitle: education)]" 
OR “[intitle: recommendation AND (intitle: exercise OR intitle: learning)]")}.

Finally, to enhance the comprehensiveness of our survey, we employed a backward snowballing strategy.  Starting from the articles selected in the third step, we traced back to the earlier influential articles they cited, ultimately forming the literature dataset for our research. 
The domain-wise distribution of the final article set is depicted in Figure~\ref{fig:bing}.


\subsection{Article Organization}
\begin{figure*}
    \centering
    \includegraphics[width=0.8\textwidth]{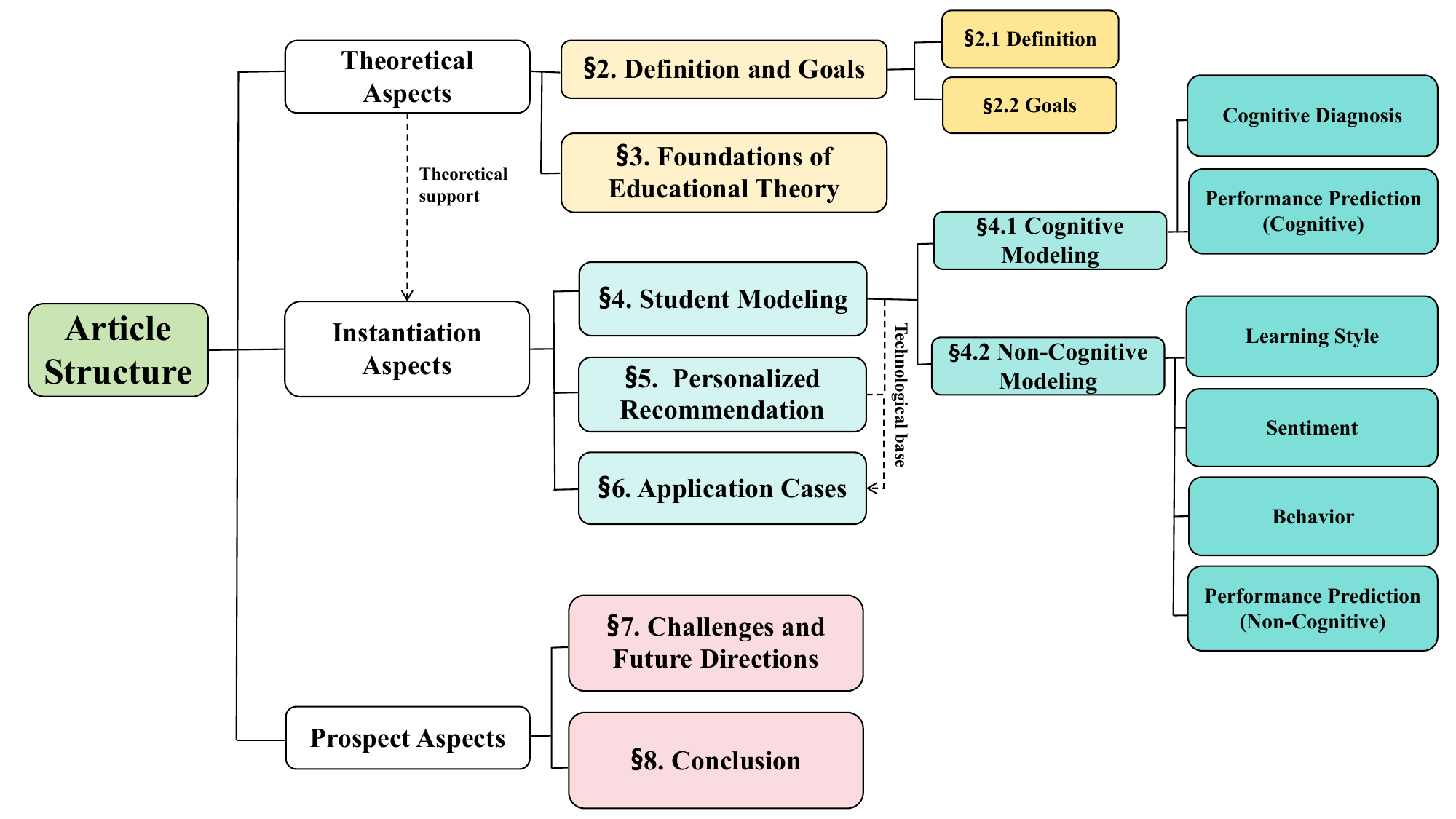}
    \caption{An overview framework of the survey}
    \label{fig:over_frame}
\end{figure*}

As illustrated in Figure~\ref{fig:over_frame}, this survey is structured as follows. 
Chapter~\ref{sec2} provides a detailed discussion on the definition and objectives of personalized learning, establishing a foundational framework for the review and linking it to the educational theories discussed in Chapter~\ref{sec3}.
Chapter~\ref{sec3} explores the theoretical foundations of personalized learning, highlighting key educational theories that underpin its development and offering a comprehensive perspective on education to support subsequent discussions. 
Chapter~\ref{sec5} examines student modeling, covering cognitive (Section~\ref{sec5_1}) and non-cognitive modeling (Section~\ref{sec5_2}), and categorizing studies by research focus and technical methodology.
This provides a comprehensive learner profile, forming the basis for personalized recommendations in Chapter~\ref{sec6}.
Building upon this foundation, Chapter~\ref{sec6} explores the second core aspect of personalized learning: personalized recommendations.
This chapter categorizes personalized recommendations into three aspects: learning path recommendations, exercise recommendations, and course recommendations.
Chapter~\ref{sec7} presents real-world applications that integrate the previously discussed theories and methodologies, showcasing actual tools and platforms implemented in practice.
Chapter~\ref{sec8} outlines key challenges and future research directions, offering insights into unresolved issues and potential advancements in the field.
Finally, Chapter~\ref{sec9} provides a summary of the survey.
This structured approach thoroughly examines personalized learning across educational theory, technology, and practice, aiming to equip academics and practitioners with a well-rounded perspective.

\section{Definition and Goals}
\label{sec2}
\subsection{Definitions of Personalized Learning}
\label{sec2_1}
Definitions of personalized learning from different organizations demonstrate the diversity and complexity of the concept, with different emphases, scopes, and implementation strategies.
Globally, there is no standardized definition \cite{schmid2019does}, yet a comparative analysis of multiple organizations' definitions provides a clear picture of the evolution of the concept and its specific application in different educational contexts.

\begin{table*}[t!]
\centering
\footnotesize
\caption{The Goal of Personalized Learning}
\label{tab_goal}
\resizebox{\textwidth}{!}{%
\rowcolors{2}{gray!20}{white} 
\begin{tabular}{p{0.15\linewidth}p{0.3\linewidth}p{0.2\linewidth}p{0.25\linewidth}}
\toprule
Goal & Explanation & Theoretical Foundation & Educational Implications \\
\midrule
    Improving academic performance &  Deliver tailored academic assistance by offering customized learning experiences that address the student's subject proficiency, strengths, and weaknesses, to enhance their academic performance in specific subjects. & Cognitive Psychology: Emphasizes individual differences in cognitive styles and information processing, supporting the need for tailored learning experiences. & By addressing individual cognitive needs, personalized learning can enhance subject mastery and academic achievement, leading to better educational outcomes. \\ 
    Stimulating interest in learning &  Identifying students’ interests through their learning processes and designing tasks that align with those interests. Providing materials and activities that resonate with these preferences to ignite their enthusiasm and foster deeper engagement with the subject matter. &  Constructivism: Highlights the importance of engaging learners through meaningful and relevant tasks that align with their interests and experiences. &  Personalized learning fosters intrinsic motivation by making learning relevant and engaging, which can lead to sustained interest and deeper engagement with the subject matter. \\ 
    Fostering self-directed learning &  Empower students by offering them flexibility to choose their learning paths and resources. Encourage them to formulate personalized learning plans, fostering greater autonomy in managing their learning processes. & Progressivism: Advocates for student-centered learning and the development of independent learning skills. & Promoting self-directed learning fosters essential life skills, including independence, accountability, and effective time management, which are crucial for continuous learning throughout life. \\ 
    Adapting to students' pace of learning &  Tailoring the content and complexity to accommodate diverse learning speeds and comprehension levels empowers students to progress at their learning pace, ensuring an optimal level of mastery. &  Behaviorist Theory: Supports the idea of reinforcement and pacing based on individual responses to stimuli. &  Allowing students to learn at their own pace reduces frustration and anxiety, leading to better comprehension and retention of material. \\ 
    Fostering Creative Thinking &  Offer inspiring tasks that encourage students to showcase creative thinking throughout the learning journey, fostering the development of their innovation skills and independent thought. &  Constructivism: Emphasizes active knowledge construction and problem-solving, which are essential for creative thinking. & A personalized learning environment that fosters creativity enables students to develop innovative solutions and cultivate critical thinking skills, both of which are essential for success. \\ 
    Reducing the learning gap & Offer tailored instructional strategies designed to meet the diverse needs of students, ensuring comprehensive understanding and mastery of subject matter. Strive for a balanced approach that considers each student's level, minimizing learning gaps across the student body. & Sociocultural Theory: Emphasizes the role of collaborative learning and social interaction in bridging learning gaps. &  Personalized learning can address individual learning gaps by providing targeted support, ensuring that all students have the opportunity to succeed. \\ 
    Promoting cooperative learning & Encourage collaborative learning by crafting tasks that foster teamwork and mutual learning among students. &  Sociocultural Theory: Stresses the importance of social interaction and collaborative learning in cognitive development. & Cooperative learning enhances social skills, promotes deeper understanding through peer interaction, and fosters a sense of community in the classroom. \\
\bottomrule
\end{tabular}%
}
\end{table*}

The definition of personalized learning was first put forward by the Organization for Economic Co-operation and Development (OECD, 2006)\footnote{Website at www.sciencedirect.com/topics/social-sciences/organization-for\\-economic-co-operation-and-development (Last visited on January 22, 2026).} reflects its systemic and holistic nature, arguing that personalized learning is not just about individual adjustments to teaching methods but rather a redesign of education through five dimensions: learning assessments, teaching strategies, curriculum choices, school organization, and partnerships.
This definition emphasizes a holistic reform of the education system, focusing on how to meet individual student needs and promote the optimization of the education process through the integration of feedback mechanisms and personalized teaching strategies.

In contrast, the International Society for Technology in Education (ISTE, 2014)\footnote{Website at www.iste.org/ (Last visited on January 22, 2026).} puts the spotlight on the role of technology, arguing that the core of personalized learning lies in the customization and optimization of each student's learning experience through the use of technological tools.
ISTE's definition highlights how technology can be a bridge to support personalized teaching and learning, especially in the age of information technology, which has become an important means of promoting personalized learning.
In this perspective, personalized learning is more often reflected in the digital customization of educational resources and learning paths, with an emphasis on the adaptability and accessibility of tools.
Meanwhile, the International Association for K-12 Online Learning (INACOL, 2016)\footnote{Website at aurora-institute.org/blog/what-is-personalized-learning/ (Last visited on January 22, 2026).} tends to focus more on the students themselves, proposing a student-centered definition of personalized learning that emphasizes the need for education to be adapted to the interests, strengths, and needs of students.
This perspective dovetails with the U.S. Department of Education's (2017) definition in its National Education Technology Plan \cite{reimagining}, which proposes to personalize each student's pace and instructional approach, emphasizing the individualization and flexibility of education. 
Both INACOL and the U.S. Department of Education's definitions exemplify the application of personalized learning to actual teaching and learning, especially at the K-12 education level, with a focus on tailoring instruction to students' individual differences to customize instructional content and pacing.
In contrast, the Stanford Research Institute (SRI, 2018) \cite{education2018using} provides a more in-depth definition, emphasizing that personalized learning is not limited to adapting instructional strategies but should also be comprehensively adapted to the needs of the student through customized assessments, curricula, and instructional methods.
This definition reflects a deeper understanding of individual differences and promotes the incorporation of multiple instructional elements into the educational process to achieve more comprehensive personalized learning.

Taken together, these definitions suggest that personalized learning is centered on making adjustments to students' individual needs and learning progress, but the specific ways in which this is practiced vary widely.
From OECD's emphasis on system-wide reforms to ISTE's prominence of the role of technology support to INACOL's and the U.S. Department of Education's emphasis on student-centered strategies, each definition reflects the understanding and application of personalized learning in different educational contexts.
Meanwhile, the Stanford Research Institute's definition further deepens the concept by advocating for a more integrated and deeper model of personalized education. These different perspectives and foci reveal the diversity of personalized learning in theory and practice, and reflect the continuing exploration and development of this innovative concept in the field of education.

In summary, although the definitions of personalized learning vary across organizations and contexts, they all point to one core goal: to meet the unique abilities, knowledge levels, and learning needs of each student through customized teaching strategies and approaches.
As the definitions continue to evolve, personalized learning focuses more on active student engagement and the effective use of technology, reflecting a gradual shift in the educational model away from a single instructional approach to one that is diverse and flexible.

\subsection{Goals of Personalized Learning}
\label{sec2_3}
The goals of personalized learning span various dimensions, including motivation, competence, and achievement, with variations in ultimate objectives across different definitions and methods \cite{bernacki2021systematic}. 
Table~\ref{tab_goal} outlines seven potential goals associated with personalized learning.
These goals are based on established educational theories and frameworks that provide the theoretical basis for their formulation and justification.

We can see that the goal of personalized learning is not just a simple adjustment to the individual learning needs of students but is also based on existing theoretical foundations and practical experience, aiming to improve learning outcomes, cultivate intrinsic interest in learning, develop self-directed learning skills and creative thinking, and provide effective empowerment strategies in a variety of areas.
The achievement of these goals will help drive the development of personalized learning, providing students with a more personalized and efficient learning experience while improving educational outcomes and laying a theoretical foundation for the full implementation of personalized education.

\section{Foundations of Educational Theory}
\label{sec3}
\subsection{Educational Theory}
Before entering the adaptive personalization stage, numerous educators have proposed many educational theories aimed at shaping and advancing personalized learning.
Each theory offers unique insights into the design and implementation of personalized learning and provides important theoretical support for its development.
Next, we present an overview of educational theories closely associated with personalized learning, highlighting their contributions to personalized learning and exploring practical applications in educational settings.
We organize these theories into three categories: educational philosophy, learning psychology, and sociocultural perspectives.
\begin{itemize}[leftmargin=*]
    \item [(1)] \textbf{Educational Philosophy}:
    \item \textbf{Progressivism} \cite{dewey2013my,dewey2023democracy,dewey1986experience}
    \begin{itemize}
        \item \textbf{Introduction}: Progressivism, a pivotal paradigm in 20th-century Western education, challenged conventional norms. 
    It focused on problem-centered learning to cultivate independent learning, critical thinking, and heightened engagement. 
    Progressivism emphasizes acknowledging individual differences and championing student-centered teaching approaches.
     \item \textbf{Contribution}: Provides the foundational goal of personalized learning, advocating for education tailored to individual needs.
    \item \textbf{Application}: Implemented through project-based learning, where students choose projects based on their interests.
    \end{itemize}
    \item [(2)] \textbf{Learning Psychology}:
    \item \textbf{Behaviorist Theory} \cite{skinner2019behavior,watson1913psychology}
    \begin{itemize}
        \item \textbf{Introduction:} Behaviorism, pioneered by psychologist Watson, sees behavior as observable responses to external stimuli.
    Modified by neo-behaviorists like Tolman, it introduces intervening variables between stimuli and responses. 
    Skinner and followers emphasize reinforcement as the key mechanism, shaping behavior through environmental influences.
    In essence, behaviorism underscores the importance of observing, strategizing, and controlling learner behavior, as well as shaping behavior through environmental influences. 
        \item \textbf{Contribution}: Informs the design of feedback systems in personalized learning, enabling immediate and tailored feedback to reinforce desired behaviors.
    \item \textbf{Critique}: Limited by its reliance on external stimuli, necessitating integration with other theories to address internal cognitive processes.
    \item \textbf{Application}: Applied in adaptive quizzing systems (e.g., Khan Academy), gamified learning platforms (e.g., Duolingo), and Intelligent Tutoring Systems (ITS) that provide personalized feedback based on student responses.
    \end{itemize}
    \item \textbf{Constructivism} \cite{piaget1952origins,vygotsky1978mind,ebbinghaus2013memory}
    \begin{itemize}
        \item \textbf{Introduction:} 
    Constructivism, a cognitive psychology branch, emerged from studying children's cognitive development.
    It explores principles governing learning processes, emphasizing active learner interaction with the environment. Learning involves participation, critical thinking, hands-on experiences, and the construction of knowledge. 
    The theory recognizes diverse learner backgrounds and asserts that knowledge construction depends on individual differences.
    \item \textbf{Contribution}: Supports the design of collaborative and exploratory learning environments in personalized learning.
    \item \textbf{Application}: Implemented through online discussion forums, collaborative projects, and problem-based learning activities.
    \end{itemize}
    \item \textbf{Cognitive Psychology} \cite{piaget1952origins,bruner2009process,bandura1986social}
    \begin{itemize}
        \item \textbf{Introduction}: 
    Cognitive theory, rooted in Gestalt psychology, focuses on information processing and mental processes.
    Like constructivism, it sees learning as an active process involving interactions with the environment and cognitive functions such as perception and memory.
    Emphasizing problem-solving, it highlights using cognitive structures like schemas for adequate comprehension.
    The theory acknowledges diverse cognitive styles, influencing learning styles, subject interests, and mastery abilities.
        \item \textbf{Contribution}: Guides the design of personalized learning paths and strategies, addressing individual cognitive needs.
        \item \textbf{Application}: Used in adaptive learning systems that adjust content based on learners’ cognitive profiles and performance.
    \end{itemize}
    \item [(3)] \textbf{Sociocultural Perspectives}: 
    \item \textbf{Sociocultural Theory} \cite{vygotsky1978mind,vygotsky2011interaction}
    \begin{itemize}
        \item \textbf{Introduction:} The theory, proposed by Soviet psychologist Lev Vygotsky, presents a theoretical framework that underscores the pivotal influence of sociocultural factors on human cognitive development.
        It posits that mental functioning is shaped through cultural mediation, primarily involving language.
        The theory introduces the Zone of Proximal Development (ZPD), stressing the importance of collaborative learning.
        Social interactions play a crucial role in individual cognition and learning.
        \item \textbf{Contribution}: Supports the implementation of collaborative learning activities in personalized learning.
        \item \textbf{Application}: Implemented through social learning platforms that facilitate peer interaction.
    \end{itemize}
\end{itemize}

\begin{table*}[t!]
\centering
\footnotesize
\renewcommand{\arraystretch}{0.9} 
\setlength{\tabcolsep}{3pt}       
\caption{Educational theories as foundations for technological modeling}
\label{tab_edu_theo}
\begin{adjustbox}{width=\textwidth}
\rowcolors{2}{gray!20}{white} 
\begin{tabular}{p{0.1\linewidth}p{0.3\linewidth}p{0.3\linewidth}p{0.35\linewidth}}
\toprule
\textbf{Educational Theory} & \textbf{Core Idea} & \textbf{Key Technical Points} & \textbf{Educational Implications} \\
\midrule
    Progressivism &  
    Emphasizing individual differences, learner-centered, and problem-based learning. & Guiding system design toward adaptive, interest-driven, and project-based personalization. & 
    Supporting learning path recommendation (Sec. \ref{sec6_1}) and course recommendation (Sec. \ref{sec6_2}). \\ 
    
    Behaviorist Theory &  
    Viewing learning as a process of stimulus–response reinforcement; emphasizing external feedback and behavior shaping through environmental control. & 
    Supporting behavior-driven modeling and reinforcement-based feedback mechanisms in learning systems. & 
    Predicting behavioral sequences in student behavior analysis (Sec. \ref{sec5_2_3}); providing the theoretical foundation for RL-driven exercise recommendation and adaptive feedback systems (Sec. \ref{sec6_3}). \\ 
    
    Constructivism &  
    Arguing that learners construct knowledge through active engagement, exploration, and collaboration. & 
    Inspiring collaborative and self-regulated modeling; emphasizing learning context and interaction patterns in model design. & 
    Modeling state transitions in cognitive diagnosis (Sec. \ref{sec5_1_1}); constructing processes in student performance prediction (Sec. \ref{sec5_1_2}); supporting knowledge graph–based recommendation (Sec. \ref{sec6_1}).  \\ 
    
    Cognitive Psychology & 
    Focusing on internal cognitive processes such as perception, attention, and memory; highlighting schema-based information processing and cognitive diversity. & 
    Modeling cognitive features; estimating ability parameters. & 
    Estimating ability parameters in cognitive diagnosis (Sec. \ref{sec5_1_1}); capturing students' knowledge states and cognitive processes in student performance prediction (Sec. \ref{sec5_1_2}); matching based on individual knowledge proficiency (Sec. \ref{sec6_3}). \\ 
    
    Sociocultural Theory &  
    Emphasizing the role of social interaction, language, and culture in learning; introducing the concept of ZPD. & 
    Modeling peer collaboration features; providing a foundation for recommendation mechanisms that incorporate social context and collaboration. & 
    Applying graph-based or community-aware recommendation models in course and exercise recommendation (Sec.\ref{sec6_2} and Sec. \ref{sec6_3}). \\ 
\bottomrule
\end{tabular}%
\end{adjustbox}
\end{table*}

In conclusion, progressivism advocates for fostering autonomous learning, behaviorist theory offers valuable insights into designing effective feedback mechanisms, constructivism underscores the importance of active knowledge construction, cognitive theory focuses on addressing learners' cognitive needs, and sociocultural theory emphasizes the critical role of collaboration and social context in student development.
Despite the inherent limitations of each theoretical framework, personalized learning adopts an integrative approach that synthesizes these perspectives to create a holistic educational paradigm.
By leveraging the foundational insights provided by these educational theories, personalized learning enables the design of flexible learning environments tailored to individual differences.
The integration of these theoretical perspectives not only deepens the understanding of students' diverse needs but also drives the development and continuous innovation of personalized learning within the dynamic landscape of modern education.

\subsection{Educational Theories for Technological Modeling}
Educational theories not only establish the fundamental principles of personalized learning but also provide direct design foundations and interpretive frameworks for subsequent student modeling and recommendation strategies.
Specifically, these theories guide the selection of core modeling variables (e.g., prior knowledge, emotional states), inform the underlying design logic of computational frameworks, and ensure the pedagogical validity of recommendation strategies.
Table \ref{tab_edu_theo} clearly illustrates the supporting relationships between key educational theories and corresponding technical approaches.
Overall, these theories function as a critical bridge connecting pedagogical philosophy with technological realization, thereby establishing a robust theoretical foundation for the technical explorations presented in the subsequent sections.

\section{Student Modeling}
\label{sec5}
Effective personalized learning relies on two fundamental components: student modeling and personalized recommendations.
Student modeling aims to construct a comprehensive representation of the learner by capturing both cognitive and non-cognitive factors, providing a foundation for adaptive learning.
Cognitive modeling focuses on inferring students’ knowledge states through techniques like cognitive diagnosis, enabling adaptive instructional content for deeper comprehension.
In contrast, non-cognitive modeling considers learning styles, behaviors, and emotional states to provide a broader perspective on individual learning needs. 
By integrating these dimensions, student modeling facilitates a nuanced learner profile that informs personalized recommendations.
Based on student modeling, personalized recommendations improve learning outcomes by selecting appropriate resources, exercises, and feedback strategies based on inferred learner characteristics.
This synergy between modeling and recommendation enhances learning outcomes by ensuring that instructional content is tailored to student's unique skills and challenges.

The following sections will provide an in-depth analysis and summary of the relevant literature on cognitive modeling, non-cognitive modeling, and personalized recommendations, exploring the advantages, challenges, and interactions among these methods, as well as their synergistic effects.

\subsection{Cognitive Modeling}
\label{sec5_1}
Cognitive modeling refers to the use of mathematical, computational, or cognitive theory methods to simulate an individual's cognitive processes in order to understand and predict his or her learning, reasoning, and other behaviors \cite{anderson2009can,vanlehn2006behavior}.
In the field of personalized learning, cognitive modeling is mainly used to portray students' knowledge states, learning behaviors, and ability levels to support personalized learning recommendations and instructional interventions.
Specifically, cognitive modeling focuses on key factors such as students' knowledge acquisition, academic performance, and cognitive abilities, which are essential for accurately assessing students' learning progress and optimizing instructional strategies.
As shown in Figure \ref{fig:cd_class}, this chapter will systematically explore the research progress of cognitive modeling around two core aspects: cognitive diagnosis and student performance prediction.
\subsubsection{\textbf{Cognitive Diagnosis}}
\label{sec5_1_1}

Rooted in Constructivism and Cognitive Psychology, cognitive diagnosis conceptualizes learning as an active process in which individuals construct and refine internal knowledge representations.
These theories posit that knowledge mastery can be decomposed into latent cognitive attributes, with observable responses serving as indirect evidence of their mastery levels.
Building on this theoretical foundation, cognitive diagnosis models (e.g., Item Response Theory (IRT), Deterministic Inputs, Noisy “And" gate model (DINA)) operationalize the estimation of these latent abilities by linking response patterns to underlying cognitive skills.

As a core component of personalized learning, cognitive diagnosis provides fine-grained assessments of students’ knowledge states.
The diagnostic results provide educators with valuable insights to customize downstream personalized learning materials and assignments, including course or question recommendations.
In this section, we will delve into cognitive diagnosis for personalized learning, covering its task formulation, background, and a literature review of recent valuable works through a fine-grained taxonomy.

\textbf{Task Formulation.}
\begin{figure}
    \centering
    \includegraphics[width=1.0\linewidth]{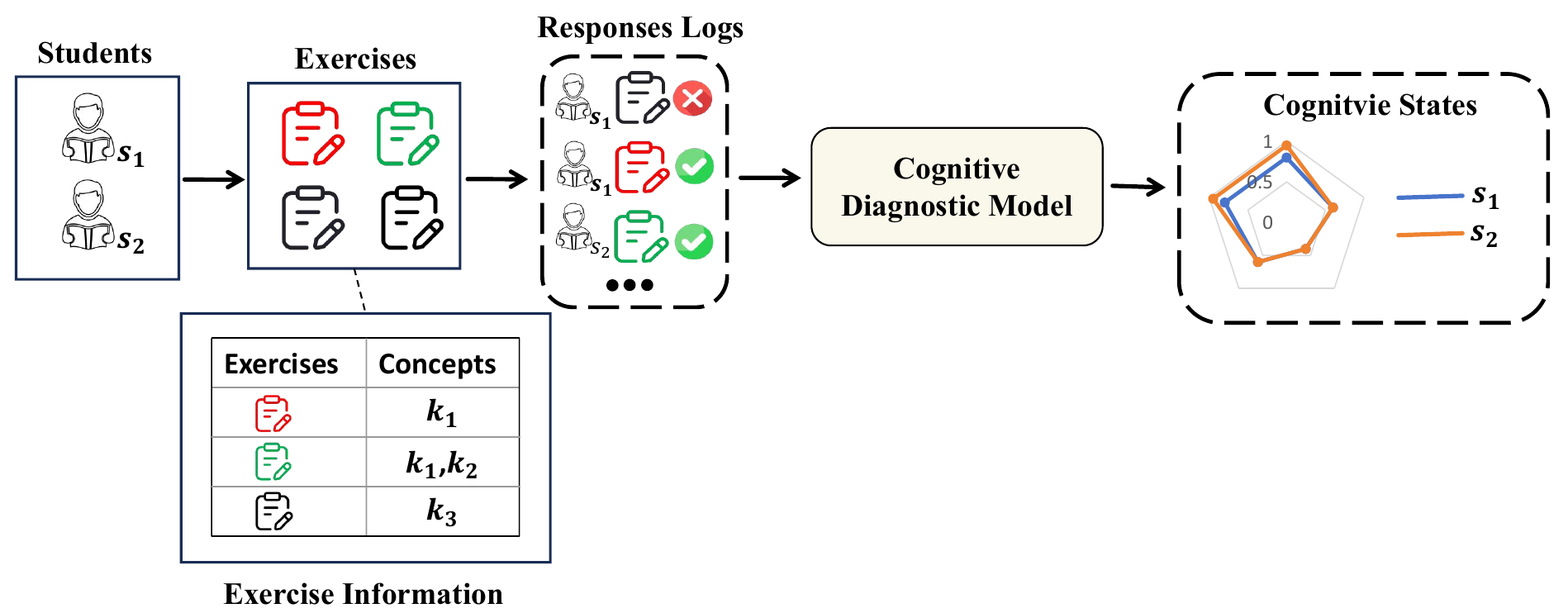}
    \caption{The essence of cognitive diagnosis}
    \label{fig:cog_dia_formu}
\end{figure}
Advancements in information technology provide educators with abundant student learning data for cognitive diagnosis, aiming to estimate proficiency in specific knowledge concepts. 
As shown in Figure \ref{fig:cog_dia_formu}, this task aims to estimate students' proficiency in a specific knowledge concept by giving them a series of test questions with their responses. 
Consider a student set $\mathcal{U}$ and a test question set $\mathcal{P}$. 
Let $r^u_i=(q^u_i, a^u_i)$ represent the $i^{\text{th}}$ response of student $u\in\mathcal{U}$. 
This response comprises the answered question $q_i^u$ and binary correctness indicator $a^u_i\in\{0, 1\}$, where $a^u_{i}=1$ denotes a correct response.
Given the historical responses of a student as a set $\mathcal{H}^u=\{r^u_{1},r^u_{2},\cdots, r^u_{|\mathcal{H}^u|}\}$, the Cognitive Diagnosis Model (CDM) captures their latent proficiency $\theta_u$ regarding the target concept. 
CDMs are designed to estimate the probability of correct responses ($p(a^u_i = 1|\theta_u, q^u_i)$) without explicit proficiency annotations, learned by maximizing the likelihood of observed responses:
\begin{equation}
\label{eq:cd_MLE}
\max\prod_{u\in\mathcal{U}}\prod_{r_i^u\in\mathcal{H}^u}p(a^u_i = 1|\theta_u, q^u_i).
\end{equation}


With the proliferation of personalized learning data, the Cognitive Diagnosis (CD) task has evolved to assess proficiency across multiple knowledge concepts.
This multi-dimensional scenario introduces a core challenge: how to formally represent the relationship between test questions and the multiple concepts they assess.
This challenge is addressed by a fundamental construct in CDMs known as the Q-matrix.
The Q-matrix, denoted as $\textbf{Q}\in\mathbb{R}^{|\mathcal{P}|\times|\mathcal{C}|}$ where $\mathcal{C}$ is the set of knowledge concepts, is a binary matrix that encodes the designer's hypothesis about which concepts are required to solve each question.
Its element $Q_{pc}=1$ if question $p$ requires concept $c$, and $0$ otherwise.
The critical assumption here is that the Q-matrix accurately reflects the cognitive attributes necessary for solving each item.
Given the Q-matrix, a student's proficiency is represented as a multi-dimensional vector $\boldsymbol{\theta}_u\in\mathbb{R}^{|\mathcal{C}|}$.
The CDM's objective is then to estimate $\boldsymbol{\theta}u$ by modeling the probability of a correct response conditioned not only on the student's proficiency and the question but also on the question's required concepts specified by $\textbf{Q}$. 
Thus, we optimize the following revised objective:

\tikzset{
    Cognitive-model/.style={
    minimum height=1.5em,
    draw=gray!70, 
    fill=gray!15,  
    text=black, font=\normalsize,
    inner xsep=2pt,
    inner ysep=4pt,
    line width=0.8pt,
    }
}
\tikzset{
    Data-driven/.style={
        minimum height=1.5em,
        draw=red!80,
        fill=white!30,
        text=black, font=\normalsize,
        inner xsep=2pt,
        inner ysep=4pt,
        line width=0.8pt,
    }
}
\tikzset{
    Cognit-diag/.style={
        minimum height=1.5em,
        draw=red!80,
        fill=white!30,
        text=black, font=\normalsize,
        inner xsep=2pt,
        inner ysep=4pt,
        line width=0.8pt,
    }
}
\tikzset{
    Pedag-theory-driven/.style={
        minimum height=1.5em,
        draw=red!80,
        fill=white!30,
        text=black, font=\normalsize,
        inner xsep=2pt,
        inner ysep=4pt,
        line width=0.8pt,
    }
}
\tikzset{
    Stu-perfor/.style={
        minimum height=1.5em,
        draw=cyan!80,
        fill=white!30,
        text=black, font=\normalsize,
        inner xsep=2pt,
        inner ysep=4pt,
        line width=0.8pt,
    }
}


\tikzset{
    Pedagogical-driven/.style={
        minimum height=1.5em,
         draw=green!80,
        fill=white!30,
        text=black, font=\normalsize,
        inner xsep=2pt,
        inner ysep=4pt,
        line width=0.8pt,
    }
}
\tikzset{
    ACC-Data/.style={
        minimum height=1.5em,
        draw=red!70,
        fill=red!15,
        text=black, font=\normalsize,
        inner xsep=2pt,
        inner ysep=4pt,
        line width=0.8pt,
    }
}

\tikzset{
    ACC-Peda/.style={
        minimum height=1.5em,
        draw=red!70,
        fill=red!15,
        text=black, font=\normalsize,
        inner xsep=2pt,
        inner ysep=4pt,
        line width=0.8pt,
    }
}
\tikzset{
    ACC-Ped/.style={
        minimum height=1.5em,
        draw=ACC-Ped-Co!80,
        fill=ACC-Ped-Co!30,
        text=black, font=\normalsize,
        inner xsep=2pt,
        inner ysep=4pt,
        line width=0.8pt,
    }
}
\tikzset{
    Int-Data/.style={
        minimum height=1.5em,
        draw=Int-Data-Co!80,
        fill=Int-Data-Co!30,
        text=black, font=\normalsize,
        inner xsep=2pt,
        inner ysep=4pt,
        line width=0.8pt,
    }
}
\tikzset{
    Int-Ped/.style={
        minimum height=1.5em,
        draw=cyan!70,
        fill=cyan!15,
        text=black, font=\normalsize,
        inner xsep=2pt,
        inner ysep=4pt,
        line width=0.8pt,
    }
}
\begin{equation}
\label{eq:cd_mult_MLE}
\max\prod_{u\in\mathcal{U}}\prod_{r_i^u\in\mathcal{H}^u}p(a^u_i = 1|\boldsymbol{\theta}_u, q^u_i, \textbf{Q}).
\end{equation}

\begin{figure}
    \centering
    \includegraphics[width=1.0\linewidth]{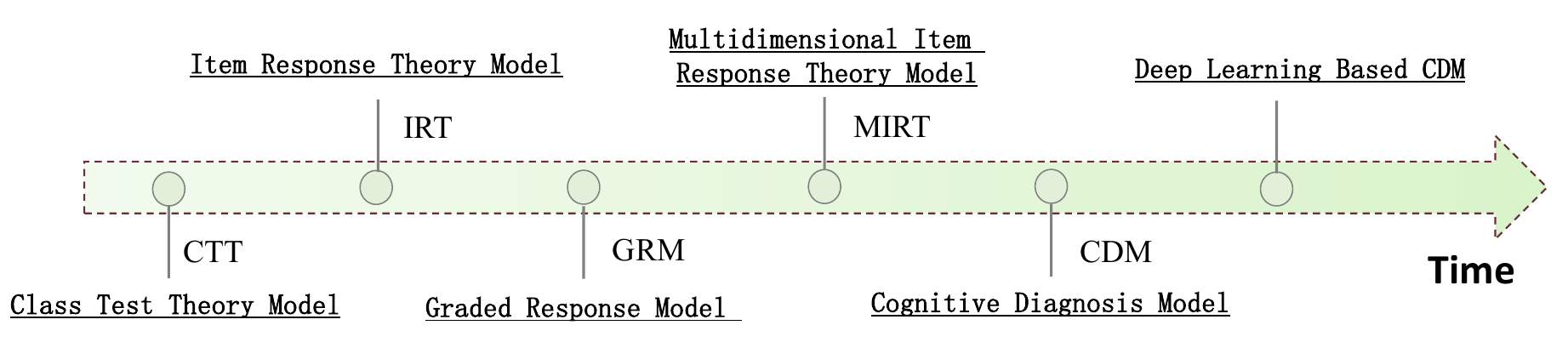}
    \caption{The development of cognitive diagnosis models}
    \label{fig:cog_diag_devel}
\end{figure}

\textbf{Background.}
The development of the cognitive diagnostic model is shown in Figure \ref{fig:cog_diag_devel}.
As a means of assessing psychological attributes, researchers historically employed psychometric methods for the CD task.
One prevalent method, Classical Test Theory (CTT) \cite{traub1997classical}, posits that an observed score $X$ comprises a true score $T$ (representing knowledge mastery, i.e., $\theta$) and random error $E$:
\begin{equation}
\label{eq:cd_ctt}
X = T + E.
\end{equation}
Despite its simplicity, CTT has limitations; for instance, it does not explicitly model key question attributes like difficulty, nor does it naturally handle multiple proficiency dimensions.
Item Response Theory (IRT) \cite{embretson2013item} provides a more powerful framework by directly modeling the functional relationship between a student's single latent proficiency $\theta_u$ and their probability of correctly answering a question $q^u_i$:
\begin{equation}
\label{eq:cd_irt}
p(a^u_i = 1|\theta_u, q^u_i) = c_i + \frac{1 - c_i}{1+e^{-a_i(\theta_u-b_i)}}.
\end{equation}
Here, $c_i$, $a_i$, and $b_i$ are the item parameters representing the question's guessability, discrimination, and difficulty, respectively. 
The foundational IRT model was later extended to multidimensional settings.
Multidimensional Item Response Theory (MIRT) \cite{reckase2009multidimensional} models proficiency as a vector $\boldsymbol{\theta}_u$ and defines the response probability as a function of a linear combination of the proficiencies, often as $p(a^u_i = 1|\boldsymbol{\theta}_u, q^u_i) = f(a_i, b_i, \boldsymbol{\theta}_u, \textbf{w}_i)$, where $\textbf{w}_i$ is a vector weighting the relevance of each concept to question $i$.
The Q-matrix can be viewed as a logical and binary precursor or special case of these weight vectors.
Building on MIRT, more complex models, such as the Graded Response Model (GRM), were developed to handle polytomous (multi-level) responses.

Later, as neural network technology gained prominence, the effectiveness of CD tasks in predicting student outcomes achieved notable levels.
Nevertheless, this progress also resulted in a reduction in interpretability \cite{liang2021explaining}.
Researchers are currently concentrating on the challenge of preserving high accuracy in forecasting student performance while simultaneously enhancing the interpretability of the models.

\begin{figure*}[!th]
    \centering
    \resizebox{0.9\textwidth}{!}{
\begin{forest}
    for tree={
        grow=east,
        reversed=true,
        anchor=base west,
        parent anchor=east,
        child anchor=west,
        base=left,
        font=\normalsize,
        rectangle,
        rounded corners,
        align=left,
        minimum width=1em,
        edge+={darkgray, line width=1pt},
        s sep=3pt,
        inner xsep=0pt,
        inner ysep=3pt,
        line width=0.8pt,
        ver/.style={rotate=90, child anchor=north, parent anchor=south, anchor=center},
    }, 
    [
        Cognitive Modeling, Cognitive-model
        [
            Cognitive Diagnosis\\(\S\ref{sec5_1_1}), Cognit-diag, text width=8em
            [
                Pedagogical theory-driven, Pedag-theory-driven, text width=8.5em
                [ 
                    \parbox{10em}{\centering
                        \cite{embretson2013item,LiuWCXSCH18,MaHTZZL23,Zhou0WWHT0CM21, guo2025multidimensional,wang2024unified,de2009dina}
                    }, ACC-Peda, text width=8em
                ]
            ]
            [ 
                Data-driven, Data-driven, text width=8.5em
                [ 
                    Deep-learning, Data-driven, text width=6em
                    [
                        \parbox{6em}{\centering
                            \cite{Gao0HYBWM0021,MaLWZC0Z22,WangHCC21,0008ZY023,SongHSYLYL23,GaoWLWLYZL023,YangQLGRZW22,LiGFX0HL22,YaoLHTHCS023,SuCWDHWCWX22,PeiYHX22,LiuYMWQ023,shen2023symbolic,huang2024interpretable,gao2024zero,ma2024enhancing,li2024towards,yu2024rdgt,liu2024inductive,ma2024dgcd,qian2024orcdf}
                        }, ACC-Data, text width=5em
                    ]
                ]
                [
                    Non-deep-learning, Data-driven, text width=6em
                    [
                        \parbox{6em}{\centering
                            \cite{lin2017adaptive,ZhuLHCLSH18,LiuQLZ23,BiCH0Z0W23,LiW00HHC0022,Tong0YHHPJ21,zhang2024path}
                        }, ACC-Data, text width=5em
                    ]
                ]
            ]
            [
                Integration methods, Data-driven, text width=8.5em
                [
                    \parbox{10em}{\centering
                        \cite{ChengLCHHCMH19,HuangLWHFWC0021,WangLCHCYHW20,WangLCHYWS23, zhou2023causality,zhang2024understanding,wang2024boosting}
                    }, ACC-Data, text width=8em
                ]
            ]
        ]
        [
            Student Performance \\Prediction (Cognitive)\\(\S\ref{sec5_1_2}), Stu-perfor, text width=8em
            [
                \parbox{10em}{\centering
                    \cite{li2025interpretable,su2024global,wang2023multivariate,MaHTZZL23}
                }, Int-Ped, text width=8.5em
            ]
        ]
    ]
\end{forest}
    }
    \caption{A taxonomy of cognitive modeling}
    \label{fig:cd_class}
\end{figure*}

\textbf{Taxonomy.}
In existing research, scholars have not reached a complete consensus on the classification of cognitive diagnosis models.
Early reviews classified models from psychometric and cognitive-theoretical perspectives, emphasizing underlying cognitive assumptions and parameter interpretability~\cite{dibello200631a}.
More recently, Liu et al.~\cite{liu2023new} distinguished probabilistic cognitive diagnosis models from those based on deep learning, while Wang et al.~\cite{wang2024survey} proposed a taxonomy encompassing both psychometric and machine learning approaches.
This evolution from traditional methods to machine learning and then to deep learning highlights the technological progression of cognitive diagnosis methodologies and the field’s broader shift from theory-driven to data-driven paradigms.
By drawing on the general taxonomy of educational technology research methodologies (data-driven and theory-driven approaches), a complementary perspective is proposed \cite{romero2010educational}.
Representative cognitive diagnosis studies are categorized into theory-driven, data-driven, and integration methods, as illustrated in Figure~\ref{fig:cd_class}.
Theory-driven approaches rely on cognitive psychology or educational theory to ensure the theoretical interpretability of students’ knowledge states, whereas data-driven approaches leverage large-scale analytics and machine learning to infer learning capabilities from interaction data.
Integration methods integrate both perspectives, for instance, by embedding cognitive theory as structural constraints within data-driven architectures.
Compared with previous classification systems, the proposed framework emphasizes the interaction between model-driven mechanisms and cognitive interpretability.
It retains the theoretical coherence of traditional classifications while reflecting emerging trends that integrate deep learning with cognitive modeling, thereby providing a unified and systematic perspective on the evolution of cognitive diagnostic models.
Appendix Table~\ref{cog_diag_sum} summarizes representative studies under this framework, and the following is a detailed description of the listed research work.

\begin{itemize}
\item[(1)] \textbf{Pedagogical theory-driven methods:}
The pedagogical theory-driven methods in cognitive diagnosis integrate educational theories and psychometric principles to model students' cognitive states, focusing on teaching and learning processes, instructional design, and cognitive assessment.
By leveraging well-established theories, this approach enhances both the accuracy and interpretability of cognitive diagnosis models.
A key aspect of this paradigm is the estimation of students' overall abilities and cognitive profiles based on their interactions with learning materials. 
For instance, the \textbf{monotonicity assumption}, a fundamental principle in instructional theory, posits that as a student's mastery of a knowledge concept improves, their probability of correctly answering related questions should increase. 
This assumption underpins classic cognitive diagnosis frameworks such as the DINA \cite{de2009dina} and IRT \cite{embretson2013item}, ensuring that cognitive assessments align with established learning progression theories. 

Beyond the monotonicity assumption, \textbf{other pedagogical theories} have been incorporated into cognitive diagnosis to refine knowledge assessment.
For instance, Ma et al. \cite{MaHTZZL23} applied the Neutral Set (NS) theory, which comprehensively assesses students' cognitive states by categorizing their understanding into three dimensions: understanding, misunderstanding, and uncertainty.
Liu et al. \cite{LiuWCXSCH18} combined fuzzy set theory and educational assumptions to model test takers and introduced sliding and guessing factors into the mix.
Further extending theory-driven approaches, Zhou et al. \cite{Zhou0WWHT0CM21} proposed a situationally aware cognitive diagnostic framework, which integrates educational context characteristics into the diagnostic process. 
Recently, Wang et al. \cite{wang2024unified} focused on the underexplored issue of uncertainty in cognitive diagnosis and proposed a novel parameterization technique based on traditional psychometrics.
This technique can be adapted to various cognitive diagnostic models across different domains, enhancing their robustness and applicability.
Meanwhile, Guo et al. \cite{guo2025multidimensional} also addressed uncertainty in cognitive diagnosis and further introduced an enhanced framework to tackle two major limitations: single-dimensional modeling and data sparsity.
They developed a multidimensional latent cognitive diagnosis model that integrates Rough Concept Analysis (RCA) and IRT.
This model effectively captures multiple cognitive dimensions simultaneously, enabling a more accurate and comprehensive assessment of students' cognitive states.

\item[(2)] \textbf{Data-driven methods:} 
The rise of information technology has provided a wealth of data for cognitive diagnosis. 
Initiatives are examining the challenges in cognitive diagnostic scenarios related to data and proposing effective data-driven approaches.
These studies, which rely on large data sets, use techniques such as machine learning and data mining to extract patterns from learner behavior, performance, and responses to facilitate cognitive diagnosis.
As shown in Figure \ref{fig:cd_class}, we have a rough categorization of existing work.

\textbf{Deep-learning methods.}
Deep learning has emerged as a prominent and rapidly evolving research direction in cognitive diagnosis, offering powerful data processing capabilities and flexible network architectures.
This section categorizes existing deep learning-based approaches based on model architectures and problem-solving strategies, providing a systematic review of recent advancements.
    
\textbf{Data sparsity and the cold-start problem} remain critical challenges in cognitive diagnosis.
To address these issues, recent studies have explored various strategies. 
For instance, EIRS \cite{YaoLHTHCS023} and CMES \cite{ma2024enhancing} improved diagnostic accuracy by designing rational sampling strategies and incorporating un-interacted exercises as auxiliary training signals. 
Gao et al. \cite{gao2024zero} introduced the DZCD framework, which employs a dual regularizer to train the diagnostic model and categorizes student states into shared and specific components to enhance prior knowledge utilization. 
Additionally, Liu et al. \cite{liu2024inductive} developed an inductive cognitive diagnostic model that significantly enhances cold-start performance by constructing a bipartite graph, modeling student proficiency using a Graph Convolutional Network (GCN), and inferring knowledge states through neighborhood information aggregation.
    
Beyond individual cognitive diagnosis, \textbf{group-level cognitive diagnosis} has gained growing research attention.
RDGT \cite{yu2024rdgt} integrates transformer-enhanced Graph Neural Networks (GNNs) to effectively capture both inter and intra-group relationships. 
DGCD \cite{ma2024dgcd} introduces adaptive denoising and entropy-weighted balancing to refine information aggregation within group-student-exercise structures. 
Meanwhile, HomoGCD \cite{LiuYMWQ023} employs a multi-granularity framework to jointly model cognitive states at both individual and group levels, enhancing diagnostic stability.
    
Beyond group-level modeling, substantial progress has been made in utilizing GNNs to \textbf{capture complex relationships among students, exercises, and knowledge concepts}, enhancing cognitive diagnosis \cite{Gao0HYBWM0021}.
Existing studies primarily focus on modeling these intricate relationships to gain a more comprehensive understanding of students' cognitive states \cite{SongHSYLYL23,GaoWLWLYZL023,YangQLGRZW22,LiGFX0HL22,MaLWZC0Z22,WangHCC21}. 
These approaches incorporate key factors such as knowledge structure, skill levels, and prerequisite dependencies to refine cognitive modeling.
    
In addition to advancements in cognitive modeling, \textbf{interpretability} has become a crucial research focus in cognitive diagnosis.
Shen et al. \cite{shen2023symbolic} proposed a symbolic tree-based approach that alternates between optimal symbolic representations and parameter mixing, effectively bridging the gap between discrete symbolic representations and continuous parameter optimization while improving both generalization and interpretability.
Similarly, Huang et al. \cite{huang2024interpretable} introduced a multi-dimensional feature extraction framework that integrates a multi-channel attention mechanism to capture complex student-exercise interactions, enhancing both model expressiveness and interpretability.
    
\textbf{Non-deep-learning methods.}
A variety of non-deep learning approaches have been explored in cognitive diagnosis, leveraging statistical modeling \cite{BiCH0Z0W23,LiW00HHC0022}, causal inference \cite{LiuQLZ23,zhang2024path}, and optimization techniques \cite{lin2017adaptive} to accurately assess students' knowledge states.
These methods aim to enhance diagnostic accuracy by explicitly modeling knowledge structures, incorporating prior knowledge, and improving assessment efficiency.
One major line of research focuses on \textbf{causal inference} to refine cognitive diagnosis. 
Liu et al. \cite{LiuQLZ23} incorporates Structural Causal Models (SCM) to capture causal relationships among students' mastery levels, enhancing the Q-matrix with an artificial Q-matrix as a prior.
This enables a comprehensive assessment of students' abilities by inferring relationships between exercises and explicit as well as latent knowledge attributes.
Causal reasoning was further extended through the proposal of a path-specific causal inference framework for fairness-aware cognitive diagnosis \cite{zhang2024path}.

Another critical research direction involves \textbf{Bayesian hierarchical modeling}.
Building on the Bayesian Network, prior knowledge and uncertainty are incorporated into Bayesian hierarchical modeling \cite{BiCH0Z0W23}.
Similarly, the Bayesian Network-based Hierarchical Cognitive Diagnostic Model (HierCDF) \cite{LiW00HHC0022} fuses traditional models (IRT, MIRT, Matrix Factorization [MF]) to capture students’ cognitive states.

In addition to probabilistic modeling, \textbf{optimization-based approaches} have been explored for improving diagnostic efficiency. 
An adaptive ant colony optimization algorithm (ACO-TC) to construct cognitive diagnostic tests \cite{lin2017adaptive}.
This method optimizes the selection of test items using historical pheromone-based search strategies and heuristic information derived from item discrimination indices.

\item [(3)] 
\textbf{Integration methods.}
Despite significant progress in both theory-driven and data-driven cognitive diagnosis, balancing interpretability and predictive accuracy remains a challenge.
Integrating theoretical insights with data-driven modeling offers a promising direction to enhance diagnostic effectiveness by leveraging both interpretability and flexibility.
A notable early effort in hybrid modeling is Deep Learning Enhanced Item Response Theory (DIRT) \cite{ChengLCHHCMH19}, which extends traditional IRT by incorporating deep learning techniques. 
Building on this foundation, NeuralCDM \cite{WangLCHCYHW20} introduces a more generalized framework that integrates neural networks to model complex student-exercise interactions while maintaining interpretability by applying monotonicity constraints.
The article was later extended by introducing variations based on content awareness and knowledge associations, focusing on the relationships between knowledge concepts \cite{WangLCHYWS23}.
Extending neural cognitive diagnosis to group-level assessments, MGCD \cite{HuangLWHFWC0021} employs a multi-task learning framework to jointly model individual and group-level cognitive states.
Beyond accuracy and interpretability, fairness is a critical concern in cognitive diagnosis. 
FairCD \cite{zhang2024understanding} addresses fairness issues by decomposing student abilities into bias-sensitive and fair components, using adversarial learning to mitigate the influence of sensitive attributes (e.g., socioeconomic status) on assessments.
Additionally, affective factors have been integrated into cognitive diagnosis, further expanding the scope of hybrid models \cite{wang2024boosting}.

\end{itemize}

This chapter has provided a comprehensive overview of cognitive diagnosis.
Our analysis reveals a predominant focus on data-driven methods to enhance accuracy, yet the educational implications of all approaches are significant.
Pedagogically, theory-driven methods are fundamentally centered on interpretability and trustworthiness, which are crucial for informing instructional decisions.
For instance, the DINA model's explicit parameterization of guessing and slipping probabilities helps teachers determine whether errors stem from fundamental misconceptions or accidental slips, thereby enabling targeted feedback.
Empirical studies demonstrate the practical impact of such approaches. 
Huang et al. \cite{huang2022multi} applied a DINA-based cognitive diagnosis in a high-school physics course on electromagnetic induction and designed multi-level remedial instruction based on the results. Their study confirmed significantly greater learning gains in the experimental group compared to the control. 
From a long-term perspective, assessments based on IRT or DINA can reveal individual knowledge structures and uncover specific conceptual gaps masked by aggregate scores, supporting differentiated instruction and mastery learning \cite{you2019feedback}.
Such diagnostic insights enable educators to design targeted exercises, ensuring students master core proficiency before tackling more complex topics.

In contrast, data-driven methods derive pedagogical value from uncovering complex learning patterns in large-scale behavioral data, thus enabling fine-grained diagnosis and personalized support.
For instance, GNN-based models analyze intricate relational networks among students, exercises and knowledge concepts to identify latent associations (e.g., between mastery of Concept A and common errors in Concept B) \cite{Gao0HYBWM0021}.
These insights lay an empirical foundation for instructional refinement.
Models such as EIRS \cite{YaoLHTHCS023}, which mitigate data sparsity, further boost student engagement and motivation by enabling reliable diagnosis for learners with sparse interaction records.

Integration-based approaches mark the maturation of cognitive diagnosis research by balancing predictive accuracy with interpretability and fairness, thereby enabling more trustworthy and comprehensive educational AI systems.
NeuralCDM \cite{WangLCHCYHW20}, for instance, leverages the expressive power of neural networks while incorporating monotonicity constraints to uphold fundamental educational principles. 
This enhances the trustworthiness of the method among both teachers and students. 
Meanwhile, FairCD \cite{zhang2024understanding} addresses fairness concerns to ensure that assessments are not biased by sensitive attributes, thereby fostering educational equity and enhancing learning motivation among disadvantaged student groups. 
Furthermore, recent models that incorporate affective factors extend beyond diagnosing what students know to understanding how they feel, paving the way for genuinely individualized and affect-sensitive learning environments \cite{wang2024boosting}.

In summary, although current research remains largely dominated by data-driven paradigms, practical educational applications demand equal attention to feasibility, transparency, and interpretability.
Future work should focus on improving the interpretability of deep learning models, mitigating data sparsity, and ensuring the fairness and robustness of cognitive diagnoses.
Furthermore, the deeper integration of domain knowledge and educational theories, the application of causal inference for informed decision-making, and the development of adaptive models capable of generalizing across diverse learning environments will be essential to advancing the field.

\subsubsection{\textbf{Student Performance Prediction (Cognitive)}}
\label{sec5_1_2}
Grounded in Constructivism and Cognitive Psychology, cognitive student performance prediction models conceptualize learning as an active process of information acquisition, organization, and schema adaptation.
Within this theoretical framework, student performance is not merely a behavioral outcome but a reflection of underlying cognitive processes such as attention, memory, and comprehension.
Consequently, predictive models aim to infer latent cognitive states from observable learning interactions, aligning model representations with the cognitive mechanisms articulated in these theories.
Student Performance Prediction (SPP) has received considerable attention due to its critical role in educational research and practice.
Accurate performance prediction enhances learning outcomes by tailoring instructional strategies to individual needs, addressing weaknesses, and fostering potential abilities.
Moreover, these predictions enable early identification of at-risk students, allowing timely interventions to prevent dropouts.
Building upon a comprehensive literature review, existing SPP studies can be broadly categorized into two dimensions: cognitive \cite{su2024global,wang2023multivariate} and non-cognitive \cite{fan2024feature,khairy2024prediction,zhang2018grade}.
This review systematically examines related studies from these two dimensions, providing a detailed analysis of representative work within each category.
This section focuses on cognitive modeling approaches, which aim to capture students' knowledge states and cognitive processes through the analysis of assessment results, exercise interaction data, and learning trajectories.
Non-cognitive modeling approaches, which account for behavioral and emotional factors, are discussed in detail in section \ref{sec5_2_4}.

Cognitive modeling leverages CDMs to estimate students' mastery of various knowledge components, enabling the prediction of their future academic performance.
A significant advancement in this area is the PCDF model \cite{li2025interpretable}, which enhances CDMs by incorporating the Cumulative Category Response Function (CCRF) to process multistage rating data, improving compatibility with binary CDMs while preserving interpretability.
Building on this, GLNC \cite{su2024global} integrates cognitive diagnosis and knowledge tracing, adaptively combining local (current state) and global (learning trajectory) information for improved performance prediction.
This fusion approach provides a new way of thinking about student performance prediction.
Similarly, Ma et al. \cite{MaHTZZL23} introduced the innovative concept of the Neutral Set (NS), which combines cognitive diagnostics, NS similarity measures, and collaborative filtering with probability matrix decomposition.
This method comprehensively assesses students' cognitive states from three dimensions, enabling accurate prediction of future test scores in personalized e-learning environments.
While these models significantly advance cognitive modeling, they primarily focus on students' ability levels and overlook the influence of other factors.
To address this limitation, the MvRCF model \cite{wang2023multivariate} introduces a compensatory mechanism that integrates the student's ability profile and effort profile to further enhance the comprehensiveness of the prediction.

In summary, cognitive modeling approaches have achieved a more comprehensive understanding of student performance and have significantly advanced the field of student performance prediction. 
Beyond algorithmic progress, these methods hold profound educational significance. 
Accurate performance prediction enables the early identification of learning difficulties, facilitating timely feedback and targeted interventions that sustain engagement and confidence \cite{yaugci2022educational}.
Adaptive models that dynamically update students’ ability profiles can further enhance motivation by offering personalized and attainable learning trajectories.
Moreover, incorporating cognitive and emotional cues from multimodal data fosters more supportive and responsive learning experiences \cite{akccapinar2019using}.
In the long term, precise and adaptive prediction establishes the foundation for continuous, individualized learning cycles within students’ ZPD. 
Future research should therefore not only pursue methodological innovation but also emphasize how predictive models can be better integrated into teaching practice to promote equity, sustained motivation, and long-term learning success.

\subsection{Non-Cognitive Modeling}
\label{sec5_2}
\tikzstyle{my-box}=[
    rectangle,
    draw=hidden-draw,
    rounded corners,
    align=left,
    text opacity=1,
    minimum height=1.5em,
    minimum width=5em,
    inner sep=2pt,
    fill opacity=.8,
    line width=0.8pt,
]

\tikzstyle{leaf-head}=[my-box, minimum height=1.5em,
    draw=gray!80, 
    fill=gray!15,  
    text=black, font=\normalsize,
    inner xsep=2pt,
    inner ysep=4pt,
    line width=0.8pt,
]

\tikzstyle{leaf-style}=[my-box, minimum height=1.5em,
    draw=red!70, 
    fill=red!15,  
    text=black, font=\normalsize,
    inner xsep=2pt,
    inner ysep=4pt,
    line width=0.8pt,
]

\tikzstyle{leaf-sent}=[my-box, minimum height=1.5em,
    draw=cyan!70, 
    fill=cyan!15,  
    text=black, font=\normalsize,
    inner xsep=2pt,
    inner ysep=4pt,
    line width=0.8pt,
]
\tikzstyle{leaf-perach}=[my-box, minimum height=1.5em,
    draw=green!70, 
    fill=green!15,  
    text=black, font=\normalsize,
    inner xsep=2pt,
    inner ysep=4pt,
    line width=0.8pt,
]

\tikzstyle{leaf-beha}=[my-box, minimum height=1.5em,
    draw=orange!70, 
    fill=orange!15,  
    text=black, font=\normalsize,
    inner xsep=2pt,
    inner ysep=4pt,
    line width=0.8pt,
]

\tikzstyle{modelnode-style}=[my-box, minimum height=1.5em,
    draw=red!80, 
    fill=white!30,
    text=black, font=\normalsize,
    inner xsep=2pt,
    inner ysep=4pt,
    line width=0.8pt,
]

\tikzstyle{modelnode-sent}=[my-box, minimum height=1.5em,
    draw=cyan!100, 
    fill=white!30,  
    text=black, font=\normalsize,
    inner xsep=2pt,
    inner ysep=4pt,
    line width=0.8pt,
]
\tikzstyle{modelnode-beha}=[my-box, minimum height=1.5em,
    draw=orange!100, 
    fill=white!30,  
    text=black, font=\normalsize,
    inner xsep=2pt,
    inner ysep=4pt,
    line width=0.8pt,
]
\tikzstyle{modelnode-perach}=[my-box, minimum height=1.5em,
    draw=green!100, 
    fill=white!30,  
    text=black, font=\normalsize,
    inner xsep=2pt,
    inner ysep=4pt,
    line width=0.8pt,
]

\begin{figure}
    \centering
    \includegraphics[width=0.8\linewidth]{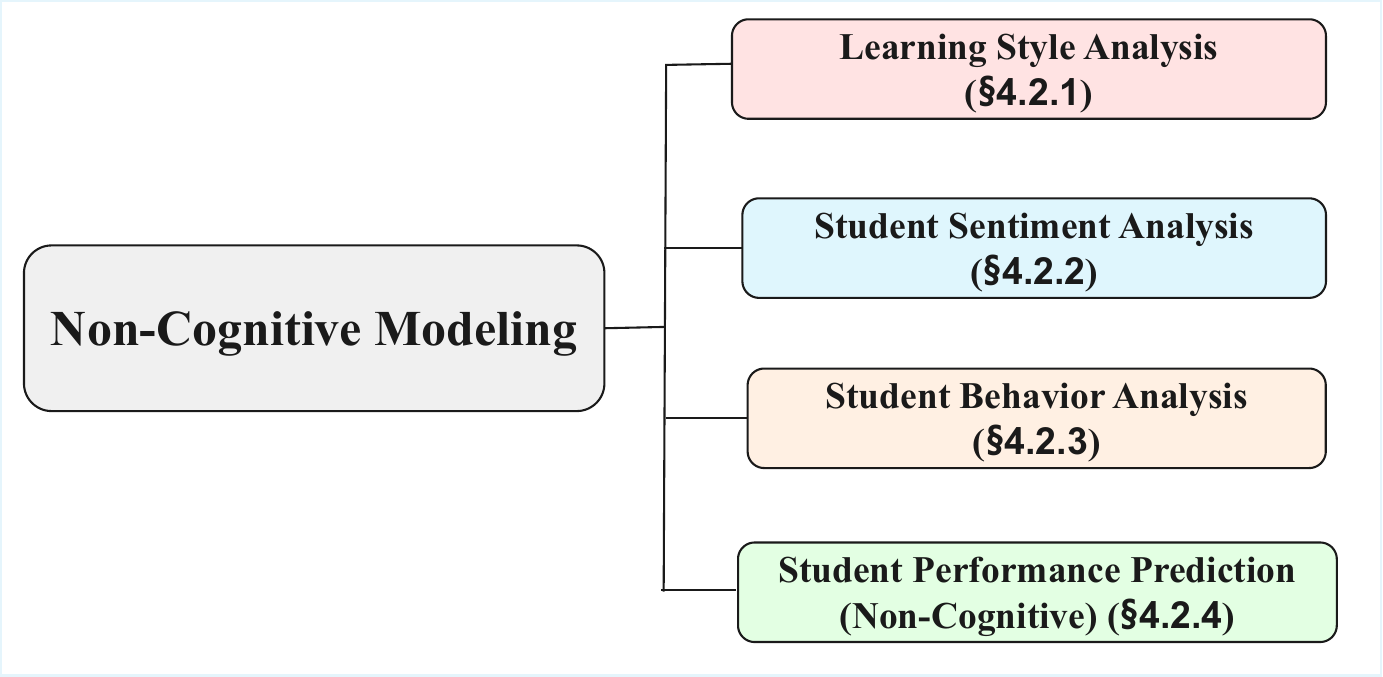}
    \caption{Non-cognitive modeling classification map}
    \label{fig:non_cog_class}
\end{figure}

Non-cognitive modeling aims to predict or optimize students' learning behaviors, academic performance, and overall learning experience by systematically collecting, analyzing, and modeling various data generated throughout the learning process.
This type of modeling focuses on factors such as learning styles, emotions, motivation, and behavioral patterns, which play a crucial role in the learning process. 
Unlike traditional cognitive modeling, which primarily focuses on students' knowledge mastery and cognitive abilities, non-cognitive modeling emphasizes exploring non-intellectual factors that influence the learning process.
It deepens understanding through a detailed analysis of students' learning states and quantitative analysis of the learning process. 
As shown in Figure \ref{fig:non_cog_class}, this study provides a comprehensive exploration of four main aspects of non-cognitive modeling: learning styles analysis, student sentiment analysis, student behavior analysis, and student performance prediction.

\subsubsection{\textbf{Learning Style Analysis}}
\label{sec5_2_1}
Learning style analysis systematically examines and categorizes learning styles to gain deeper insights into individual learner differences. 
A systematic analysis of learning styles enables personalized learning systems to precisely adjust teaching methods and provide customized resources, thereby enhancing knowledge mastery and learning efficiency. 
Building on this insight, the system can personalize teaching strategies and content based on learners' cognitive preferences and learning tendencies, ultimately optimizing the learning experience.

\begin{figure}[!th]
    \centering
    \resizebox{0.4\textwidth}{!}{
        \begin{forest}
            for tree={
                grow=east,
                reversed=true,
                anchor=base west,
                parent anchor=east,
                child anchor=west,
                base=left,
                font=\scriptsize, 
                rectangle,
                draw=none,
                rounded corners,
                align=left,
                minimum width=1em,
                edge+={darkgray, line width=0.8pt},
                s sep=4pt,
                inner xsep=0pt,
                inner ysep=2pt,
                line width=0.6pt,
            }, 
            [
                Learning Style \\Analysis \\(\S\ref{sec5_2_1}), modelnode-style, text width=5.5em
                [
                    FSLSM, modelnode-style, text width=4em
                    [\parbox{4em}{\centering 
                        \cite{anantharaman2018modelling,bernard2022improving,bernard2017learning,el2019combining,el2018integrating,gomede2020use,hidayat2021determine,hmedna2020predictive,hmedna2017identifying,kolekar2017prediction,rishard2022adaptivo,dominguez2025data,muhammad2024evolving,jebbari2024identifying,ayyoub2024learning}}, leaf-style]
                ]
                [
                    VARK, modelnode-style, text width=4em
                    [\parbox{5em}{\centering 
                        \cite{kuttattu2019analysing,nguyen2024model,abomelhaN24}}, leaf-style]
                ]
                [
                    MI Theory, modelnode-style, text width=4em
                    [\parbox{4em}{\centering 
                        \cite{rasheed2021learning}}, leaf-style]
                ]
                [
                    Kolb, modelnode-style, text width=4em
                    [\parbox{4em}{\centering 
                        \cite{ni2023design,hidalgo2024mapping}}, leaf-style]
                ]
            ]
        \end{forest}
    }
    \caption{Classification of studies on learning style analysis}
    \label{lear_style_class}
\end{figure}

Early reviews primarily focused on the diversity of learning style theories and the reliability and validity of their measurement instruments~\cite{coffield2004learning,pashler2008learning}.
More recent reviews have shifted toward data-analytic perspectives, highlighting the potential of data-driven methods for learning style identification \cite{clinton2024really}. 
Existing studies typically analyze learning styles by mapping learners’ behaviors, preferences, or cognitive characteristics onto predefined theoretical frameworks.
Accordingly, the present study adopts a classification framework grounded in learning style theories themselves, emphasizing the applicability of different theoretical models within adaptive learning contexts.
This perspective places greater emphasis on the adaptability and application trends of different learning style theories in model construction and personalized design, rather than being limited to validity verification or data method analysis.
As shown in Figure \ref{lear_style_class}, to systematically synthesize the literature, studies are categorized according to the learning style theories they adopt.
The Felder-Silverman Learning Style Model (FSLSM) is the most widely used, followed by the VARK model \cite{kuttattu2019analysing}, known for its simplicity.
Although Gardner’s Multiple Intelligences (MI) Theory and Kolb’s Learning Style Model are less frequently applied, they are still explored in some studies and are therefore included in this discussion.
We detail the research work based on different learning style theories in the following section, with Appendix Table \ref{lear_style_sum} summarizing their core characteristics for reference.

\begin{itemize}[leftmargin=*]
    \item [(1)] \textbf{FSLSM:} 
    Traditional learning style assessment relied on questionnaires, but their static nature and potential errors have driven a shift toward machine learning and data mining techniques.
    Early studies focused on fusing clustering techniques and machine learning to infer learning styles.
    For instance, mining techniques paired with the K-means algorithm captured learner behaviors \cite{el2018integrating}, while an enhanced K-means method, revising centroids via learning style vectors, was designed for more effective style determination \cite{hidayat2021determine}.

    Building on these foundational approaches, subsequent research has expanded the application of Machine Learning (ML) techniques to learning style classification.
    Studies such as~\cite{hmedna2020predictive,el2019combining,rasheed2021learning} employed data mining methods in conjunction with classical classification algorithms, including Decision Tree (DT), Random Forest (RF), Naïve Bayes (NB), and Support Vector Machine (SVM), to automate the identification of individual learning styles.
    Deep learning (DL) methods have also been explored in this context. 
    For instance, Anantharaman et al.~\cite{anantharaman2018modelling} integrated Convolutional Neural Networks (CNNs), Random Forests, and Long Short-Term Memory (LSTM) networks for learning style recognition. 
    Additionally, Kolekar et al. \cite{kolekar2017prediction} uses the Fuzzy Mean algorithm for FSLSM clustering and presents the Gravitational Search-Based Backpropagation Neural Network for classification.
    
    Recent studies have enhanced FSLSM-based learning style analysis with dynamic and adaptive approaches. 
    Muhammad et al. \cite{muhammad2024evolving} employed a bipartite graph with LSTM autoencoders and k-means clustering to capture evolving student-resource interactions. 
    In MOOCs, Jebbari et al. \cite{jebbari2024identifying} leveraged machine learning models, including neural networks and RF, for large-scale learning style classification. 
    Ayyoub et al. \cite{ayyoub2024learning} introduced a semi-supervised self-training SVM that iteratively expands its labeled dataset, improving classification under data scarcity.
    Beyond mere classification, the effectiveness of FSLSM-based instructional adaptations in improving learning outcomes has been evaluated \cite{dominguez2025data}.
    
    \item [(2)] \textbf{VARK:} In the literature, Kuttattu et al. \cite{kuttattu2019analysing} employed K-means, SVM, and DT methods to correlate learning styles with VARK styles. 
    The approach utilizes SVM for individual learning style predictions and decision trees for predicting combinations of learning styles.
    Nguyen et al. \cite{nguyen2024model} extended this work by integrating machine-learning-based VARK style recognition as an automated plugin into Moodle and generating responsive, personalized learning paths from predicted learning styles.
    Abomelha et al. \cite{abomelhaN24} developed an adaptive e-learning system that classifies students via VARK questionnaires and uses predictive analytics to recommend personalized learning resources.
    \item [(3)] \textbf{MI Theory:} 
    Research that integrates Multiple Intelligences Theory into learning style modeling remains limited. A notable study addresses this gap by proposing a neural network to identify learning styles based on both the FSLSM and MI Theory \cite{rasheed2021learning}.
    The study reveals that the accuracy of the approach varies across different dimensions of the combined theoretical framework.
    \item [(4)] \textbf{Kolb:}
    The SRGSML \cite{ni2023design}, a two-layer ensemble model for the recognition of Kolb learning style that uses resampling and hybrid labeling (rule-based classification + k-means) to address class imbalance and subjective bias.
    Hidalgo et al. \cite{hidalgo2024mapping} mapped Kolb’s learning styles to software development roles, highlighting their relevance in engineering education.
\end{itemize}

Learning style analysis is a critical component of personalized learning and has received increasing attention in recent research.
Machine learning and deep learning have significantly enhanced the precision and scalability of learning style identification, while dynamic and semi-supervised approaches have improved model adaptability in evolving learning contexts.
Beyond these methodological advances, the educational implications are substantial. 
For example, adaptive models that capture evolving styles ensure the learning experience remains aligned with the learner's developing profile, which is crucial for sustaining long-term engagement \cite{muhammad2024evolving}.
By identifying and responding to individual preferences, such as providing infographics for visual learners and designing interactive simulations for kinesthetic learners, we can create more intuitive learning environments.
In a quasi-experimental study of an adaptive e-learning system, the experimental group exhibited significantly higher levels of engagement across behavioral, cognitive, and affective dimensions than the control group \cite{el2021adaptive}.
In the long term, effective learning style analysis is profoundly significant for cultivating self-directed learning capabilities. 
When learners consistently engage with materials tailored to their unique traits, they not only grasp complex concepts more readily but also develop metacognitive awareness of their own learning processes.
Systematic reviews corroborate that responsiveness to learning-style differences enhances both system adaptability and the overall user experience \cite{oviedo2025exploring}.
Ultimately, aligning instructional strategies with students’ preferred learning patterns transforms the long-standing ideal of teaching in accordance with individual aptitude into evidence-based practice.

Despite these advances and the broad application prospects, certain challenges remain.
Many studies assume that learning styles are static, overlooking their potential evolution over time due to contextual and experiential factors. 
While machine learning has enhanced classification accuracy, most studies focus on testing, comparing, or combining existing models rather than developing fundamentally novel approaches.
Moreover, research often prioritizes algorithmic improvements over theoretical rigor and lacks interdisciplinary perspectives and age-diverse samples.
Future research should focus on developing more comprehensive learning style prediction models by integrating contextual factors and reinforcing theoretical foundations, thereby enhancing the validity and applicability of learning style analysis in personalized learning environments.

\subsubsection{\textbf{Student Sentiment Analysis}}

\begin{figure*}[!th]
\centering
\resizebox{0.9\textwidth}{!}{
    \begin{forest}
        for tree={
            grow=east,
            reversed=true,
            anchor=base west,
            parent anchor=east,
            child anchor=west,
            base=left,
            font=\normalsize,
            rectangle,
            draw=none,
            rounded corners,
            align=center,
            minimum width=1em,
            edge+={darkgray, line width=1pt},
            s sep=3pt, 
            inner xsep=0pt,
            inner ysep=2pt,
            line width=0.8pt,
        }, 
        [
            Student Sentiment \\ Analysis (\S\ref{sec5_2_2}), modelnode-sent, text width=7em
            [
                Technical \\ Methodology, modelnode-sent, text width=5em
                [
                    Machine learning, modelnode-sent, text width=8em
                    [\parbox{18em} {\centering \cite{chanaa2022sentiment,hew2020predicts, hixson2019reactions,jena2019sentiment,liu2016sentiment, kasumba2024practical,dake2023using}}, leaf-sent]
                ]
                [
                    Deep learning, modelnode-sent, text width=8em
                    [\parbox{18em} {\centering \cite{kastrati2020weakly,li2019shallow,sindhu2019aspect, yu2018improving,ren2023automatic,shaik2022educational}}, leaf-sent]
                ]
                [
                    Hybrid and advanced \\ techniques, modelnode-sent, text width=8em
                    [\parbox{18em} {\centering \cite{liu2021sentiment,mrhar2021bayesian,Barron-EstradaZ20,baqach2024new, ashwin2024identifying,shaikh2023exploring,liu2019temporal}}, leaf-sent]
                ]
            ]
            [
                Application\\ Domain, modelnode-sent, text width=5em
                [
                    Educational environ-\\ments, modelnode-sent, text width=8em
                    [\parbox{18em} {\centering \cite{onan2020mining,pong2019sentiment, rani2017sentiment,sindhu2019aspect,tubishat2023sentiment,yu2018improving,dake2023using}}, leaf-sent]
                ]
                [
                    MOOCs / Online \\ learning, modelnode-sent, text width=8em
                    [\parbox{18em} {\centering \cite{moreno2018learning,Barron-EstradaZ20,chen2022understanding,hew2020predicts,kastrati2020weakly,li2019shallow,baqach2024new,li2022key,chanaa2022sentiment}}, leaf-sent]
                ]
            ]
        ]
    \end{forest}
}

\caption{Classification of student sentiment analysis}
\label{sent_analy_class}
\end{figure*}
\label{sec5_2_2}
As an essential component of non-cognitive modeling, sentiment analysis captures and interprets learners' emotional states, providing crucial insights for precisely adjusting instructional strategies and optimizing learning experiences. 
Sentimental factors such as interest, motivation, and frustration significantly influence learners' learning processes and outcomes. 
This section systematically reviews 27 relevant studies (as shown in Figure \ref{sent_analy_class}) and introduces a dual-dimensional classification framework encompassing both technological approaches and application domains.
The technical dimension builds upon prior reviews~\cite{yadegaridehkordi2019affective} by grouping studies according to their underlying computational paradigms, ranging from traditional machine learning techniques to recent deep learning–based sentiment recognition approaches, thus ensuring both technological progression and methodological continuity.
The application dimension categorizes studies by their pedagogical contexts, such as classroom interaction and online learning, aligning emotional modeling techniques with their educational functions. 
The aim is to establish a comprehensive research framework to reveal the latest advancements and future directions in the field of sentiment analysis.
Appendix Table \ref{sent_ana_sum} shows a list of studies related to sentiment analysis.
The following is a detailed description of the research work in Appendix Table \ref{sent_ana_sum}.

\textbf{Technical methodology:}
We have broken down the references in this section into three main categories: traditional machine learning methods, deep learning methods, and studies combining deep learning with other techniques.
    \begin{itemize}
        \item \textbf{Traditional machine learning.} Traditional machine learning methods have been widely applied in sentiment analysis \cite{hew2020predicts,hixson2019reactions,jena2019sentiment,liu2016sentiment,chanaa2022sentiment,kasumba2024practical,dake2023using}. These studies employ various supervised algorithms, including SVM, NB, logistic regression, and RF \cite{hew2020predicts,chanaa2022sentiment}, often integrating them with multifactor analytics to achieve a comprehensive understanding of student sentiment.
        Technical approaches have included integrating NB and SVM with Hadoop to improve processing efficiency \cite{jena2019sentiment}, and employing particle swarm optimization for feature selection to enhance emotion recognition accuracy \cite{liu2016sentiment}. Furthermore, the analysis of qualitative student feedback has utilized multiple classifiers, including NB, SVM, J48 decision trees, and RF, to predict classroom labels and inform educational strategies \cite{dake2023using}. 
        Collectively, these studies highlight the effectiveness and versatility of traditional machine learning methods in student sentiment analysis, showcasing their potential to provide deeper insights into educational experiences.

        \item \textbf{Deep learning.} In recent years, deep learning methods have been widely used in student sentiment analysis \cite{sindhu2019aspect,yu2018improving,mrhar2021bayesian,li2019shallow,ren2023automatic} by employing CNN \cite{yu2018improving,kastrati2020weakly}, LSTM and its variants \cite{sindhu2019aspect,mrhar2021bayesian,shaik2022educational}, and Bidirectional Encoder Representation from Transformers (BERT) \cite{li2019shallow} and other models, which effectively enhance the accuracy and robustness of sentiment analysis.
        For example, Yu et al. \cite{yu2018improving} combined SVM with CNN to recognize the sentiment information in students' self-evaluations, thus improving the accuracy of students' performance prediction. 
        The LSTM model and its variants perform well in sentiment polarity determination and deep-level feature extraction, which effectively alleviates the uncertainty problem in sentiment analysis \cite{sindhu2019aspect,mrhar2021bayesian}.
        In particular, Shaik et al. \cite{shaik2022educational} used a Bi-directional Long Short-Term Memory (Bi-LSTM) deep learning model to construct a multi-label classification system for the higher education domain to assist in educational decision-making and course optimization.

        \item \textbf{Hybrid and advanced techniques.}
        Beyond traditional and deep learning methods, hybrid models and advanced techniques have been explored to improve sentiment analysis accuracy.  
        This includes integrating deep learning with Bayesian Neural Networks \cite{mrhar2021bayesian}, capsule networks, and attention mechanisms \cite{liu2021sentiment}. 
        Models like BERT-LSTM-CNN leverage multiple architectures for improved performance \cite{baqach2024new}.
        Ashwin et al. \cite{ashwin2024identifying} addressed bias in sentiment analysis by combining the High-Speed Sentiment Recognition Library (HSEmotion) with Multi-task Cascaded Convolutional Networks (MTCNN) to promote equitable educational practices.
        In addition, Shaikh et al. \cite{shaikh2023exploring} investigated Large Language Model (LLM) applications in sentiment analysis, using ChatGPT-3.5 to directly generate category labels and comparing its performance with LSTM- and Transformer-based models. 
        Their findings highlighted LLMs’ considerable potential to boost the accuracy and efficiency of student sentiment analysis.
        Other studies explored the combination of multiple techniques \cite{Barron-EstradaZ20,liu2019temporal,mrhar2021bayesian} and the use of BERT-based fusion models to capture contextual nuances and elevate prediction accuracy \cite{liu2021sentiment,li2019shallow}. 
        These innovations underscore the value of advanced integration strategies for optimizing sentiment analysis algorithms.
    \end{itemize}
     Categorizing existing studies by research methodology reveals the relative strengths of each technological approach: traditional machine learning serves as the foundational baseline, deep learning enhances the modeling of complex sentiment patterns, and the integration of advanced techniques further improves the overall performance of sentiment analysis algorithms.
     
\textbf{Application domain:} A review of existing literature reveals a prominent focus on applying sentiment analysis in educational settings, reflecting the diversity of learning environments and instructional methods.
These studies primarily explore its usage in conventional educational environments and MOOCs/Online learning.
    \begin{itemize}
        \item \textbf{Traditional educational environments. }
        Within this context, sentiment analysis is applied to various multi-channel data sources, including course information \cite{pong2019sentiment,sindhu2019aspect,yu2018improving}, tweets \cite{tubishat2023sentiment}, forums, and teacher evaluation websites \cite{onan2020mining}.
        Sentiment analysis methods for teaching evaluation examine teaching process feedback, providing valuable guidance to optimize teaching methodologies \cite{pong2019sentiment,rani2017sentiment,dake2023using}. 
        In social media, Tubishat et al. \cite{tubishat2023sentiment} explore ChatGPT’s educational applications, using a tweet sentiment analysis model to identify dominant sentiments and opinions toward ChatGPT.
        \item \textbf{MOOCs/Online learning.}
        The evolving landscape of learning styles has led to an increased focus on sentiment analysis in online learning environments \cite{Barron-EstradaZ20,hew2020predicts,chen2022understanding,kastrati2020weakly,li2019shallow}. 
        These studies leverage user-generated data to identify factors influencing learner satisfaction and evaluate overall perceptions of MOOCs \cite{hew2020predicts,kastrati2020weakly,li2019shallow,baqach2024new,li2022key}.
        For instance, forum activities provide insights into learners' social, emotional, and skills (3S) dimensions, as demonstrated by the 3S learning analytics approach proposed in \cite{moreno2018learning}, which introduces the visualization tool LAT$\exists$S for sentiment analysis through forum comments extraction.
        Natural Language Processing (NLP) and sentiment analysis techniques have been applied to Coursera reviews to explore key factors influencing MOOC success \cite{chen2022understanding}. 
    \end{itemize}

A comprehensive review of the existing literature on student sentiment analysis reveals a clear evolution from traditional machine learning to deep learning and hybrid approaches, greatly improving the ability to capture complex affective patterns.
These technical advancements enhance not only analytical accuracy and scalability but also hold profound implications for educational practice. 
By automatically detecting student emotions (such as confusion, frustration, or interest) from forum posts or feedback texts, these systems facilitate a shift away from a one-size-fits-all approach toward a more responsive, student-centered paradigm.
For instance, when widespread confusion is detected in MOOC discussion forums, instructors can intervene promptly with targeted clarifications \cite{ashwin2024identifying}, thereby sustaining engagement and preventing learner dropout.
In the long term, integrating affective insights into personalized learning systems facilitates the early identification of and proactive response to negative states such as anxiety and burnout \cite{ashwin2024identifying}.
For example, systems can provide personalized encouragement or adapt task difficulty, thereby helping students navigate learning challenges while maintaining their self-efficacy.
Notwithstanding the noteworthy advancements, several challenges persist, including the reliance on manual annotation for sentiment labels, the inadequate exploration of domain generalization, and the challenges in managing unstructured text complexities.
Future research endeavors should prioritize the development of adaptive and generalized sentiment analysis models, the incorporation of unsupervised or weakly supervised learning, and the exploration of affective dynamics across diverse educational contexts.

\subsubsection{\textbf{Student Behavior Analysis}}
\label{sec5_2_3}
 \begin{figure}[!th]
    \centering
    \resizebox{0.45\textwidth}{!}{
        \begin{forest}
            for tree={
                grow=east,
                reversed=true,
                anchor=base west,
                parent anchor=east,
                child anchor=west,
                base=left,
                font=\normalsize,
                rectangle,
                draw=none, 
                rounded corners,
                align=left,
                minimum width=1em,
                edge+={darkgray, line width=1pt},
                s sep=5pt, 
                inner xsep=0pt,
                inner ysep=3pt,
                line width=0.8pt,
            }, 
            [
                Student Behavior Analysis \\(\S\ref{sec5_2_3}), modelnode-beha, text width=9em
                [
                    Cluster Algorithm, modelnode-beha, text width=6.5em
                    [\parbox{5em}{\centering \cite{bao2021analysis,delgado2021analysis,li2021unsupervised,shen2021college,shi2024characteristics}}, leaf-beha]
                ]
                [
                    Association Rule \\ Mining, modelnode-beha, text width=6.5em
                    [\parbox{4em}{\centering \cite{bao2021analysis,WangXM22}}, leaf-beha]
                ]
                [
                    Machine Learning, modelnode-beha, text width=6.5em
                    [\parbox{5em}{\centering \cite{0008FJXDHLZ22,li2020university,shi2023analysis,shen2021college,villalobos2024learning,abhirami2022student}}, leaf-beha]
                ]
                [
                    Integrated Study, modelnode-beha, text width=6.5em
                    [\parbox{5em}{\centering \cite{cantabella2019analysis,liu2017mining,kamberovic2023personalized,yu2024raw}}, leaf-beha]
                ]
            ]
        \end{forest}
    }
    \caption{Classification of student behavior analysis}
    \label{beha_anal_class}
\end{figure}

Rooted in Behaviorist Theory, student behavior analysis conceptualizes learning as a process shaped by external stimuli, responses, and reinforcement.
This perspective emphasizes observable actions over internal cognition, suggesting that consistent behavioral patterns, such as participation frequency, interaction intensity, and response timing, can serve as reliable indicators of engagement and learning outcomes.
In personalized learning, these principles underpin the design of feedback and reinforcement mechanisms, enabling adaptive systems to monitor, interpret, and guide student behaviors through data-driven strategies.
Behavioral engagement is thus recognized as a key indicator of student participation and attention, directly influencing academic performance \cite{appleton2008student}.
As a critical component of non-cognitive modeling, student behavior analysis supports performance prediction and personalized recommendations. 
By analyzing behavioral data (e.g., online activities, learning history, and interaction patterns), this approach provides real-time feedback, dynamically optimizes models, and enhances instructional design and resource allocation to foster personalized learning experiences.
As shown in Figure \ref{beha_anal_class}, we systematically categorized the literature into four categories based on the analytical methods employed: clustering algorithms, association rule mining, machine learning, and integrated studies.
This taxonomy builds upon prior reviews in educational data mining, which commonly organize studies by analytical techniques \cite{baker2016educational}.
This classification helps to sort out the characteristics and application scenarios of different research methods more clearly, thus providing a structured framework for subsequent literature analysis.
The relevant research work is summarized in Appendix Table \ref{stu_beha_ana_sum}.
   
\textbf{Clustering algorithm:} Clustering algorithms have been widely used to analyze student behavior and uncover learning patterns.  
    Shen et al. \cite{shen2021college} applied K-means to categorize behaviors into study, life, and Internet usage, while Li et al. \cite{li2021unsupervised} combined Density-Based Spatial Clustering of Applications with Noise (DBSCAN) and K-means to explore the link between behavior and Grade Point Average (GPA), improving clustering accuracy.  
    Complementing these findings, Delgado et al. \cite{delgado2021analysis} adopted a Self-Organizing Map (SOM) to capture diverse behavioral patterns at the Universidad Internacional de La Rioja (UNIR). 
    By considering SOM distances in two phases, their method offered a more granular depiction of students' online learning activities. 
    Recent research extends the underlying clustering algorithm by using Latent Dirichlet Allocation (LDA) to mine the learning preferences and underlying logical relationships of student groups \cite{shi2024characteristics}.
    It is also shown that the introduction of the LDA model can efficiently extract keywords and fuzzy identify student groups at different levels, providing valuable insights for personalized instructional design.

\textbf{Association rule mining:} This technique is commonly used to discover correlations among various aspects of student behavior, offering insights into learning influences and outcomes.
In typical work, a four-layer association architecture paired with a three-step mining process has been proposed, covering stages from data preprocessing to knowledge acquisition \cite{WangXM22}.
Similarly, the Eclat association rule algorithm integrated with a clustering algorithm has been employed to scrutinize associations among diverse student data categories sourced from information centers and grades, thereby elucidating the primary factors influencing academic performance \cite{bao2021analysis}.
    
\textbf{Machine learning:} 
    Machine learning is widely used in student behavior analysis, with classifiers detecting key patterns \cite{shi2023analysis,shen2021college}.
    Enhancing accuracy, Chen et al. \cite{0008FJXDHLZ22} integrated forest optimization with NB, DT, and RF in a blended learning model. 
    Li et al. \cite{li2020university} integrated neural networks, NB, and DT algorithms using a Spark-based platform, showcasing the scalability and real-time processing capabilities of distributed computing.
    Although these traditional machine learning approaches effectively identify behavior patterns, they often overlook the temporal dynamics of learning processes.
    To address this, recent studies have explored dynamic behavioral modeling to capture temporal learning patterns. 
    For example, Villalobos et al. \cite{villalobos2024learning} used sequence analysis and temporal dynamics metrics to characterize individual learning behaviors in Blended Learning (BL) environments. 
    By applying Inverse Probability Weighting (IPW), the study also evaluated the causal impact of learning dashboards on student behavior and achievement, offering new insights into personalized interventions.
    
\textbf{Integrated study:} Integrated studies combine multiple analytical methods to provide a more comprehensive understanding of student behavior.
    Early implementations in big data environments utilized frameworks like Hadoop MapReduce alongside statistical and association rule techniques for in-depth analysis \cite{cantabella2019analysis}. 
    More recently, the rise of Large Language Models (LLMs) has enabled new integrative frameworks.
    For instance, end-to-end approaches have been proposed that integrate temporal motion detection with LLMs to analyze classroom videos and automatically generate detailed behavior reports \cite{yu2024raw}. 
    Such work exemplifies the growing influence of LLMs in augmenting AI-assisted educational assessment through the detailed capture and interpretation of behavioral sequences.

Student behavior analysis employs clustering algorithms, association rule mining, machine learning, and integrated studies to gain comprehensive insights into learning patterns.
Clustering algorithms identify commonalities and disparities, while association rule mining uncovers connections between behavior and academic achievement. 
Machine learning methods enhance prediction accuracy and identify complex patterns.
Comprehensive research combining multiple approaches and the utilization of LLMs has come to the fore in learning analytics, providing richer educational assessments.
However, although current research integrates machine learning and other techniques for student behavior analysis, it predominantly emphasizes data analysis without effectively translating findings into actionable insights. 
Additionally, limited integration of psychological and educational theories, inadequate use of multimodal data, and shallow exploration of social factors constrain the depth of these studies. 
Furthermore, most experiments rely on proprietary datasets that are not publicly accessible due to privacy concerns, hindering reproducibility and slowing the field's advancement.

\subsubsection{\textbf{Student Performance Prediction (Non-Cognitive)}}
\label{sec5_2_4}
\begin{figure}[!th]
    \centering
    \resizebox{0.45\textwidth}{!}{
        \begin{forest}
            for tree={
                grow=east,
                reversed=true,
                anchor=base west,
                parent anchor=east,
                child anchor=west,
                base=left,
                font=\normalsize,
                rectangle,
                draw=none, 
                rounded corners,
                align=left,
                minimum width=1em,
                edge+={darkgray, line width=1pt},
                s sep=5pt, 
                inner xsep=0pt,
                inner ysep=3pt,
                line width=0.8pt,
            }, 
            [
                Student Performance \\
                Prediction (Non-Cognitive) \\ (\S\ref{sec5_2_4}), modelnode-perach, text width=9em
                [
                    Machine Learning, modelnode-perach, text width=6.5em
                    [\parbox{5em}{\centering \cite{feng2022analysis,yaugci2022educational,verma2022prediction,zhang2018grade,khairy2024prediction,ni2023leverage,xu2019prediction,priyambada2023two}}, leaf-perach]
                ]
                [
                    Ensemble Learning, modelnode-perach, text width=6.5em
                    [\parbox{5em}{\centering \cite{asselman2023enhancing,joshi2021catboost,kukkar2023prediction,xu2017progressive,fan2025complementary,fan2024feature,baig2023prediction}}, leaf-perach]
                ]
                [
                    Deep Learning, modelnode-perach, text width=6.5em
                    [\parbox{5em}{\centering \cite{giannakas2021deep,kusumawardani2023transformer,yang2024framelet}}, leaf-perach]
                ]
                [
                    Other, modelnode-perach, text width=6.5em
                    [\parbox{5em}{\centering \cite{oh2024language,ni2024enhancing}}, leaf-perach]
                ]
            ]
        \end{forest}
    }
    \caption{Classification of student performance prediction (non-cognitive)}
    \label{per_pre_class_non}
\end{figure}

The detailed methodology of cognitive modeling is presented in Section \ref{sec5_1_2}. 
This section shifts the focus to non-cognitive modeling in student performance prediction, which explores the influence of behavioral factors, such as internet usage and study time, on academic outcomes. 
These factors, closely tied to students' learning habits, engagement, and external environments, capture non-cognitive influences beyond traditional knowledge assessments \cite{khairy2024prediction,kukkar2023prediction,asselman2023enhancing}.
Unlike cognitive modeling, which estimates knowledge states, non-cognitive approaches primarily employ data-driven machine learning techniques to predict academic performance rather than infer knowledge mastery.
By integrating these factors, non-cognitive modeling offers insights into their underlying influence on academic success \cite{ni2023leverage}.
Such studies provide new perspectives and a theoretical basis for the design of personalized interventions and learning support systems.
In order to systematically review the existing work, we summarize the techniques, datasets, evaluation metrics, etc. of the existing work in Appendix Table \ref{stu_per_pre_sum}.
Building on established review frameworks, in Educational Data Mining (EDM) and Learning Analytics (LA) \cite{romero2010educational,baker2016educational}, this study classifies existing literature into our methodological categories, as illustrated in Figure \ref{per_pre_class_non}: machine learning-based models, ensemble learning-based models, deep learning-based models, and other predictive models. 
This framework encapsulates the methodological evolution of student performance prediction, from traditional machine learning to ensemble and deep learning approaches, consistent with the method-oriented classification schemes adopted in recent reviews, while offering a finer distinction between ensemble and deep learning paradigms.

\textbf{Machine learning-based:}
In student performance prediction, many studies directly apply traditional machine learning algorithms, such as DT, RF, SVM, NB, and logistic regression, to compare their predictive effectiveness \cite{yaugci2022educational,verma2022prediction,zhang2018grade,khairy2024prediction}, or integrate multiple machine learning methods \cite{priyambada2023two}. 
These approaches primarily focus on evaluating the accuracy and efficiency of various classifiers without extensive model customization or adaptation to educational contexts. 
Additionally, some studies explore alternative strategies to enhance prediction reliability. 
A common approach involves utilizing the K-means algorithm to cluster student performance. The predictive validity of this method is often optimized by determining the k-value through objective quantitative analysis \cite{feng2022analysis}.
These methods provide baseline evaluations of machine learning models in educational settings but lack advanced optimization tailored to the unique characteristics of student learning data.

\textbf{Ensemble learning-based:}
Literature in this category leverages ensemble learning methods to improve the accuracy of student performance predictions. 
For example, the Student Academic Performance Predicting (SAPP) system \cite{kukkar2023prediction}, the Categorical Boosting (CatBoost) model \cite{joshi2021catboost,fan2025complementary,fan2024feature}, and applications using ensemble learning techniques \cite{asselman2023enhancing,xu2017progressive,baig2023prediction}.
The SAPP system \cite{kukkar2023prediction} integrates a 4-layer LSTM network, RF, and Gradient Boosting (GB) within an ensemble learning framework for academic performance prediction. 
CatBoost \cite{joshi2021catboost} enhances model transparency and accuracy by categorizing reasons for improved performance into low, medium, and high levels. 
Building on this, C-CatBoost \cite{fan2025complementary} introduces a residual error model to address prediction inaccuracies by supplementing the base model’s outputs, effectively capturing complex data relationships.
Similarly, the Feature Importance-Based Multi-Layer CatBoost (DCatBoostF) \cite{fan2024feature} improves accuracy by progressively introducing features ranked by importance using RF. 
Xu et al. \cite{xu2017progressive} proposes a stepwise prediction algorithm, integrating learning techniques in a two-layer structure.
In contrast, the Performance Factors Analysis (PFA) method employs RF, AdaBoost, and XGBoost to model learners in an e-learning system, effectively improving performance prediction accuracy \cite{asselman2023enhancing}.

\textbf{Deep learning-based:} 
Deep learning techniques have been widely used in educational data analytics in recent years due to their superior ability to capture complex data patterns.
This section systematically reviews deep learning models, including Deep Neural Networks (DNNs) \cite{giannakas2021deep}, Hypergraph Neural Networks (HGNN) \cite{yang2024framelet}, and other variants \cite{kusumawardani2023transformer}, for modeling student learning behaviors and predicting academic performance.
Giannakas et al. \cite{giannakas2021deep} introduces a binary classification deep neural network with two hidden layers, emphasizing deep learning's crucial role in predicting team performance.
FD-HGNN \cite{yang2024framelet} integrates a dual hypergraph neural network with different channels for low-pass and high-pass components and incorporates an attention mechanism for multi-layer feature learning and complex relationship modeling.

\textbf{Others:} 
The rise of LLMs in AI has created new opportunities to enhance student performance prediction through better semantic understanding and contextualization. These approaches leverage the powerful representation capabilities of LLMs to address complex educational challenges, such as cold-start problems and multimodal data integration.
For example, LmgMF \cite{oh2024language} addresses the cold-start problem by leveraging a large language model (GPT-J and Llama 2) as a multimodal information integrator to generate semantically rich embedding vectors, which are then combined with matrix factorization for enhanced semantic embedding and prediction.
Ni et al. \cite{ni2024enhancing} integrates LLM-generated semantic embeddings with Signed Graph Neural Networks (SGNNs) to model student response accuracy and enhance noise robustness.
Combining the contextual embedding capabilities of LLMs with graph-based structural learning effectively addresses cold-start issues and improves predictive accuracy.

\begin{figure*}[t]
    \centering
    \includegraphics[height=7cm, width=0.65\textwidth]{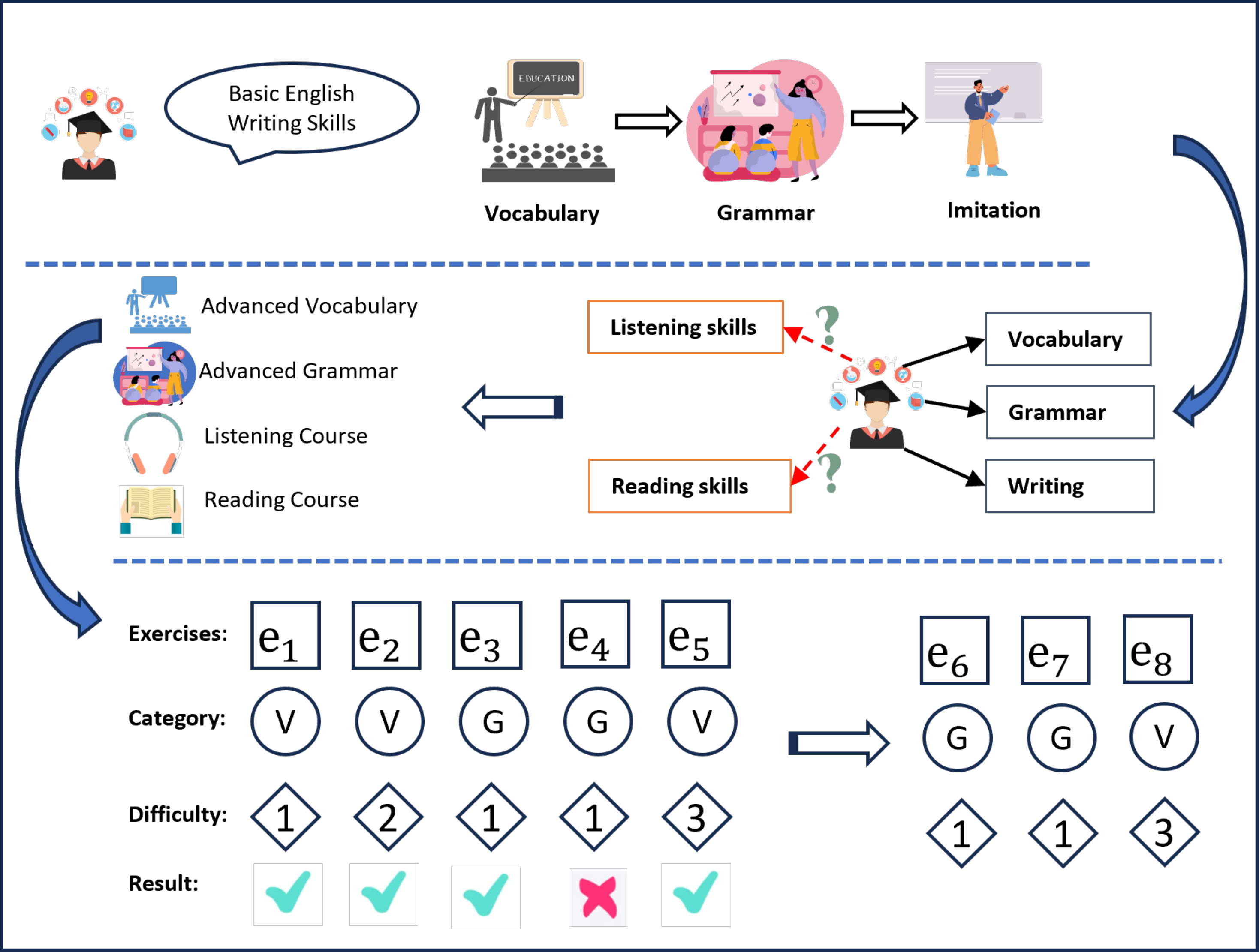}
    \caption{A three-stage personalized learning recommendation framework for English proficiency enhancement}
    \label{fig:RS_Introduction}
\end{figure*}

Non-cognitive modeling has emerged as a crucial complement to cognitive approaches in student performance prediction, providing deeper insights into the influence of behavioral, emotional, and social factors on academic outcomes.
While traditional machine learning and ensemble techniques have established solid baselines, deep learning has markedly enhanced the ability to capture complex behavioral patterns, and LLMs have further expanded semantic understanding and contextual reasoning.
Despite these advancements, persistent challenges include data imbalance, limited model interpretability, and the nascent integration of BERT- and LLM-based architectures.
Beyond algorithmic optimization, non-cognitive modeling carries profound implications for educational practice.
Predictive models can identify students at risk of academic difficulty at an early stage by analyzing behavioral data. 
For example, ensemble-based systems like SAPP \cite{kukkar2023prediction} leverage multi-source data to achieve high-precision predictions, thereby enabling timely, data-driven interventions that help reduce dropout rates.
Empirical research on early-warning systems confirms the feasibility of detecting engagement declines and dropout risks from digital footprints such as interaction logs and submission patterns \cite{hlosta2022predictive}.
In the long term, incorporating interpretable non-cognitive indicators into learning analytics dashboards allows educators to provide feedback on learning habits, prompting students to reflect on and refine their strategies \cite{cobos2023self}.
Recent empirical analyzes suggest that predictive models yield greater educational impact when integrated with pedagogically grounded interventions and longitudinal support, rather than functioning as isolated alert systems \cite{huang2023take}.
The introduction of LLMs further enhances the semantic understanding of students’ behavioral patterns.
Through these integrated approaches, students can receive personalized guidance that aligns with their non-cognitive traits, which not only enhances academic achievement but also cultivates the self-directed learning skills essential for navigating future challenges.

Future research should therefore move beyond predictive accuracy to explore how non-cognitive insights can be operationalized into practical teaching interventions.
Key directions include multimodal data integration, algorithmic optimization, and improving model adaptability and generalization.
Moreover, enhancing interpretability and transparency remains essential for effective deployment in real-world educational contexts.
Only by integrating computational modeling with teaching practices can we fully realize the educational value of predicting noncognitive performance.


\section{Personalized Recommendation}
\label{sec6}

\tikzstyle{my-box}=[
    rectangle,
    rounded corners,
    align=left,
    text opacity=1,
    minimum height=1.5em,
    minimum width=5em,
    inner sep=2pt,
    fill opacity=.8,
    line width=0.8pt,
]

\tikzstyle{leaf-head}=[my-box, minimum height=1.5em,
    draw=gray!80, 
    fill=gray!15,  
    text=black, font=\normalsize,
    inner xsep=2pt,
    inner ysep=4pt,
    line width=0.8pt,
]

\tikzstyle{leaf-style}=[my-box, minimum height=1.5em,
    draw=red!70, 
    fill=red!15,  
    text=black, font=\normalsize,
    inner xsep=2pt,
    inner ysep=4pt,
    line width=0.8pt,
]

\tikzstyle{leaf-sent}=[my-box, minimum height=1.5em,
    draw=cyan!70, 
    fill=cyan!15,  
    text=black, font=\normalsize,
    inner xsep=2pt,
    inner ysep=4pt,
    line width=0.8pt,
]
\tikzstyle{leaf-perach}=[my-box, minimum height=1.5em,
    draw=green!70, 
    fill=green!15,  
    text=black, font=\normalsize,
    inner xsep=2pt,
    inner ysep=4pt,
    line width=0.8pt,
]

\tikzstyle{leaf-beha}=[my-box, minimum height=1.5em,
    draw=orange!70, 
    fill=orange!15,  
    text=black, font=\normalsize,
    inner xsep=2pt,
    inner ysep=4pt,
    line width=0.8pt,
]

\tikzstyle{modelnode-style}=[my-box, minimum height=1.5em,
    draw=red!80, 
    fill=white!30,
    text=black, font=\normalsize,
    inner xsep=2pt,
    inner ysep=4pt,
    line width=0.8pt,
]

\tikzstyle{modelnode-sent}=[my-box, minimum height=1.5em,
    draw=cyan!100, 
    fill=white!30,  
    text=black, font=\normalsize,
    inner xsep=2pt,
    inner ysep=4pt,
    line width=0.8pt,
]
\tikzstyle{modelnode-beha}=[my-box, minimum height=1.5em,
    draw=orange!100, 
    fill=white!30,  
    text=black, font=\normalsize,
    inner xsep=2pt,
    inner ysep=4pt,
    line width=0.8pt,
]
\tikzstyle{modelnode-perach}=[my-box, minimum height=1.5em,
    draw=green!100, 
    fill=white!30,  
    text=black, font=\normalsize,
    inner xsep=2pt,
    inner ysep=4pt,
    line width=0.8pt,
]



Personalized learning recommendations encompass optimizing learning routes, courses, and exercises for individual students.
Learning paths aim to minimize costs and achieve learning goals, personalized course recommendations provide suitable courses, and personalized exercise recommendations target knowledge gaps.
Figure~\ref{fig:RS_Introduction} illustrates three use cases of personalized learning recommendations for English proficiency.

Initially, to enhance the learner's writing skills, the system provides a learning path: “Vocabulary → Grammar → Imitation," forming a comprehensive personalized learning path recommendation. 
Following this sequence, the system utilizes the knowledge graph derived from the student's learning history to pinpoint additional skills that the student may desire to cultivate, such as listening and reading.
\begin{figure}[htbp]
    \centering 
    \includegraphics[width=0.45\textwidth]{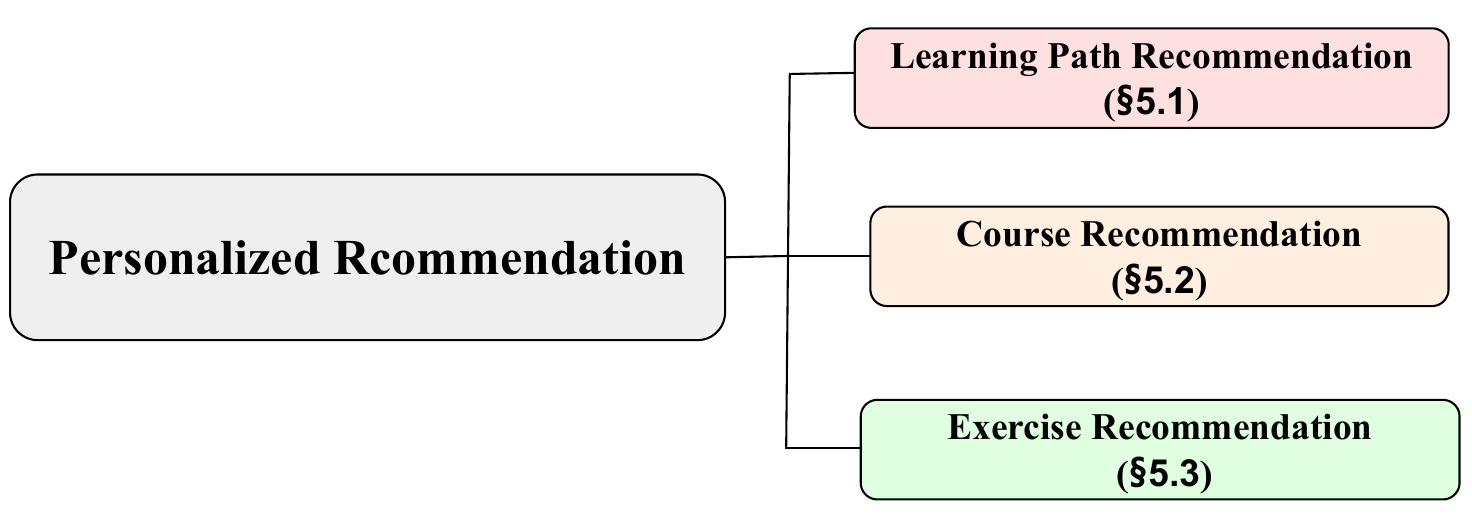} 
    \caption{Taxonomy of personalized learning recommendations} 
    \label{per_recom} 
\end{figure}
Consequently, the personalized course recommendation algorithm recommends advanced courses encompassing aspects of vocabulary, grammar, listening, and reading. 
Simultaneously, each course incorporates diverse exercises, and the subsequent personalized exercise recommendation is derived from the user's history of answer sequences.
It suggests exercises tailored to the current learning category, exercise difficulty, and the student's mastery level.

As illustrated in Figure~\ref{per_recom}, our research on personalized learning recommendation systems adopts a three-dimensional taxonomy comprising personalized learning path recommendation, personalized course recommendation, and personalized exercise recommendation.
This classification schema is derived from the hierarchical recommendation needs across various learning stages. 
Path recommendation involves holistic learning planning by analyzing learning objectives, knowledge graphs, and prerequisite relationships to devise optimal learning sequences. 
At the course recommendation level, appropriate course content is filtered based on learners' interest preferences, knowledge mastery levels, and course relevance to assist in resource selection.
Exercise recommendation focuses on specific skill consolidation by providing targeted practice problems through real-time knowledge state diagnosis and forgetting curve analysis. 
This hierarchical framework not only aligns with the pedagogical logic of “goal-content-practice” but also corresponds to the technical implementation pathway of educational data mining from coarse-grained to fine-grained processing, thereby systematically encompassing the learners' complete cognitive chain.

Existing surveys on personalized learning recommender systems have primarily organized the field along technical methodologies, input data types, and evaluation dimensions, rather than based on the specific learning tasks being recommended.  
For example, Urdaneta-Ponte et al.~\cite{urdaneta2021recommendation} classified educational recommender systems according to how recommendations are generated, presented, and evaluated, emphasizing hybrid approaches and accuracy-centered assessments.  
Similarly, Da Silva et al.~\cite{da2023systematic} conducted a systematic review of recommendation methods, collaborative filtering, content-based, and hybrid, across educational contexts, maintaining a methodological and system-level orientation.  
More recent reviews focusing on personalized learning structured the field around algorithmic pipelines, including learner and resource modeling, dataset usage, and algorithmic trends, thereby reinforcing an algorithm-centered taxonomy~\cite{bin2024comprehensive,gm2024digital}.  
Overall, these surveys mainly focus on “which techniques are used, with which inputs, and how accuracy is measured,” while paying less attention to the pedagogical function or granularity of the recommended learning entities.
In contrast, this paper adopts a task-oriented taxonomy that decomposes personalized learning recommendations into three functional strata—\textbf{learning path} (goal sequencing and prerequisite planning), \textbf{course} (resource selection based on interests and mastery), and \textbf{exercise} (fine-grained practice aligned with learners’ knowledge states).  
This “goal→content→practice” hierarchy aligns with learners’ cognitive workflows and serves as a classification lens that re-indexes existing research according to the recommended object and decision process rather than algorithmic family alone.  
Specifically, our taxonomy emphasizes: (i) task granularity (path vs.\ course vs.\ exercise) as the primary axis of comparison; and (ii) a technique-to-task mapping that positions common methods, such as knowledge graphs for structural constraint modeling, reinforcement learning for sequencing decisions, knowledge tracing for mastery estimation, and large language models for semantic enhancement, within their corresponding decision stages.

\subsection{Personalized Learning Path Recommendation}
\label{sec6_1}
Grounded in Progressivism and Constructivism, the evolution of personalized learning path recommendation systems reflects a theoretical shift toward learner-centered and knowledge-constructive paradigms.
Progressivism advocates individualized and interest-driven learning trajectories, offering a pedagogical rationale for adaptive path design.
Meanwhile, Constructivism emphasizes that knowledge is actively constructed through interaction with learning contexts, providing a theoretical foundation for dynamic and semantically linked path generation based on knowledge graph modeling.
Building on these theoretical underpinnings, the development of learning path recommendation systems has exhibited distinct phase-specific characteristics in both research paradigms and technical approaches.
The initial phase relied predominantly on clustering analysis and historical data mining to establish basic recommendation frameworks by identifying behavioral patterns among similar learners.
With the maturation of knowledge graphs and reinforcement learning technologies, subsequent research shifted focus to dynamic competency modeling and semantic relationship mining, marking a paradigm transition from static recommendations to adaptive learning paths.
In the most recent phase, the convergence of knowledge graphs with deep reinforcement learning, real-time analytics, and large language models has propelled the development of intelligent recommendation systems featuring real-time adaptability and explainability.
These three stages collectively illustrate the technological progression from basic data-driven methods to sophisticated multitechnology integration.
As shown in Figure~\ref{per_recom_learning_path}, we have classified the relevant literature into three parts based on the discussion aforementioned.
The technical approaches, datasets, evaluation metrics, and other elements of the related work are summarized in Appendix Table~\ref{path.}.
\begin{figure}[htbp]
    \centering
    \resizebox{0.45\textwidth}{!}{
        \begin{forest}
            for tree={
                grow=east,
                reversed=true,
                anchor=base west,
                parent anchor=east,
                child anchor=west,
                base=left,
                font=\normalsize,
                rectangle,
                draw=hidden-draw,
                rounded corners,
                align=left,
                minimum width=1em,
                edge+={darkgray, line width=1pt},
                s sep=3pt,
                inner xsep=0pt,
                inner ysep=3pt,
                line width=0.8pt,
                ver/.style={rotate=90, child anchor=north, parent anchor=south, anchor=center},
            }, 
            [
                Learning \\Path (\S\ref{sec6_1}),modelnode-style,text width=3.5em
                [
                    Clustering and Histori-\\cal Data,modelnode-style,text width=8em
                    [
                        \begin{varwidth}{5em}
                            \centering
                            \cite{KraussSM18,ZhouHHZT18,supic2018case,li2019personalized,mansur2019personalized,nabizadeh2019estimating,vanitha2019collaborative,xia2019peerlens}
                        \end{varwidth},leaf-style,text width=4em
                    ]
                ]
                [
                    Knowledge Graphs and \\Reinforcement Learning,modelnode-style,text width=8em
                    [
                        \begin{varwidth}{4em}
                            \centering
                            \cite{vanitha2019collaborative,liu2020learning, LiuDW20, mansur2019personalized,shi2020learning,sun2021personalized,li2021optimal}
                        \end{varwidth},leaf-style,text width=4em
                    ]
                ]
                [
                    Intelligent,modelnode-style,text width=8em
                    [
                        \begin{varwidth}{4em}
                            \centering
                            \cite{vanitha2019collaborative,ChenWTZ23,KrahnKC23,LiXYSR0CT023,ZhangLW23,li2023personalized,frej2024finding,abu2024knowledge,yekollu2024ai,xiao2024highlighting,raj2024improved}
                        \end{varwidth},leaf-style,text width=4em
                    ]
                ]
            ]
        \end{forest}
    }
    \caption{Classification of studies on personalized learning path recommendation}
    \label{per_recom_learning_path}
\end{figure}

\textbf{Clustering and historical data: early approaches to learning path personalization.}  
    Early research primarily focused on learning path recommendations by leveraging \textbf{clustering methods} to identify similar learners based on historical data. 
    Studies such as~\cite{ZhouHHZT18,supic2018case,li2019personalized,xia2019peerlens} proposed various algorithms, including an LSTM model \cite{ZhouHHZT18} and personalized paths derived from individual learning histories \cite{supic2018case,li2019personalized}. 
    Some approaches, like~\cite{li2019personalized}, further prioritized paths from influential learners for users with similar profiles.  
    
    However, recommendations based solely on historical data lack sufficient personalization. 
    Addressing learners' adaptability and ability differences is critical, as highlighted by Mansur et al.~\cite{mansur2019personalized}. 
    For instance, Vanitha et al.~\cite{vanitha2019collaborative} proposed aligning learning content with personal characteristics through ant colony optimization and genetic algorithms to generate more tailored learning paths.  
    
    Additionally, early studies incorporated time constraints of online learning resources into path recommendations \cite{KraussSM18,nabizadeh2019estimating}. 
    While these methods focused on assessing similarities between learners or resources, their reliance on historical data limited their ability to capture personalized needs, and clustering techniques often struggled to handle diverse cognitive levels effectively.  

\textbf{Knowledge graphs and reinforcement learning: dynamic modeling for adaptive paths.}  
    With the rise of deep learning, personalized path recommendation methods have increasingly leveraged more personalized learner features, as highlighted in~\cite{LiuDW20,mansur2019personalized}.
    These approaches integrate individual learning needs, habits, and preferences into comprehensive \textbf{knowledge graphs} tailored to specific problems. 
    For example, Shi et al. \cite{shi2020learning} constructed multidimensional knowledge graphs to model diverse learning relationships for various objectives, while Liu et al.\cite{liu2020learning} addressed the challenges of low-engaged users by building interaction networks between courses and learners. 
    Similarly, Sun et al. \cite{sun2021personalized} developed personalized knowledge graphs with extensive English practice questions, aiming to create unique learning paths for individual learners.  
    
     \textbf{Reinforcement learning} techniques further advance these methods by integrating learner mastery levels. For instance, cognitive diagnostic models analyze user proficiency, with hierarchical reinforcement learning refining mastery-level-based recommendations \cite{li2021optimal}.
    
    These methods are designed to address the diverse and dynamic needs of learners by combining knowledge graphs and reinforcement learning. 
    However, despite advancements, the full potential of these technologies in personalized learning path recommendation remains underexplored, as clustering-based methods continue to dominate much of the research landscape.  

\textbf{Integrated intelligent systems: multi-technology convergence for real-time learning paths.}  
    Recent research has significantly advanced personalized learning path recommendations by integrating \textbf{knowledge graphs, deep reinforcement learning, and other cutting-edge technologies}. 
    Major advances include the development of multidimensional knowledge graphs with higher-order relevance modeling to capture learner preferences more accurately \cite{ZhangLW23}, as well as the creation of dedicated knowledge graphs to address diverse learning needs \cite{vanitha2019collaborative}. 
    Further work has combined temporal and graph attention networks to model dynamic learner-resource interactions for reinforcement learning \cite{ChenWTZ23}. 
    To address the oversimplification of learner abilities, graph-based genetic algorithms have also been proposed to optimize feature alignment and generate diverse learning paths \cite{li2023personalized}.
    These methods fully leverage the capabilities of knowledge graphs and deep reinforcement learning, incorporating knowledge-tracking models to account for learners' evolving proficiency levels.
    For example, Frej et al.~\cite{frej2024finding} enhanced explainability in MOOC recommendations by using graph reasoning to generate interpretable path-based explanations. 
    Knowledge graphs have also been utilized to provide factual context for LLMs, reducing hallucinations and improving the precision of personalized explanations \cite{abu2024knowledge}.
    In addition, end-to-end planning frameworks have been proposed that integrate prompt engineering with models like GPT-4 to generate pedagogically grounded and adaptive learning paths, demonstrating improvements in accuracy and learner satisfaction \cite{ng2024educational}. 
    The integration of real-time learning analytics, as demonstrated by~\cite{raj2024improved}, further refines these methods by dynamically adjusting learning paths based on learner performance and preferences.
    Meanwhile, Xiao et al.~\cite{xiao2024highlighting} highlighted the importance of extracting logically coherent learning pathways using large language models, optimizing recommendations by preserving the contextual flow of learners' enrollment histories.

    \textbf{Practical implementation challenges in personalized learning path systems.}  
    Although personalized learning path systems have evolved from clustering-based analytics to LLM-driven adaptive frameworks, their large-scale deployment still faces notable practical challenges.  
    First, data fragmentation across learning platforms and repositories hinders the creation of unified learner profiles essential for continuous adaptation.  
    Second, an alignment gap often exists between algorithmic recommendations and curricular objectives, raising concerns about pedagogical validity and accountability.  
    Third, ensuring real-time adaptivity demands high computational efficiency and robust system design, which can be difficult to achieve in large or resource-limited environments.  
    Finally, transparency, ethics, and privacy remain critical issues, particularly as LLM-based models rely on multimodal behavioral data.  
    Addressing these challenges requires moving beyond algorithmic performance toward \textbf{system-level design}, emphasizing interpretable AI, data governance, and teacher-in-the-loop frameworks to ensure pedagogical coherence and sustainable integration.

    \textbf{Educational implications of personalized learning path systems.}  
    Personalized learning path systems are not only technical solutions but also pedagogical tools that reshape how learning takes place.  
    By dynamically adjusting the sequence and difficulty of learning materials to match individual progress, these systems create adaptive environments that foster autonomy and sustained engagement.
    For example, Ouyang~\cite{ouyang16self} revealed that AI-driven adaptive features enhance educational quality through self-regulated learning and engagement, while Hariyanto~\cite{hariyanto2025artificial} found that multimodal and reinforcement learning methods improve personalization and educational equity.  
    Further supporting this, Du Plooy et al.~\cite{du2024personalized} confirmed in a higher education context that adaptive learning pathways correlate with higher academic performance and motivation.  
    Students are thus encouraged to take ownership of their trajectories, linking effort with visible progress, which strengthens both motivation and self-efficacy.  
    For teachers, this shift means moving from delivering fixed content to guiding and supporting learners through individualized pathways.  
    From a long-term educational perspective, adaptive path systems can promote motivation, independence, and continuous engagement, making learning more personal and meaningful rather than purely algorithm-driven.

\subsection{Personalized Course Recommendation}
\label{sec6_2}
Progressivism and Sociocultural Theory provide the pedagogical foundation for personalized course recommendation systems.
Rooted in learner-centered and experiential learning principles, Progressivism informs adaptive recommendation mechanisms that dynamically adjust course sequences through reinforcement learning and generative modeling.
Meanwhile, Sociocultural Theory emphasizes socially mediated knowledge construction, supporting the use of attention-based knowledge graphs that model contextual and relational dependencies among learning resources.
Technically, the development of personalized course recommendation systems has exhibited a clear phased evolution in research paradigms and methodologies. 
The initial phase predominantly addressed fundamental challenges such as data mining and sparsity issues, establishing basic recommendation frameworks through distributed computing and matrix factorization techniques. 
With the advent of knowledge graphs and attention mechanisms, the research focus shifted toward mining semantic course relationships and modeling dynamic learner competencies, marking a paradigm transition from static to context-aware recommendations. 
Most recently, the deep integration of knowledge graphs with reinforcement learning and engagement prediction technologies has propelled the field toward intelligent and holistic recommendation systems, capable of real-time adaptation to learners' evolving needs.
As shown in Figure~\ref{per_recom_course}, we have categorized the relevant literature into three parts based on the aforementioned discussion. 
The technical approaches, datasets, evaluation metrics, and other elements of related works are summarized in Appendix Table~\ref{course.}.
\begin{figure}[htbp]
    \centering
    \resizebox{0.42\textwidth}{!}{
        \begin{forest}
            for tree={
                grow=east,
                reversed=true,
                anchor=base west,
                parent anchor=east,
                child anchor=west,
                base=left,
                font=\normalsize,
                rectangle,
                rounded corners,
                align=left,
                minimum width=1em,
                edge+={darkgray, line width=1pt},
                s sep=3pt,
                inner xsep=0pt,
                inner ysep=3pt,
                line width=0.8pt,
                ver/.style={rotate=90, child anchor=north, parent anchor=south, anchor=center},
            }, 
            [
                Course\\(\S\ref{sec6_2}),modelnode-beha,text width=2em
                [
                    Data-Driven,modelnode-beha,text width=7em
                    [
                        \begin{varwidth}{5em}
                            \centering
                            \cite{HouZXW18,zhang2018mcrs,symeonidis2019multi,YangJ19,ZhangHCL0S19}
                        \end{varwidth},leaf-beha,text width=4em
                    ]
                ]
                [
                    Semantic and \\ Dynamic Modeling,modelnode-beha,text width=7em
                    [
                        \begin{varwidth}{5em}
                            \centering
                            \cite{ZhuLQSCN20,Tan0LZ20,MaWCS21,TianL21,XuJSZ21,ZhaoYLN21}
                        \end{varwidth},leaf-beha,text width=4em
                    ]
                ]
                [
                    Towards \\ Holistic Intelligence,modelnode-beha,text width=7em
                    [
                        \begin{varwidth}{5em}
                            \centering
                            \cite{BanWHLZH22,ChenMJF22,JungJKK22,LinLYZLW22,HaoLB23,LinLZXLWLM22,ZhangSYWF23,Zhang23a,ZhangYSZY23,amin2024adaptable,li2024quantification,gharahighehi2025enhancing}
                        \end{varwidth},leaf-beha,text width=4em
                    ]
                ]
                [
                    LLM-enhanced,modelnode-beha,text width=7em
                    [
                        \begin{varwidth}{5em}
                            \centering
                            \cite{wang2025learnmate, van2024interests, li2025improving, jin2025personalized, ma2025good, beutling2024personalised}
                        \end{varwidth},leaf-beha,text width=4em
                    ]
                ]
            ]
        \end{forest}
    }
    \caption{Classification of studies on personalized course recommendation}
    \label{per_recom_course}
\end{figure}

\textbf{Foundations of data-driven recommendations: mining large-scale patterns and addressing sparsity.} 
    With the rise of online learning platforms like MOOCs, traditional recommendation methods face challenges due to the rapid growth of learners and course information.
    Initially, large-scale pattern mining was tackled through distributed computational frameworks and specialized algorithms for extracting course association rules \cite{zhang2018mcrs}.
    Concurrently, the challenge of user heterogeneity and sequential decision-making led to the proposal of methods like Hierarchical Bandits for recommending high-reward courses \cite{HouZXW18}.
    To combat sparsity, Yang et al.~\cite{YangJ19} constructed learner and course networks using the Hyperlink-Induced Topic Search algorithm to extend the user's rating matrix. 
    Beyond sparsity, modeling multi-faceted learner-course interactions has been a focus. This includes employing multi-dimensional matrix factorization models to capture the diverse attributes of courses \cite{symeonidis2019multi}.
    Furthermore, to model the dynamic nature of learning, hierarchical reinforcement learning methods have been introduced to determine when and how a learner's focus should shift among different interests during the learning process \cite{ZhangHCL0S19}.

    In the initial stages, pioneering efforts were made to extract valuable insights from extensive course data, addressing the issue of interaction sparsity to a certain extent. 
    However, early studies didn’t consider the connections and prerequisites between courses, which are important for personalized course suggestions.

\textbf{Semantic and dynamic modeling: knowledge graphs and attention mechanisms for context-aware recommendations.}  
    Subsequent research has increasingly focused on incorporating learner preferences and course characteristics into knowledge graphs to uncover semantic associations between courses. 
    For instance, precedence maps have been developed by analyzing differences in learners' knowledge backgrounds to identify significant precedence relations between concepts and courses \cite{ZhaoYLN21}. 
    Similarly, graph-based networks have been proposed to evaluate instructional quality by modeling multiple influences on learners, such as individual interests, peer interactions, and instructor guidance \cite{ZhuLQSCN20}. 
    To overcome the limitations of traditional collaborative filtering in capturing semantic correlations, knowledge graph representation learning has been employed to embed course semantics into a unified vector space, enabling the calculation of semantic similarity for recommendations \cite{XuJSZ21,MaWCS21}.  
    Attention mechanisms have also been widely adopted to model correlations between courses more effectively. 
    For example, course correlations have been constructed from textual descriptions using an Attention Manhattan Siamese Long Short-Term Memory network combined with an autoencoder and self-attention mechanisms, allowing recommendations to be adaptively tailored to student preferences \cite{Tan0LZ20}. 
    Furthermore, researchers have investigated the dynamic modeling of learners' proficiency levels to enhance recommendation accuracy.
    Tian et al. \cite{TianL21} proposed a knowledge tracking model that updates learners' abilities in real-time by incorporating multi-dimensional factors, integrating these estimates as attributes into a collaborative filtering framework for course recommendations.  

    Despite these advancements, challenges remain in handling the interconnected nature of courses and adapting to learners' evolving abilities and skills in real time.
    Future research must focus on developing more robust methods to address these limitations and further improve the personalization of course recommendations.  

\textbf{Towards holistic intelligence: knowledge graphs, reinforcement learning, and engagement prediction.}  
Recent research has focused on extracting personalized information from courses and learners and integrating it into knowledge graphs to model complex relationships.
For instance, Jung et al.~\cite{JungJKK22} created a hierarchical map of learners and courses using MOOC data and external knowledge, while Ban et al.~\cite{BanWHLZH22} built a course sequence diagram across learners, analyzing web-learning course representations through a graph structure.
To more deeply model learning processes and relationships, a Collaborative Sequence Graph constructed with Graph Convolutional Network (GCN) layers has been proposed \cite{ChenMJF22}.
Further, knowledge graphs have been employed to project course-learner interactions into unified embedding spaces \cite{ZhangSYWF23}.
For handling intricate semantic information, self-symmetric meta-paths and meta-graphs within course knowledge graphs have been introduced, utilizing specialized graph embedding techniques \cite{HaoLB23}.
Additionally, Zhang et al.~\cite{ZhangYSZY23} proposed a heterogeneous information network for personalized course recommendations using a factorial memory network and graph neural network.  

Reinforcement learning has also been widely adopted to enhance course recommendations.
For sequential decision-making, methods including policy gradients \cite{LinLZXLWLM22}, attention mechanisms with recursive reinforcement learning \cite{LinLYZLW22}, and a deep reinforcement learning multi-agent framework incorporating learner sentiment and competency have been proposed \cite{amin2024adaptable}. 
Efforts have also been made to incorporate language models and deep knowledge tracking techniques.
For instance, Zhang et al.~\cite{Zhang23a} leveraged unstructured textual data by extracting word vectors using BERT and analyzing their semantic content. 
Relatedly, Ban et al.~\cite{BanWHLZH22} addressed the forgetting problem in knowledge tracking by developing a personalized controller that simulates learner forgetting and individual differences based on cognitive psychology principles.
Additionally, Gharahighehi et al.~\cite{gharahighehi2025enhancing} integrated Survival Analysis (SA) methods with collaborative filtering to model time-to-dropout and time-to-completion, significantly improving recommendation performance by considering time-to-event information.  
Moreover, engagement prediction has emerged as a critical factor in reducing dropout rates.
Li et al.~\cite{li2024quantification} proposed a Quantified Engagement (QE) method and an Engagement Neural Network (ENN) model to predict learners' engagement in unenrolled courses.
By applying QE and ENN to personalized course recommendations, the study demonstrated a significant reduction in dropout rates, validating the effectiveness of engagement-based recommendations.

\textbf{LLM-enhanced course recommendation: explainability, goal alignment, and adaptive personalization.} 
Recent research increasingly leverages LLMs to enhance adaptability, interpretability, and goal alignment in course recommendation.  
Systems such as LearnMate~\cite{wang2025learnmate} and Retrieval-Augmented Generation (RAG)-based course discovery models~\cite{van2024interests} demonstrate how prompt engineering and semantic retrieval enable context-aware and conversational recommendations.  
Benchmarking studies further show that LLMs achieve comparable accuracy with greater diversity~\cite{ma2025good}, while ontology-integrated frameworks connect course selection to career competencies~\cite{beutling2024personalised}.  
Overall, these advances mark a transition from static filtering toward adaptive, semantically grounded, and learner-centered recommendation paradigms.


\textbf{Practical implementation challenges in personalized course recommendation systems.}  
Despite significant advances in integrating knowledge graphs, reinforcement learning, and large language models, personalized course recommendation systems still face major implementation barriers.   
First, the heterogeneity of course data across disciplines and institutions complicates the construction of unified semantic representations.  
Second, the alignment issue persists, as automatically generated recommendations may conflict with curricular structures or accreditation requirements.  
Third, maintaining scalable personalization across thousands of evolving courses demands intensive computation and learner modeling.  
Finally, ensuring explainability and fairness is crucial—recommendations must transparently link to learner goals while avoiding bias from uneven data distributions.  
Future efforts should prioritize \textbf{curriculum-aware, interpretable, and fairness-driven system design} to enable educationally valid and institutionally integrated course recommendation frameworks.

\textbf{Educational implications of personalized course recommendation.}  
Course recommendation systems do more than suggest the next item in a learner’s queue; they actively shape learners’ planning, motivation, and long-term trajectories.  
By aligning recommended courses with learners’ prior knowledge, interests, and vocational aspirations, these systems enhance the perceived relevance of learning and foster intrinsic motivation, thereby promoting sustained engagement and informed, autonomous learning.  
Recent studies substantiate these pedagogical effects: Cha et al.~\cite{cha2024impact} demonstrated that AI-based recommenders reduce uncertainty during course selection and strengthen students’ confidence in their academic decisions, while Zhang et al.~\cite{zhang2025optimizing} showed that graph-based and learning-style-aware modeling improves recommendation accuracy and learning engagement.  
For educators and institutions, such systems enable more flexible curriculum design and data-informed academic advising.  
However, to realize these benefits in practice, recommendation mechanisms must remain curriculum-aligned, interpretable to educators and learners, and designed to avoid biases and narrow personalization.  
When appropriately implemented, course recommendations can support self-directed learning and empower students to make informed, confident choices about their educational journeys.

\subsection{Personalized Exercise Recommendation}
\label{sec6_3}

\begin{figure}[htbp]
    \centering
    \resizebox{.45\textwidth}{!}{
        \begin{forest}
            for tree={
                grow=east,
                reversed=true,
                anchor=base west,
                parent anchor=east,
                child anchor=west,
                base=left,
                font=\normalsize,
                rectangle,
                rounded corners,
                align=left,
                minimum width=1em,
                edge+={darkgray, line width=1pt},
                s sep=3pt,
                inner xsep=0pt,
                inner ysep=3pt,
                line width=0.8pt,
                ver/.style={rotate=90, child anchor=north, parent anchor=south, anchor=center},
            }, 
            [
                Exercise (\S\ref{sec6_3}),modelnode-perach,text width=5em
                [
                    Foundations of \\Knowledge Modeling, modelnode-perach,text width=8em
                    [
                        \begin{varwidth}{5em}
                            \centering
                            \cite{diao2018personalized,lv2018utilizing,XiaLC18}
                        \end{varwidth},leaf-perach,text width=4em
                    ]
                ]
                [
                    Semantic Understading \\and Mastery Tracking, modelnode-perach,text width=8em
                    [
                        \begin{varwidth}{5em}
                            \centering
                            \cite{zhu2020study,wu2020exercise,ZhouW21a,li2021exercise}
                        \end{varwidth},leaf-perach,text width=4em
                    ]
                ]
                [
                    Dynamic \\Personalization, modelnode-perach,text width=8em
                    [
                        \begin{varwidth}{5em}
                            \centering
                            \cite{huang2019exploring,chuang2021moocers,he2022exercise,li2022personalized,guan2023kg4ex,li2023knowledge,ren2023muloer,wu2023contrastive}
                        \end{varwidth},leaf-perach,text width=4em
                    ]
                ]
                [
                    LLM-enhanced, modelnode-perach,text width=8em
                    [
                        \begin{varwidth}{5em}
                            \centering
                            \cite{lai2025using, li2025large, ta2023exgen, meissner2024automated, zeng2024instruction, luo2024bpe }
                        \end{varwidth},leaf-perach,text width=4em
                    ]
                ]
            ]
        \end{forest}
    }
    \caption{Classification of studies on personalized exercise recommendation}
    \label{per_recom_exercise}
\end{figure}

Behaviorist Theory, Cognitive Psychology, and Sociocultural Theory jointly underpin personalized exercise recommendation systems.
Behaviorism informs reinforcement-based feedback optimization, Cognitive Psychology supports knowledge tracing through cognitive state modeling, and Sociocultural Theory emphasizes social interaction, inspiring peer- and community-aware recommendation mechanisms.
Building on these theoretical insights, the evolution of personalized exercise recommendation systems has exhibited distinct phase-specific characteristics.
The initial research phase primarily focused on the structural modeling of knowledge points to establish foundational frameworks for recommendation systems. 
Subsequent advancements in deep learning and knowledge graph technologies prompted a paradigm shift toward semantic comprehension and dynamic knowledge tracing, thereby enhancing recommendation precision. 
In recent years, the convergence of multiple technologies (e.g., reinforcement learning, deep knowledge tracing, and knowledge graphs) has further propelled the development of systems with enhanced adaptability and intelligent capabilities. 
The following three stages reflect the temporal progression of technological evolution, illustrating the transformation of research paradigms from static recommendation approaches to dynamic personalization frameworks.
As shown in Figure~\ref{per_recom_exercise}, we have categorized the relevant literature into three parts based on the aforementioned discussion. The technical approaches, datasets, evaluation metrics, and other elements of related works are summarized in Appendix Table~\ref{exercise}.

\textbf{Foundations of knowledge modeling: structuring knowledge points for basic exercise recommendations.}
    Early research has established the importance of modeling the relationships between knowledge points and topics, as well as accounting for variations in learners' knowledge levels.
    For instance, Xia et al.~\cite{XiaLC18} identified two critical criteria for personalized exercise recommendations: (1) comprehensive coverage of knowledge points relevant to the learner's objectives, and (2) exercises with difficulty levels tailored to the learner's current knowledge proficiency.
    Building on this, Lv et al.~\cite{lv2018utilizing} constructed a knowledge graph to explicitly capture dependencies between knowledge points, while Diao et al.~\cite{diao2018personalized} utilized a knowledge graph enriched with exercise topics, employing tree search algorithms to generate recommendations.  

    Despite these advancements, a significant limitation of early studies lies in their superficial exploration of the connections between knowledge points. 
    These approaches often lacked depth in incorporating learners' dynamic knowledge mastery, which is essential for truly personalized and adaptive exercise recommendations.  

\textbf{Semantic understanding and mastery tracking: deep knowledge tracing for context-aware recommendations.}
    The significance of knowledge graphs in modeling the relationships between knowledge points and learners has gained increasing attention, particularly in the context of personalized exercise recommendations powered by deep knowledge tracking and knowledge diagnostic models.
    To address the limitations of earlier approaches in capturing semantic relationships between exercises and knowledge points, knowledge graphs incorporating extensive practice questions and knowledge points have been developed.
    These models leverage techniques such as word2vec and graph convolutional networks to extract and represent semantic information \cite{zhu2020study,guan2023kg4ex}.
    Furthermore, to overcome the shortcomings of traditional methods in accurately assessing learners' dynamic knowledge states, deep knowledge tracking models have been introduced.
    These models predict a learner’s mastery of specific knowledge concepts based on their interaction history \cite{wu2020exercise,ZhouW21a,li2021exercise}.
    These predictions were subsequently integrated with collaborative filtering or clustering techniques to generate personalized exercise recommendations tailored to individual learners' needs.  

    However, several challenges remain unexplored, including the development of more sophisticated deep knowledge tracking models, the integration of advanced technologies, and the real-time adaptation of recommendations to reflect the dynamic learning trajectories of individual students.
    These areas represent critical directions for future research in the field of personalized exercise recommendations.  
    
\textbf{Dynamic personalization: knowledge graphs, deep tracing, and reinforcement learning.}
    Recent research has significantly advanced personalized exercise recommendations by focusing on the construction of knowledge graphs and the application of deep knowledge tracking models.
    This includes the development of comprehensive knowledge graphs that integrate knowledge points, student profiles, and exercise topics to create a unified resource framework \cite{guan2023kg4ex}. 
    Building on this, He et al.~\cite{he2022exercise} and Ren et al.~\cite{ren2023muloer} further utilized deep knowledge tracking models to dynamically assess and monitor students' mastery of knowledge concepts, providing real-time insights into their learning progress. 
    Incorporating principles from constructivist learning theory, Yan et al.~\cite{yan2023personalization} combined knowledge structure maps with deep knowledge tracking to identify and recommend diverse, challenging, and innovative topics tailored to individual learners. Additionally, Li et al.~\cite{li2023knowledge} enhanced existing knowledge graphs by mining personalized topic information, transforming them into multidimensional structures that better capture the complexity of learner-exercise interactions.
    Furthermore, Li et al.~\cite{li2022personalized} proposed a syllabus-aligned recommendation method that considers selected knowledge points and expected score ranges, ensuring that exercises are closely aligned with teaching objectives.  

    Reinforcement learning techniques have also been extensively explored to further enhance exercise personalization.
    Early applications focused on optimizing core learning dynamics, with reinforcement learning frameworks balancing objectives (review, exploration, smoothness, learner engagement) \cite{huang2019exploring}. 
    This multi-objective approach was further extended to explicitly model exercise difficulty, integrating review schedules, difficulty levels and learning efficacy to recommend topics at suitable challenge levels \cite{chuang2021moocers}.
    Moreover, to go beyond immediate recommendations, research has targeted long-term learning needs, using reinforcement learning to adapt and sequence recommendations over time to sustain learner progress \cite{wu2023contrastive}.

    \textbf{From recommendation to generation: LLM-powered evolution of exercise personalization.}  
    Distinct from learning path and course recommendation, where personalization focuses on sequencing and selection, recent studies reveal a conceptual shift in exercise recommendation—from selecting pre-existing tasks to \textbf{generating} new and adaptive exercises using LLMs.  
    A growing trend is the development of \textbf{exercise generation systems} that integrate pedagogical reasoning with automated content creation.  
    For instance, Ta et al.~\cite{ta2023exgen} introduced ExGen, which leverages LLMs to automatically generate personalized, ready-to-use programming exercises, while {Mei{\ss}ner et al.~\cite{meissner2024automated} compared GPT-4 and LaMDA for software engineering tasks, emphasizing that LLMs can greatly reduce authoring effort, though human oversight remains essential.  
    Further, Zeng et al.~\cite{zeng2024instruction} demonstrated that instruction-tuned LLMs using the Evol-Instruct and Adaptive Curriculum for Exercise Synthesis (ACES) algorithms can enhance the quality and diversity of generated exercises, and Luo et al.~\cite{luo2024bpe} applied Bloom’s Taxonomy to prompt design, producing physics problems aligned with cognitive levels and learning goals.  
    Additionally, Li et al.~\cite{li2025cikt} proposed CIKT, a collaborative and iterative knowledge tracing framework that integrates LLMs to enhance both prediction accuracy and interpretability in exercise-based learning environments, providing a bridge between explainable student modeling and adaptive content generation.
    Collectively, these advancements signify a shift from static exercise recommendations toward intelligent content generation, where LLMs act not only as recommenders but also as creative co-designers capable of generating pedagogically coherent, context-aware, and cognitively aligned exercises.


    \textbf{Practical implementation challenges in personalized exercise recommendation and generation.}  
    Unlike path or course recommendation, exercise personalization operates at the item level and increasingly involves on-the-fly generation, introducing distinct deployment challenges.  
    First, ensuring the quality and validity of generated items remains difficult; errors in question stems, solutions, or distractors can compromise reliability and pedagogy.  
    Second, difficulty calibration and alignment with curriculum standards and cognitive levels (e.g., Bloom’s taxonomy) are nontrivial, as improper mapping undermines mastery estimation.  
    Third, coupling with knowledge tracing and diagnostics is fragile due to noisy metadata and shifting concept taxonomies that propagate errors through adaptive feedback loops.  
    Fourth, academic integrity and provenance demand safeguards against prompt inversion, plagiarism, and data leakage, requiring traceable item generation pipelines.  
    Fifth, real-time efficiency and fairness pose additional constraints—generation must balance latency, accuracy, and inclusivity across linguistic and cultural contexts.  
    Addressing these issues requires \textbf{generation-aware MLOps} that integrate instructor-in-the-loop validation, provenance metadata, fairness auditing, and curriculum-aligned difficulty modeling, ensuring that LLM-driven exercise systems evolve from recommendation toward pedagogically sound, reliable generation at scale.

    \textbf{Educational implications of personalized exercise recommendation and generation.}  
    The evolution from recommending to generating exercises represents a paradigm shift in how students practice and consolidate knowledge.   
    LLM-powered systems can now produce exercises tailored to individual proficiency levels and deliver real-time, adaptive feedback, rendering practice sessions both highly efficient and deeply engaging.
    This pedagogical potential is increasingly supported by empirical research.
    For adaptive learning, Cui and Sachan~\cite{cui2023adaptive} demonstrated that combining knowledge tracing with controlled text generation produces adaptive language exercises aligned with learners’ knowledge states.
    In programming education, Rytilahti et al.~\cite{rytilahti2024exploring} found that LLM-generated programming exercises diversify practice and reduce teachers’ workload.
    When effectively implemented, such systems do more than bolster conceptual understanding and self-efficacy; they also free educators from routine task creation, allowing them to dedicate more attention to high-level instructional guidance.
    Thus, automated exercise generation transcends its technical role to become a vital pedagogical bridge—seamlessly connecting targeted practice, personalized feedback, and sustained learning growth.

\section{Application Case}
\label{sec7}
In recent years, the application of personalized learning in the classroom has been quite extensive. 
Based on our previous categorization, these real-world application cases can be reviewed from three aspects: cognitive modeling, non-cognitive modeling, and personalized recommendations, depending on the types of personalized learning technologies employed in them.

\subsection{Concrete Examples}
\label{sec7_1}

We summarize recent applications of personalized learning in teaching and learning, as shown in Table~\ref{tab:prac}.

Examples focusing on \textbf{cognitive diagnostics and non-cognitive modeling} include \cite{bellarhmouch2023proposed}, who proposed a learner model combining stereotyping, fuzzy logic, and similarity techniques to personalize learning paths based on user interactions (e.g., page visits, mouse clicks), enhancing self-regulation.
However, in practice, it relies heavily on questionnaire-based initialization, introducing subjective bias.
Moreover, the emotional dimension is limited to two coarse variables (“motivation” and “mood”), which fail to capture nuanced affective states.
Yang et al. \cite{yang2023personalized} implemented a personalized learning analytics intervention in an undergraduate accounting course using the 'BookRoll' e-book system and recommender tools.
Their findings showed improved academic performance and behavioral engagement in the personalized group.

\begin{table*}[!t]
\centering
\footnotesize
\caption{Practical examples of the use of personalized learning}
\label{tab:prac}
\resizebox{\textwidth}{!}{%
\rowcolors{2}{gray!20}{white} 
\begin{tabular}{@{\extracolsep\fill}p{2.5cm} p{3cm} p{3.5cm} p{3.5cm} p{3.5cm} p{3cm}}
\toprule
\textbf{Background} & \textbf{Implementation} & \textbf{Students' Participation} & \textbf{Educators' Participation} & \textbf{Data-Driven Feedback} & \textbf{Challenges} \\
\midrule

Taiwanese university, systems programming course, flipped classroom & 
AI-enabled personalized video recommendations & 
Improved engagement, motivation, and learning outcomes & 
Facilitation of AI-enhanced learning & 
AI algorithms for content recommendation & 
Effective integration of AI in education, addressing diverse student needs \\

German universities, Bachelor's and Master's students, various majors & 
Real time analytics-based scaffolds, online learning tools, rule-based AI & 
Influence on self-regulated learning activities, performance, engagement with learning strategies & 
Provided scaffolds, structured environment, focus on technology-driven personalization & 
Real time data collection and analysis, personalized feedback & 
Addressed technical challenges, refinement based on empirical findings and AI analysis \\

Broad application in e-learning, not specific to any institution & 
Learner modeling with learning styles, domain data, assessment data, affective data; fuzzy logic and similarity analysis & 
Personalized experiences based on individual characteristics, improving engagement and outcomes & 
Involvement in content and assessment provision, integrated into personalized system & 
Continuous adaptation using personal, performance, and emotional data & 
Addressing complexity in learner modeling, technical implementation; combination of modeling techniques \\

Higher education, blended learning, unspecified institution & 
E-book and recommendation system, personalized learning analytics & 
Improved academic performance and increased learning interest & 
Integration of personalized system into learning & 
Analysis of student performance for personalized feedback & 
Addressing complexity in personalization and technical aspects \\

Systematic review on K-12 e-learning, not specific to any institution & 
Development of a personalized recommender system for e-learning, modules for student profiling, material collection, and recommendation & 
Enhancement of engagement, performance, and knowledge through personalized learning materials & 
Provision of course materials, validation of recommended materials, monitoring, and feedback & 
Use of machine learning and data mining for profiling and recommendations, teacher validation & 
Complexity of personalized e-learning recommendations, comprehensive framework for implementation \\

Turkish university, flipped classroom, pre-service teachers & 
Learning analytics for personalized recommendation and feedback & 
Focus on engagement, learning outcomes, self-assessment & 
Facilitate personalized learning, guide students & 
Learning analytics for data collection, analysis, personalized feedback & 
Navigating personalized learning complexities, data privacy concerns \\

Cochin University, India, online learning & 
Rule-based system, Felder-Silverman Model, IEEE Standard & 
Enhanced engagement and learning outcomes & 
Guide in personalized learning process & 
Learner profiling, content alignment & 
Learner modeling, content recommendation \\

San Agustin University, Peru, virtual learning & 
CBR, Honey-Alonso Model for learning styles & 
Improved engagement and learning outcomes & 
Guiding the use of the personalized learning model & 
CBR for adapting learning content to styles & 
Adapting to diverse learning styles, efficiency in personalization \\

\bottomrule
\end{tabular}%
}
\end{table*}

In the \textbf{recommendation domain}, Huang et al. \cite{huang2023effects} studied AI-driven personalized video recommendations in a Taiwanese flipped classroom, reporting increased engagement and learning outcomes, particularly for moderately motivated students.
However, the recommendation process depends on IBM Watson, which suffers from limited interpretability. Moreover, the lack of diversity metrics in the recommendation mechanism raises concerns about potential information silos.
Lim et al. \cite{lim2023effects} applied real-time analytics and rule-based AI to deliver personalized scaffolding to German university students, supporting self-regulated learning via real-time behavioral data analysis.
However, because the scaffolding rules are manually defined by experts and relatively sparse, the system exhibits limited adaptability to complex or dynamic learning scenarios.
Similarly, another system leverages the Learning Object Metadata Standard for content delivery, with its logic also dependent on preset rules \cite{raj2019rule}.
To enhance adaptability, intelligent systems integrating Case-Based Reasoning with learning style models have been designed to tailor resources to learner preferences.
Nevertheless, the limited scale and domain diversity of their case libraries constrain cross-context transferability \cite{maraza2019model}.
Karaoglan Yilmaz et al. \cite{karaoglan2020student} investigated learning analytics-based personalized recommendation and feedback in a Turkish flipped classroom, showing that customized feedback improved students' self-assessment, motivation, and learning attitudes.
However, the study relies primarily on subjective satisfaction measures, lacking a quantitative evaluation of actual learning gains.
Furthermore, a personalized recommendation framework for K-12 education has been proposed, emphasizing adaptation based on individual characteristics and learning styles while discussing integration challenges in practice.
Although this framework provides a valuable design reference, it lacks empirical validation and concrete implementation details, leaving its effectiveness in real-world settings to be verified \cite{zayet2023needed}.

Overall, most applications currently focus primarily on behavioral and real-time testing, lacking objective, multidimensional variables such as emotional cognition and sociocultural factors. 
Moreover, the relatively short duration of experimental studies limits the evaluation of long-term persistence and temporal decay of personalization effects. 
In addition, the interpretability of recommendation mechanisms and the management of data privacy remain critical yet underexplored dimensions that warrant greater attention in future research.

\subsection{Intelligent Educational Software}
\label{sec7_2}

Intelligent educational software utilizes modern technology, particularly artificial intelligence and data analytics, to enhance the educational experience by offering personalized teaching and learning support based on students' habits, performance, and preferences. 
This software can be categorized into six main types: adaptive learning systems, virtual education assistants, learning analytics tools, online learning platforms, learning management systems, and intelligent educational games. 
Below, we describe each type, highlighting their features, student modeling methods, personalized recommendation strategies, and implementation examples.

{\bfseries Adaptive Learning Systems (ALS)} dynamically model students’ knowledge states and learning behaviors to enable real-time adaptation of instructional content and difficulty. 
Knewton \cite{upbin2012knewton} is an example that employs advanced algorithms to analyze student behavior and performance, providing customized learning paths that align with individual abilities and interests. 
It employs knowledge modeling to assess conceptual mastery while integrating personalized recommendation strategies, such as collaborative filtering and reinforcement learning, to generate individualized learning paths. 
In higher education, Knewton’s University Algebra course at the University of Arizona (involving approximately 2,000 students) reported a 17\% increase in pass rates, and a 56\% decrease in dropout rates\footnote{Website at pro.huxiu.com/article/188600.html (Last visited on January 22, 2026).}.
In K-12 education, a randomized controlled trial (RCT) by RAND with 1,400 middle school students using a Knewton-style ALS achieved moderate learning gains (Cohen’s $d = 0.34$) and a 4.2-percentage-point reduction in dropout rates \footnote{Website at www.rand.org/pubs/research\_reports/RRA108-1.html (Last visited on January 22, 2026).}.
Another prominent commercial platform, DreamBox, also illustrates the efficacy of adaptive learning systems at scale.
During the 2020–2021 academic year, an impact assessment spanning 8,000 schools and 3 million K–8 students revealed that learners who completed at least five lessons per week (approximately 60 minutes) improved their NWEA MAP percentile scores by an average of 9.9 points \footnote{Website at www.evidenceforessa.org/program/dreambox-learning (Last visited on January 22, 2026).}.
ALS platforms like Knewton and DreamBox are highly representative of demonstrating large-scale efficacy.
However, their reliance on well-defined learning objectives and structured assessments limits applicability to open-ended or interdisciplinary learning.
Moreover, the limited transparency of their recommendation algorithms raises concerns regarding reproducibility, generalizability, and potential instructional bias.

{\bfseries Virtual Educational Assistants (VEA)} are AI-driven learning support systems that provide personalized resource recommendations and real-time tutoring through natural language interaction and multimodal feedback. 
These systems integrate both cognitive modeling to monitor knowledge acquisition and learning trajectories and non-cognitive modeling, which analyzes motivation, attention, and engagement behaviors to enable adaptive interventions. 
The language learning platform Duolingo \cite{vesselinov2012duolingo} exemplifies a typical VEA application, employing reinforcement learning and sequence modeling to dynamically adjust content difficulty and feedback, thereby offering a highly personalized language learning experience.
Empirical studies have consistently confirmed Duolingo’s effectiveness in language acquisition. 
For instance, in the transitional preparatory English course, a blended learning approach combining Duolingo and ReadTheory increased CEFR-aligned receptive skill test scores from 8.12 ± 2.16 to 15.17 ± 1.79 ($p<0.001$), indicating substantial improvement \cite{hajira2025mobile}. 
Similarly, a study involving 48 U.S. college students reported that after three months of continuous use, language proficiency advanced from Novice Low to Novice Mid ($r = 0.61, p<0.001$), with notable gains in writing and speaking (Cohen’s $d\approx1.0$) \cite{smith2024effectiveness}.
These results demonstrate that AI-based VEAs can significantly enhance both language learning outcomes and learner motivation.
Yet, effectiveness is less documented for complex STEM subjects or project-based tasks, where real-time adaptation may require deeper cognitive modeling and domain expertise.

{\bfseries Online Learning Platforms (OLP)} such as Coursera, Khan Academy, and Udacity empower students to independently select learning content based on personal interests and needs by offering a wide variety of course resources. 
These platforms emphasize adaptive pacing and resource allocation to accommodate individual learning trajectories. 
For example, Coursera \cite{knox2012mooc} provides an extensive range of academic and professional training programs that allow learners to progress at their own pace.
Evaluation data indicate that instructional interventions—such as requiring at least one instance of human grading—can improve learner retention by approximately 6\%, while stricter control of study duration contributes an additional 5\% increase in retention \footnote{Website at about.coursera.org/press/wp-content/uploads/2022/05/Courser\\as-Drivers-of-Retention-in-Online-Degree-Programs-Report-1.pdf (Last visited on January 22, 2026).}.
In K–12 education, multiple randomized controlled trials and large-scale empirical studies involving over 11,000 students in grades 3–6 have shown that Khan Academy significantly enhances students’ mathematical achievement and learning attitudes in foundational mathematics courses \footnote{Website at blog.khanacademy.org/research/ (Last visited on January 22, 2026).}.
Although the user base of OLP has grown steadily in recent years, with increasing incorporation of artificial intelligence technologies, dropout rates remain high in fully online settings. Furthermore, these platforms often face challenges in delivering personalized learning beyond standard course pathways, and their integration of fine-grained cognitive modeling remains limited compared to ALS.

{\bfseries Learning Analytics Tools (LAT)} systematically analyze student data, such as engagement, performance, and learning progress, to generate actionable insights that support adaptive instruction and personalized feedback.
These tools also facilitate early warning systems and data-driven decision support for educators.
IBM Watson Education \cite{rajeshwari2020ibm} exemplifies this paradigm by leveraging cognitive modeling and advanced analytics to identify learning patterns, predict academic outcomes, and assist instructors in optimizing pedagogical strategies.
The system further integrates personalized recommendation techniques, including collaborative filtering and sequence modeling, to deliver targeted learning interventions tailored to individual students’ needs.
Nonetheless, dependence on institutional data quality and privacy constraints may hinder broader adoption, and predictive models can propagate existing biases in educational data.

{\bfseries Learning Management Systems (LMS)} provide a centralized platform for creating, managing, and delivering educational content. 
They support personalized curriculum and learning tracking to help students learn at their own pace.
Moodle \cite{al2008moodle}, one of the most representative open-source LMSs, empowers educators to flexibly design instructional materials and evaluation methods while fostering online collaboration and learner engagement.
Empirical evidence from a study involving Nigerian distance-learning undergraduates ($n=113$) \cite{mohammed2025effect} revealed that the use of Moodle led to significant improvements in both academic performance ($F(3,336)=193.19, p<0.001, \eta^2=0.963$) and learning satisfaction ($F(3,336)=154.11, p<0.001, \eta^2=0.957$), confirming its strong efficacy in enhancing learning outcomes and learner experience.
LMSs are highly representative in infrastructure provision and curriculum management.
Yet, their adaptability relies heavily on educator input, and personalization capabilities remain limited compared to ALS or VEA.

{\bfseries Educational Games (EG)} integrate pedagogical objectives with entertainment elements to enhance learner motivation and engagement through interactive and immersive experiences.
These systems dynamically adapt task difficulty based on player performance and progression, achieving an optimal balance between enjoyment and educational impact.  
Minecraft: Education Edition (MCEE) \cite{kuhn2018minecraft} exemplifies this approach by enabling students to engage in inquiry-based learning and collaborative problem-solving within a virtual environment. 
In an empirical study conducted in Irish primary schools ($N\approx166$), 87.5\%–96.3\% of participating students reported that MCEE improved peer collaboration, while 93.7\%–100\% indicated that it significantly fostered creativity \cite{slattery2023primary}.
These results underscore the effectiveness of educational games in cultivating students’ collaborative, problem-solving, and creative abilities.
However, measuring learning outcomes quantitatively is challenging, and long-term academic impact is less well-established. The cost and technical requirements can also restrict scalability.

Appendix Table \ref{tab:ada_lea_sys} summarizes mainstream intelligent educational software and their key features. 
These cases collectively demonstrate that educational software varies substantially in both target learning stages and technical adaptability. 
For example, Knewton is deployed across K–12 and higher education; Duolingo primarily serves higher and adult education; DreamBox and Khan Academy focus on K–12 mathematics and foundational subjects; while Coursera and Udacity are oriented toward higher education, degree programs, and vocational training.
This variation reflects fundamental differences in technological objectives, data structures, and instructional models across educational levels. 
In K–12 education, system design typically prioritizes measurable academic outcomes, such as standardized test improvement, course completion, and dropout reduction, supported by structured curricula and teacher-led implementations. 
RCTs are widely applied in this context to quantify instructional impact, as exemplified by the evaluation studies of DreamBox and Knewton. 
Data in K–12 environments are highly structured (e.g., item-level logs, assessment scores), which facilitates fine-grained modeling through CDM and Knowledge Tracing (KT) approaches.
However, privacy regulations and parental oversight often constrain large-scale data sharing.

In higher education and adult learning, technological priorities shift toward optimizing course pass rates, learner retention, and time efficiency.
Learners in these settings exhibit greater heterogeneity in prior knowledge and motivation, requiring systems with stronger generalization and semantic reasoning capabilities. 
Consequently, adaptive frameworks increasingly integrate knowledge graphs, LLMs, and personalized recommender systems to support open-ended, project-based, and self-directed learning.
At the same time, the technical adaptability of cognitive diagnostic models varies across different educational stages.
K–12 education places greater emphasis on precise assessment and cognitive level identification, while higher education focuses more on understanding complex learning behaviors and providing personalized feedback.
Although intelligent educational software can deliver targeted learning experiences and provide real-time feedback, enhancing efficiency and interactivity, it remains limited in transparency, cross-domain adaptability, long-term efficacy evaluation, and incorporation of non-cognitive factors.
Future work should address these gaps by developing interpretable adaptive algorithms, enhancing longitudinal evaluation, and expanding personalization to complex, interdisciplinary learning contexts.

\subsection{Paradigm Breakthroughs of LLM in Personalized Learning}
In recent years, LLMs have brought paradigm-level transformations to personalized learning.
Unlike traditional models that rely on structured learning logs and handcrafted features, LLM-based systems leverage natural language interaction and contextual reasoning to dynamically capture learners’ cognitive, affective, and behavioral states.
This section reviews representative studies and systems to elucidate the core breakthroughs achieved by LLMs in personalized learning and the primary challenges they face.

The pioneering EduChat system exemplifies the deep integration of large-scale educational dialogue with cognitive-affective modeling \cite{dan2023educhat}.  
Trained on massive instructional corpora, it interprets learners' utterances and feedback to infer their mastery levels and learning intentions in real time, thereby transforming static learner profiles into dynamic, dialogue-based models.  
Through natural language interaction, it generates personalized learning paths and exercises aligned with learners’ self-stated goals, thereby redefining the logic of recommendation. 
Nevertheless, its dependence on substantial computational resources and potential privacy risks in cloud-based deployment remain key challenges.
Mitigating these issues requires deploying private education servers and optimizing real-time inference pipelines.
Similarly, SocraticLM \cite{liu2024socraticlm}, based on the Socratic Dialogue Model, adopts a multi-agent architecture (Dean–Teacher–Student model) to promote reflective reasoning and structured inquiry rather than providing direct answers. 
Recent research has extended this approach to dialogue-based knowledge tracing.
For instance, Scarlatos et al. \cite{scarlatos2025exploring} introduced a framework that uses LLM prompting to identify knowledge components and misconceptions within tutor–student dialogues, achieving superior predictive accuracy in open-ended conversational learning. 
By simulating strategy-driven, personalized instruction grounded in constructivist learning theory, these systems identify cognitive misconceptions through multi-turn reasoning, while supervised agents enhance content reliability and mitigate biases.

A second line of research leverages LLMs to enhance diagnostic precision and interpretability. 
Models such as KCD \cite{dong2025knowledge} and MalAlgoPy \cite{sonkar2024llm} integrate LLM-based semantic reasoning with quantitative cognitive diagnosis frameworks. 
Building on this foundation, Wang et al. \cite{wang2025llm} proposed LLM-KT, a plug-and-play alignment framework that connects LLMs with knowledge tracing (KT) tasks through multi-modal context adapters, achieving state-of-the-art performance across benchmark datasets. 
Similarly, Yang et al. \cite{yang2025difficulty} introduced a difficulty-aware programming knowledge tracing model (DPKT) that extracts both textual and conceptual difficulty dimensions of programming problems to refine student knowledge estimation.
Furthermore, the DDKT framework incorporates retrieval-augmented generation and dual-channel difficulty modeling, combining subjective and objective difficulty perceptions to enhance interpretability and mitigate cold-start problems \cite{cen2025llm}.
Complementarily, Lee et al. \cite{lee2024language} present a simpler integration of pre-trained language models (PLMs) into KT, showing that semantic representations from textual data can significantly improve prediction accuracy and interpretability.
Collectively, these works mark a paradigm shift from modeling purely observable behavior to developing semantically grounded, interpretable, and context-rich diagnostic frameworks that balance quantitative precision with human-readable reasoning.

Additionally, the study presented in \cite{zhang2023exploring} conceptualizes LLMs as diagnostic subjects in educational assessments.
By applying Bloom’s Taxonomy to examine their internal knowledge structures, the study reveals that LLMs can exhibit overconfidence and conceptual inconsistencies even in fundamental topics. 
These findings provide critical insights for calibrating LLM-based student models and preventing biased diagnostic inferences.
Meanwhile, frameworks such as LLM-SS \cite{nguyen2023large} leverage prior code attempts and expert examples within a contextual learning paradigm to predict learners’ behavioral trajectories in visual programming tasks, thereby enabling adaptive, task-level feedback generation.
Recent meta-analyses such as the systematic review by \cite{sharma2025role} further synthesize over fifty studies, underscoring LLMs’ impact on learner engagement, affective development, and ethical challenges, and calling for stronger integration between AI-driven personalization and educational validity.
However, maintaining temporal consistency in synthesized learning trajectories remains a major challenge.
Future research could address this by developing hybrid architectures that combine LLMs with recurrent student modeling networks, thereby enhancing stability and coherence over time.

Collectively, these studies highlight a paradigm shift in personalized learning from explicit feature engineering and static response modeling to language-driven, adaptive, and semantically grounded personalization. 
LLMs are redefining student modeling through dialogue-based cognitive inference, semantic-behavioral integration, and simulation-based augmentation, while also reshaping recommendation logic via natural language-driven learning path generation.
Despite these advances, key challenges persist.
First, inherent knowledge update latency in pre-trained models restricts integration of evolving curricula and emerging learning materials, which can be alleviated via Retrieval-Augmented Generation (RAG) and continuous fine-tuning for timely incorporation of up-to-date educational content.  
Second, guaranteeing generated output reliability and fairness remains a critical issue, addressable through post-hoc verification pipelines and evaluation agents based on objective scoring criteria, deployed for automatic assessment of factual accuracy and pedagogical relevance.
Finally, challenges around data privacy, computational efficiency, model explainability, and reconciling generative flexibility with curricular constraints remain prominent.
Future research should prioritize developing coherent, interpretable, continuously adaptive LLM frameworks that balance educational validity with the deep personalization required for truly intelligent learning systems.

\section{Challenges and Future Directions}
\label{sec8}
Personalized learning, as a cutting-edge application of intelligent education and a leading research focus in education, has flourished but faces challenges. 
These challenges primarily manifest in  data quality, technology, assessment, and ethics, and increasingly intersect with issues introduced by the rapid adoption of Large Language Models (LLMs), including generative uncertainties, temporal dynamics, and pedagogical implications.
This chapter provides a comprehensive analysis of these challenges across four key areas, offering insights into future directions and potential innovations to guide further research in personalized learning.

\begin{itemize}
    \item [(1)] \textbf{Data:} 
    Data serves as the foundation of personalized learning, encompassing diverse information derived from learners' academic history, behaviors, and life experiences.
    This vast array of data enables systems to assess students' knowledge states, address individual learning needs, and continuously refine models based on learning styles, emotions, and behavioral patterns.
    However, existing datasets are often limited in scale, noisy, or lack essential metadata, which can lead to biased conclusions.
    In student behavior analysis, researchers frequently rely on privately collected datasets that are rarely disclosed, creating challenges in data completeness and standardization.
    The rise of multimodal data (e.g., text, speech, gaze) further increases the complexity of standardization and fusion.
    A new challenge with LLMs lies in their reliance on static, often outdated corpora, leading to knowledge cutoffs that lag behind curricula and educational standards.
    Even fine-tuned models can reproduce obsolete facts or outdated pedagogical concepts, aggravating data-quality and alignment issues.
    Additionally, continuous logging of learner interactions for adaptive feedback increases privacy and security risks through persistent data collection and implicit profiling.
    Balancing data utility, standardization, transparency, and privacy, while keeping adaptive models current, representative, and ethically grounded, remains a core challenge for next-generation personalized learning systems.
    \item [(2)] \textbf{Technology:} 
    Current personalized learning algorithms primarily focus on specific tasks, such as text assessment and recommendation systems but lack sufficient research on generalization across domains.
    Achieving holistic personalization requires models that are adaptable, interpretable, and grounded in educational and psychological theory.
    In learning analytics, effectively analyzing student data amid complexity and uncertainty remains a significant challenge.
    Despite advances in deep learning, most existing methods struggle to effectively handle multimodal signals and contextual uncertainty.
    The integration of LLMs introduces both opportunities and challenges \cite{MilanoML23}.
    LLMs enable dynamic, dialogue-driven learning support and contextual reasoning, but also raise new issues:
    \begin{itemize}
        \item Knowledge update latency, as static LLMs may rely on outdated information.
        \item Educational suitability, since generated content is not always pedagogically aligned or age-appropriate.
        \item High computational and energy costs, which hinder deployment in large-scale educational systems.
        \item Bias amplification and factual hallucination, which may mislead learners or distort educational fairness.
    \end{itemize}
    Overcoming these technological challenges requires deeper research into efficient LLM adaptation, continual fine-tuning with educational data, and the establishment of content-alignment mechanisms to ensure that generated materials support verified learning objectives.
    
    \item [(3)] \textbf{Assessment:} 
    Evaluating the effectiveness of personalized learning remains a complex challenge, as traditional metrics such as Hit Ratio (HR), Normalized Discounted Cumulative Gain (NDCG), and Recall fail to capture comprehensive learning progress and real-world impact.
    Relying solely on offline assessment methods complicates real-time performance evaluation in dynamic learning environments, delaying user feedback and overlooking a crucial aspect of personalized learning, which is enhancing the student experience and addressing individual needs.
    Existing metrics often fail to accommodate diverse learning styles, limiting the generalizability of evaluation results, while those derived from individual experiments or subjective judgments lack universal applicability and broad recognition.
    Moreover, the emergence of AI tutors and LLM-driven systems introduces new evaluation challenges, as learning outcomes become intertwined with conversational dynamics, self-directed exploration, and adaptive reasoning. 
    Traditional quantitative metrics are often insufficient to assess such open-ended, interaction-driven learning processes. 
    In summary, current evaluation approaches remain limited in timeliness, adaptability, and generalizability, relying excessively on static performance metrics while insufficiently capturing the interactive, dynamic nature of personalized and LLM-driven learning processes.
    \item [(4)] \textbf{Ethics:} 
    Ethical considerations in personalized learning are crucial due to the variability in model quality and the associated risks.
    The collection of extensive student data raises significant privacy concerns, while potential biases in training data can lead to unfair customization of learning content.
    With the rapid rise of LLMs, new ethical risks have surfaced: biased or culturally insensitive content generation, lack of interpretability, and potential dissemination of inaccurate or hallucinated information.
    Moreover, integrating LLMs into educational infrastructures amplifies privacy and accountability challenges, as personal learning records may inadvertently be exposed during model optimization or fine-tuning.
    In summary, the ethical landscape of personalized learning is becoming increasingly complex, as emerging AI-driven technologies amplify long-standing concerns about bias, transparency, and data protection.
\end{itemize}

Navigating the challenges of personalized learning requires ongoing technological advancements and in-depth research, which offer a promising pathway for future development.
From a data perspective, efforts should focus on establishing standardized collection protocols, privacy-preserving sharing frameworks, and multimodal fusion techniques to ensure data fairness, diversity, and completeness.
From a technological standpoint, research should advance lightweight model architectures, federated learning, and on-device inference to alleviate computational and deployment constraints.
Equally important is interdisciplinary collaboration, particularly the integration of technological innovation with educational theory and psychological insight, to construct a more holistic and human-centered personalized learning framework.

Meanwhile, the integration of LLMs presents new opportunities to enhance personalized learning guidance.
LLMs hold transformative potential to create interactive, adaptive, and context-aware learning environments that tailor content to individual learners’ styles, preferences, and knowledge gaps.
However, realizing this potential requires addressing enduring challenges such as curriculum misalignment, knowledge obsolescence, and the generation of hallucinated or pedagogically inappropriate content.
Future research opportunities include generating personalized educational questions, evaluating programming courses, and exploring the transformative impact of LLMs on education.
The transition of academic models and algorithms from theory to practical implementation requires collaborative efforts to drive innovation and the real-world application of personalized learning technologies.
To maintain accountability and contextual relevance, teacher-in-the-loop paradigms and transparent governance frameworks must be incorporated into AI-assisted educational systems.

In summary, despite its challenges, personalized learning offers significant opportunities for intelligent, flexible systems capable of adapting to diverse fields and subject areas, meeting students' varied needs, and advancing personalized learning to new heights in the future.

\section{Conclusion}
\label{sec9}
This paper provides a comprehensive examination of personalized learning within the domain of intelligent education.
We elucidate various definitions of personalized learning, examine its connections with educational theories, and trace its historical development, research motivations, and overarching goals. 
A systematic analysis is conducted from two key dimensions: student modeling and personalized recommendation methods.

For student modeling, we examine both cognitive and non-cognitive perspectives. 
For cognitive modeling, we review foundational algorithms and their progression, with a particular focus on cognitive diagnosis, tracing its development from early psychometric models to modern attention-based approaches.
Non-cognitive modeling includes learning styles, affect assessment, behavior analysis, and performance prediction.
For personalized recommendations, we categorize methods into learning path, course, and practice recommendations, and examine their interplay with student modeling.
Beyond theoretical discourse, we present empirical evidence of real-world applications through case studies and an overview of existing platforms, tools, and software.
However, it should be acknowledged that the practical cases reviewed in this study are predominantly concentrated in MOOCs and K-12 settings, with relatively limited coverage of vocational training, lifelong learning, and corporate learning applications.
While these examples effectively demonstrate the potential of personalized learning technologies, they may not fully capture the diversity and complexity of implementation across all educational contexts.
Future work could expand the scope of analysis by incorporating broader institutional types and diverse learner populations to ensure a more holistic understanding of personalized learning in practice.

Despite significant progress, challenges such as privacy, ethical considerations, model evaluation, and interpretability remain critical barriers to widespread adoption. 
Looking ahead, Large Language Models in education hold great potential to further enhance personalized learning experiences.
This survey aims to provide researchers and practitioners with valuable insights, stimulate academic discourse, and contribute to the continuous advancement of personalized learning in intelligent education.

\section*{Acknowledgement}
\label{ack}
This work was supported in part by National Key R\&D Program of China (No. 2023YFC3341200) and National Natural Science Foundation of China (No. 92270119 and No. 62572198).

\section*{Appendixes}
\subsection*{Appedix A: Relevant tables}
\label{appendix_sec2}
This section includes several tables referenced in the main text, including 
Table ~\ref{cog_diag_sum}, Summary of cognitive diagnosis literature;
Table~\ref{lear_style_sum}, Summary of studies on learning style analysis;
Table ~\ref{sent_ana_sum}, Summary of studies on student sentiment analysis;
Table ~\ref{stu_beha_ana_sum}, Summary of student behavior analysis literature;
Table ~\ref{stu_per_pre_sum}, Summary of studies on student performance prediction (non-cognitive);
Table~\ref{path.}, Summary of studies on personalized learning path recommendation; Table~\ref{course.}, Summary of studies on personalized course recommendation; and Table~\ref{exercise}, Summary of studies on personalized exercise recommendation. 
Additionally, Table \ref{tab:ada_lea_sys} summarizes the mainstream intelligent educational software and their characteristics beyond those listed above.




\onecolumn
\begin{landscape}
\footnotesize
\begin{longtable}{@{\extracolsep\fill}p{1cm} p{3.5cm} p{3cm} p{3.5cm} p{2cm} p{2cm} p{1cm}}
\caption{Summary of cognitive diagnosis literature}\label{tab3} \\
\toprule
References & Categories & Technique & Motivation\footnotemark[1] & Dataset & Indicator & Code\footnotemark[1] \\
\midrule
\endfirsthead

\caption[]{(continue)} \\ 
\toprule
References & Categories & Technique & Motivation\footnotemark[1] & Dataset & Indicator & Code\footnotemark[1] \\
\midrule
\endhead

\midrule
\endfoot

\endlastfoot
\cite{LiuWCXSCH18} & Pedagogical theory-driven & Fuzzy Theories, Assumptions & Partial data utilization, narrow scope of analysis & FrcSub, Math1, Math2 & RMSE, MAE & No \\
\cite{MaHTZZL23} & Pedagogical theory-driven & NS, MF, CF & Incomplete information modeling & FrcSub, Math2015, ASSISTments & RMSE, ACC, Precision, Recall, F1 & No \\
\cite{Zhou0WWHT0CM21} & Pedagogical theory-driven & Context-aware Modeling, Hierarchical Attentive Network & Introducing educational contexts & PISA 2015 & RMSE, AUC, ACC & No \\
\cite{guo2025multidimensional} & Pedagogical theory-driven & Rough Concept Analysis, IRT & Multidimensional & FrcSub, Math1, Math2 & MAE, RMSE & No \\
\cite{wang2024unified} & Pedagogical theory-driven & Psychometric-based Ability Estimation Paradigm & Uncertainty & FrcSub, Math, Eedi & Customizable & Yes \\
\cite{Gao0HYBWM0021} & Data-driven (Deep-learning) & Relation Map Driven, Hierarchical Attention Network & Exploring deeper concept relationships & Junyi, ASSISTments & RMSE, ACC, AUC & Yes \\
\cite{MaLWZC0Z22} & Data-driven (Deep-learning) & Knowledge Embedding Matrix & Exploring deeper concept relationships & Junyi, Math & RMSE, ACC, AUC & Yes \\
\cite{WangHCC21} & Data-driven (Deep-learning) & Neural Network, Graph-Based Representation & Exploring deeper concept relationships & ASSISTments, Math & RMSE, ACC, AUC & No \\
\cite{0008ZY023} & Data-driven (Deep-learning) & Self-Supervised Learning, Graph-Based & Data sparsity & ASSISTments, Junyi & RMSE, ACC & Yes \\
\cite{SongHSYLYL23} & Data-driven (Deep-learning) & Deep Neural Network, MIRT & Cross-modality, Partial data utilization & Self-built data, Junyi & RMSE, ACC, AUC, Precision, Sensitivity, F1 & No \\
\cite{GaoWLWLYZL023} & Data-driven (Deep-learning) & GCN & Zero-shot & Core Math, Advanced Math, Junyi, ASSISTments & RMSE, ACC, AUC & Yes \\
\cite{YangQLGRZW22} & Data-driven (Deep-learning) & Neural Network & Limited feature representation & FrcSub, Math1, Math2, ASSISTments & RMSE, MAE, ACC, AUC & Yes \\
\cite{LiGFX0HL22} & Data-driven (Deep-learning) & Neural Network & Exploring deeper concept relationships & ASSISTments, Math, Algebra, Bridge & RMSE, ACC, AUC & No \\
\cite{YaoLHTHCS023} & Data-driven (Deep-learning) & Information-weighted Sampling Strategies & Data sparsity, Long-tail & Junyi, ASSISTments & RMSE, ACC, AUC & Yes\\ 
\cite{SuCWDHWCWX22} & Data-driven (Deep-learning) & Heterogeneous Graph, Attention Mechanism & Incomplete information modeling & ASSISTments & RMSE, ACC, AUC & No\\
\cite{PeiYHX22} & Data-driven (Deep-learning) & Self-Attention Gating, Meta-Learning & Zero-shot & ASSSITments, FrcSub, Math1, Math2 & RMSE, Precision, F1, AUC & No\\
\cite{LiuYMWQ023} & Data-driven (Deep-learning) & Homogeneous Group Mining, Multi-Grained Modeling & Group cognitive diagnosis & ASSISTments & RMSE, MAE & No\\
\cite{huang2024interpretable} & Data-driven (Deep-learning) & NN, Multi-Task Learning & Multidimensional & PISA 2015, Math, Junyi, ASSISTments & RMSE, ACC, AUC & No\\
\cite{gao2024zero} & Data-driven (Deep-learning) & Transfer Learning, Dual Regularization & Zero-shot & iFLYTEK Learning Machine1 & RMSE, ACC, AUC & Yes\\
\cite{ma2024enhancing} & Data-driven (Deep-learning) & Clustering, Sampling Strategies, Self-Attention & Data sparsity & ASSISTments, Math & RMSE, ACC, AUC & Yes\\
\cite{li2024towards} & Data-driven (Deep-learning) & Neural Network & Interpretability & ASSISTments, Algebra, Math1, Math2 & RMSE, ACC, F1, DOA, Customizable & Yes\\
\cite{yu2024rdgt} & Data-driven (Deep-learning) & GNN, Transformer, Hierarchical Graph Construction & Group cognitive diagnosis & ASSISTments, NIPS-Edu, SLP & RMSE, MAE & No\\
\cite{liu2024inductive} & Data-driven (Deep-learning) & GCN, Induction & Zero-shot, Interpretability  & FrcSub, EdNet-1, ASSISTments, NeurIPS20 & RMSE, ACC, AUC, DOA & Yes\\
\cite{ma2024dgcd} & Data-driven (Deep-learning) & GNN & Group cognitive diagnosis & ASSISTments, NIPS\_Edu, SLPbio, SLPmath & RMSE, MAE & Yes\\
\cite{qian2024orcdf} & Data-driven (Deep-learning) & Response-aware GCN & Oversmoothing & ASSISTments, EdNet-1, Junyi, XES3G5M & ACC, AUC, DOA, Custom Oversmoothing Metric & Yes\\
\cite{lin2017adaptive} & Data-driven (Non-deep-learning) & ACO Algorithm & Limited applicability and slow convergence of heuristics & self-built & A/B testing & No\\
\cite{ZhuLHCLSH18} & Data-driven (Non-deep-learning) & Feature Extraction, Traditional Optimization & Incomplete information modeling & Math1, Math2 & RMSE, MAE & No\\
\cite{LiuQLZ23} & Data-driven (Non-deep-learning) & Causal Model & Data sparsity, Interpretability & Junyi, Math1, Math2 & DOA, AH Metric & No\\
\cite{BiCH0Z0W23} & Data-driven (Non-deep-learning) & Bayesian Network, Meta-Learning, Variational Reasoning & Interpretability & ECPE, ASSISTments, EXAM & ACC, AUC, ECE & No\\
\cite{LiW00HHC0022} & Data-driven (Non-deep-learning) & Bayesian Network & Exploring deeper concept relationships & Junyi, Math & RMSE, ACC, AUC, F1, ODA & Yes\\
\cite{Tong0YHHPJ21} & Data-driven (Non-deep-learning) & Item Response Ranking, Pairwise Learning & Incomplete information modeling & ASSISTments, Math & AUC, Precision, Recall, F1, DOA & No\\
\cite{zhang2024path} & Data-driven (Non-deep-learning) & Causal Reasoning, Statistical Learning & Fairness issues & PISA 2015 & ACC, AUC, DOA, Customized metrics & Yes\\
\cite{ChengLCHHCMH19} & Integration methods & DNN, LSTM, IRT & Improvements in traditional methods & self-built & RMSE, MAE, ACC, AUC & No\\
\cite{HuangLWHFWC0021} & Integration methods & Context-aware Attention Network, Neural Network & Group cognitive diagnosis & ASSISTMents, Math & RMSE, MAE & No\\
\cite{WangLCHCYHW20} & Integration methods & Monotonicity Assumption, Neural Network & Limited feature representation & ASSISTments, Math & RMSE, ACC, AUC, DOA & Yes\\
\cite{WangLCHYWS23} & Integration methods & Monotonicity Assumption, Neural Network & Limited feature representation, Exploring deeper concept relationship & ASSISTments, Math & RMSE, ACC, AUC & Yes\\
\cite{zhou2023causality} & Integration methods & Bayesian Network, NN, Causal Inference, Feature Engineering & Interpretability & ASSISTments, Junyi, Math1, Math2 & RMSE, ACC, AUC, DOA & No\\
\cite{zhang2024understanding} & Integration methods & Adversarial Learning, Theoretical Analysis & Fairness issues & PISA-OECD, PISA-deen & RMSE, MAE, ACC, AUC, |$F_{CD}$| & No\\
\cite{wang2024boosting} & Integration methods & Affective State Modeling, NN & Introducing the affective element & ASSISTments, Junyi & RMSE, ACC, AUC & No\\
\midrule
\label{cog_diag_sum}
\end{longtable}
\footnotetext[1]{“Motivation” refers to the primary research objective of the article, highlighting the key problem it addresses. “Code” indicates whether the article has publicly released its code, with ‘yes’ denoting availability and ‘no’ indicating non-disclosure.}
\footnotetext[2]{NS: Neutral Set Theory, CF: Collaborative Filtering, MF: Matrix Factorization, GNN: Graph Neural Network, GCN: Graph Convolutional Network, DNN: Deep Neural Networks, LSTM: Long Short-Term Memory, NN: Neural Networks}
\end{landscape}
\twocolumn

\onecolumn
\begin{landscape}
\footnotesize 
\begin{longtable}{@{\extracolsep\fill}p{1cm} p{1cm} p{2cm} p{2cm} p{2cm} p{2cm} p{2cm} p{3.5cm}}
\caption{Summary of studies on learning style analysis} \\
\toprule
References & Style & Technique & Study Type & \multicolumn{2}{c}{Dataset} & Indicator & Research Outcomes \\
\cmidrule(lr){5-6} 
 &  &  &  & Name & Link &  &  \\ 
\midrule
\endfirsthead

\caption[]{(continue)} \\ 
\toprule
References & Style & Technique & Study Type & \multicolumn{2}{c}{Dataset} & Indicator & Research Outcomes \\
\cmidrule(lr){5-6}
 &  &  &  & Name & Link &  &  \\ 
\midrule
\endhead

\midrule
\endfoot

\endlastfoot
\cite{bernard2022improving} & FSLSM & ACS, ANN, Confidence-based Segmentation & System development & Self-built data & NA & SIM, ACC, LACC, \%Match & Proposing a common architecture that can be integrated into education systems \\ 
\cite{bernard2017learning} & FSLSM & Multilayer Perceptron, DT & System development & Self-built data & NA & Correctly classified instances, Kappa statistics, Root mean squared error & Developed a tool that can consider multiple learning modalities \\
\cite{el2019combining} & FSLSM & K-means, NB & Algorithm validation & E-learning platform log files & \url{http://www.supmanagement.ma/fc/login/index.php} & Confusion matrix & Propose a method for automatic detection of learning styles \\
\cite{gomede2020use} & FSLSM & Multi-target Classification, ANN & Algorithm validation & Self-built data & NA & Sensitivity, Specificity, Prevalence, PPV, NPV, ACC, Precision, Recall, F1, ACC & Proposing a deep multi-objective prediction algorithm \\
\cite{hidayat2021determine} & FSLSM & Improved K-means & Practical experiments on moodle & Moodle learning management system & NA & NA & Practical experiments on the Moodle LMS \\
\cite{hmedna2020predictive} & FSLSM & DT, RF, KNN, NN, Cluster Analysis & Algorithm validation & Stanford edX "Statistical Learning" (Winter 2015 \& 2016), provided by CAROL & NA & Confusion matrix, ACC, Precision, Recall, F1, Macro-precision & Existing classification methodology test \\
\cite{hmedna2017identifying} & FSLSM & NN & Algorithm validation & Self-built data & NA & Verification in practical experiments & Recognizing learning styles with neural networks
 \\
\cite{kolekar2017prediction} & FSLSM & Fuzzy Mean, BPNN, GSA & Algorithm validation & Website collection & \url{http://www.mitelearning.com} & Precision, Recall, F1, ACC & Proposing a method for automatic detection of learning styles \\
\cite{rishard2022adaptivo} & FSLSM & K-means, RF, SVM, DT, LR & Algorithm validation, System development & Stanford edX "Statistical Learning" (Winter 2015 \& 2016), provided by CAROL & NA & Verification in practical experiments & A personalized adaptive e-learning system, "Adaptivo," is proposed \\
\cite{dominguez2025data} & FSLSM & RF, Voting Ensemble & Algorithm validation & Self-built data & NA & T-test, VIF & The impact of learning styles on academic achievement is explored, emphasizing that while changes in learning styles can improve short-term engagement and comprehension, effective study habits play a more critical role in ensuring long-term academic success \\
\cite{muhammad2024evolving} & FSLSM & K-means, Dynamic Bipartite Graph, LSTM & Algorithm validation & KDDCup 2015 & NA & ACC, Precision, Recall, F1 & Proposing a learning style detection method based on bipartite  graph \\
\cite{jebbari2024identifying} & FSLSM & NB, NN, DT, RF & Algorithm validation,system development & XuetangX & \url{https://www.xuetangx.com/} & ACC, Precision, Recall, Macro-averaged precision, Micro-averaged precision & A prediction system is proposed \\
\cite{ayyoub2024learning} & FSLSM & self-training, SVM & Algorithm validation & Self-built data & NA & Precision, Recall, AUC & The introduction of self-training machine learning methods \\
\cite{kuttattu2019analysing} & VARK & K-means, SVM, DT & Algorithm validation & Self-built data & NA & Verification in practical experiments & A study of the performance of classification algorithms \\
\cite{nguyen2024model} & VARK & K-means, XGBoost, SVM, LR & System development & Self-built data & NA & ACC & A Moodle-compatible plugin was implemented within the learning management system (LMS) \\
\cite{rasheed2021learning} & MI Theory & DT, SVM, K-nearest neighbors, NB, LDA, RF, LR & Algorithm validation & Self-built data & NA & ACC, AUC, Precision, Recall, F1 & Existing classification methodology test \\
\cite{ni2023design} & Kolb & Clustering, SMOTE, Semi-supervised, RF, GDBT, SVM, MLP, LR & Algorithm validation & Self-built data & NA & ACC, Recall, Precision, F1 & Proposing multiple classification model fusion methods \\
\cite{hidalgo2024mapping} & Kolb & Matching & Application & Self-built data & NA & Relevance & Application of learning styles \\
\midrule
\label{lear_style_sum}
\end{longtable}
\footnotetext{ACS: Ant Colony System, ANN: Artificial Neural Networks, NN: Neural Networks, BPNN: Back Propagation Neural Network, GSA: Gravity Search Algorithm, RF: Random Forest, SVM: Support Vector Machine, DT: Decision Tree, LR: Logistic Regression, LDA: Linear Discriminant Analysis, NB: Na\"{i}ve Bayes, KNN:k-Nearest Neighbors}
\footnotetext{SIM: Similarity, ACC: Accuracy, LACC: Lowest Accuracy, \%Match: Customized metrics}
\end{landscape}%
\twocolumn

\onecolumn
\begin{landscape}
\footnotesize 
\begin{longtable}{@{\extracolsep\fill}p{1cm} p{2cm} p{2cm} p{2.5cm} p{2cm} p{2.5cm} p{2cm} p{4.5cm}}
\caption{Summary of studies on student sentiment analysis}\label{tab3} \\
\toprule
References & Classification & Technique & Study Type & \multicolumn{2}{c}{Dataset} & Indicator & Research Outcomes \\
\cmidrule(lr){5-6} 
 &  &  &  & Name & Link &  &  \\ 
\midrule
\endfirsthead

\caption[]{(continue)} \\ 
\toprule
References & Classification & Technique & Study Type & \multicolumn{2}{c}{Dataset} & Indicator & Research Outcomes \\
\cmidrule(lr){5-6}
 &  &  &  & Name & Link &  &  \\ 
\midrule
\endhead

\midrule
\endfoot

\endlastfoot
\cite{chanaa2022sentiment} & Machine learning, MOOCs/Online learning & SVM, NB, LR, GB, RF, NN & Algorithmic model & Self-built data & NA & ACC, Precision, Recall, F1, Kappa & Sentiment analysis of comments on MOOC forums and existing classification methodology test \\ 
\cite{hew2020predicts} & Machine learning, MOOCs/Online learning & KNN, GDBT, SVM, LR, NB & Algorithmic model & Self-built data & NA & ACC, Precision, Recall, F1, Kappa & Using machine learning algorithms to analyze student satisfaction with MOOCs \\ 
\cite{jena2019sentiment} & Machine learning & NB, Multinomial NB, SVM, ME & Algorithmic model & Self-built data from Twitter, Facebook and Moodle & NA & ACC, Precision, Recall, F1 & Analyzing student sentiment using machine learning algorithms \\ 
\cite{liu2016sentiment} & Machine learning & Multi-swarm Particle Swarm Optimization & Algorithmic model & Massive open online course & \url{http://open.163.com/} & Precision, Recall, F1, AUC & Dimensionality and redundancy reduction in feature space \\ 
\cite{kasumba2024practical} & Machine learning & Crowdsourcing, SVM, RF & Algorithmic model & Self-built data & NA & Recall, Precision, F1, Negative class recall & Improving sentiment analysis through student crowdsourcing labels and analyzing the differences between crowdsourcing and expert annotations \\ 
\cite{dake2023using} & Machine learning, Educational environments & NB, SVM, DT, RF  & Algorithmic model & Self-built data & NA & Precision, Recall, F1, AUC & Existing classification methodology test \\ 
\cite{kastrati2020weakly} & Deep learning, MOOCs/Online learning & Weakly Supervised, CNN & Algorithmic model & Self-built data (Coursera) & NA & Precision, Recall, F1 & Recognizing student emotions using weakly supervised signals \\ 
\cite{sindhu2019aspect} & Deep learning, Educational environments & LSTM & Algorithmic model & Self-built data & NA & ACC, Precision, Recall, F1 & Sentiment analysis of student feedback for instructional performance evaluation \\ 
\cite{li2019shallow} & Deep learning, MOOCs/Online learning & BERT, CNN, Self-Attention & Algorithmic model & Self-built data (MOOC course reviews) & NA & ACC, F1 & Reducing the parameters of BERT by half achieves almost the same performance \\ 
\cite{yu2018improving} & Deep learning, Educational environments & SVM, CNN & Algorithmic model & Self-built data & NA & ACC, Precision, Recall, F1 & Using sentiment analysis for academic prediction \\ 
\cite{ren2023automatic} & Deep learning & Bi-LSTM, Attention, Average Pooling & Algorithmic model & Self-built data & NA & Precision, Recall, F1 & Research has shown that models that use subject dictionaries as inputs perform best in combination with attentional mechanisms \\ 
\cite{shaik2022educational} & Deep learning & TextBlob, Bi-LSTM & Application systems & Self-built data & NA & Precision, Recall, F1 & Analyzing and categorizing students' qualitative feedback using Biggs models \\ 
\cite{liu2021sentiment} & Hybrid and advanced techniques & Albert, Attention Mechanisms, BiGRU, Capsule Networks & Algorithmic model & Self-built data (MOOC course reviews) & NA & ACC, Precision, Recall, F1 & Addresses the limitation of traditional sentiment analysis in distinguishing word meanings across contexts \\ 
\cite{mrhar2021bayesian} & Hybrid and advanced techniques & BNN, CNN, LSTM & Algorithmic model & Coursera & \url{https://www.kaggle.com/septa97/100k-courseras-course-reviews-dataset} & ACC, Precision, Recall, F1 & Quantifying uncertainty in sentiment analysis tasks \\ 
\cite{Barron-EstradaZ20} & Hybrid and advanced techniques, MOOCs/Online learning & EvoMSA, NB, KNN, DT, SVC, RF, LSTM, CNN, BERT, EvoMSA & Algorithmic model & Self-built data (Udemy, Platzi, YouTube) & NA & ACC & Existing classification methodology test \\ 
\cite{baqach2024new} & Hybrid and advanced techniques, MOOCs/Online learning & BERT, LSTM, CNN & Algorithmic model & Self-built data (Coursera) & NA & ACC, Precision, Recall, F1 & Multiple model fusion modeling \\ 
\cite{ashwin2024identifying} & Hybrid and advanced techniques & Multi-Task Cascaded Convolutional Networks & Algorithmic model & Self-built data & NA & Verification in practical experiments & Algorithmic bias analysis, integration of sentiment and classroom data, Improved accuracy and fairness of sentiment identification in education \\ 
\cite{shaikh2023exploring} & Hybrid and advanced techniques & LLM & Algorithmic model & Self-built data & NA & Confusion matrix, ACC, Precision, Recall, F1, Kappa, Mathew Correlation Coefficient & Explored the potential of LLM in sentiment analysis \\ 
\cite{pong2019sentiment} & Educational environments & Relevance Mining & Algorithmic model & Self-built data & NA & ACC, Precision, Recall, F1 & Researching student attitudes toward classroom instruction and identifying strategies for instructional improvement \\ 
\cite{rani2017sentiment} & Educational environments & NA & System development & Coursera, Self-built data  & NA & Verification in practical experiments & Presentation of the SA system \\ 
\cite{tubishat2023sentiment} & Educational environments & LLM & Algorithmic mode & Self-built data (Twitter) & NA & ACC, Precision, Recall, F1 & Proposing a sentiment analysis model for tweets related to ChatGPT in education \\ 
\cite{moreno2018learning} & MOOCs/Online learning & Learning Analysis & Algorithmic model & Self-built data & \url{https://media.readthedocs.org/pdf/devdata/latest/devdata.pdf} & Verification in practical experiments & Observed a decrease in positive sentiment over time and before the deadline for open-ended assignments \\ 
\cite{chen2022understanding} & MOOCs/Online learning & Recursive CNN & Algorithmic model & Class Central & \url{https://www.classcentral.com/help/highest-rated-online-courses} & Precision, Recall, F1 & Using sentiment analysis and deep learning to identify key factors influencing learner satisfaction \\ 
\cite{li2022key} & MOOCs/Online learning & Crowdsourcing & Algorithmic model & Self-built data (Coursera) & NA & Data analysis & Identifying key factors like course design and material quality that drive MOOC success \\ 
\midrule
\label{sent_ana_sum}
\end{longtable}
\end{landscape}%
\twocolumn

\onecolumn
\begin{landscape}
\footnotesize 
\begin{longtable}{@{\extracolsep\fill}p{1cm} p{2cm} p{2cm} p{4cm} p{2cm} p{2cm} p{4.5cm}}
\caption{Summary of student behavior analysis literature}\label{tab3} \\
\toprule
References & Algorithm Category & Technique & Behavioral factor & Dataset & Indicator & Research Outcomes \\
\midrule
\endfirsthead

\caption[]{(continue)} \\ 
\toprule
References & Algorithm Category & Technique & Behavioral factor & Dataset & Indicator & Research Outcomes \\
\midrule
\endhead

\midrule
\endfoot

\endlastfoot
\cite{bao2021analysis} & Cluster algorithm, Association rule & K-Mediods, Eclat & Frequency of Consumption, Books borrowed amounts, Frequency of Library visits, Total Study Time & Self-built data & Correlation analysis & A hybrid student behavior analysis algorithm incorporating clustering and association rules is proposed \\
\cite{delgado2021analysis} & Cluster algorithm & Self-Organizing Map Clustering  & Questionnaires, Academic performance, Number of forum visits, Participation in online courses & Self-built data & Verification in practical experiments & Behavioral analysis using Self-organizing map artificial neural network \\
\cite{li2021unsupervised} & Cluster algorithm & DBSCAN, K-Means & Average consumption level , Library access, Average daily online hours & Self-built data & Verification in practical experiments & Analyzed the relationship between different behavioral patterns and students' GPA \\
\cite{shen2021college} & Cluster algorithm, Machine learning & K-Means, PCA & Average consumption level , Library access, Online activity duration, Workout log & Self-built data & Cluster analysis & Simple behavioral analysis \\
\cite{shi2024characteristics} & Cluster algorithm & LDA & Student self-evaluation text & Self-built data & Correlation analysis & The LDA model can efficiently extract keywords and fuzzy recognize different levels of student groups \\
\cite{WangXM22} & Association rule & Data Correlation Mining & Academic performance, Average consumption level, Breakfast frequency, Average daily online hours, Canteen meal frequency, Books borrowed amounts & Self-built data & Chi-square test & Designing a four-tier architecture for data correlation mining, encompassing data collection, storage, computation, and analysis \\
\cite{0008FJXDHLZ22} & Machine learning & NB, DT, RF & Basic student information, Video viewing hours, Chapter quiz scores, Programming exam scores & Self-built data & Verification in practical experiments & Proposing quantitative metrics to assess learners' motivation and the stability of blended learning behaviors. \\
\cite{li2020university} & Machine learning & NB, DT, NN & Average consumption level, Books borrowed amounts, Average daily online hours, Academic performance, Number of physical activities & Self-built data (Spark Platform) & Verification in practical experiments & Simple behavioral analysis \\
\cite{shi2023analysis} & Machine learning & NB, DT, KNN, NN, RF & Information literacy learning Data & Self-built data & PCC, ACC, Precision, Recall, F1, KIA & A significant correlation was found between information thinking traits and learning outcomes \\
\cite{villalobos2024learning} & Machine learning & IPW, SA  & Interactive information on curriculum resources & Self-built data & Customized metrics & Assessed the causal impact of the Learning Analytics Dashboard on student behavior and achievement, highlighting the interplay between motivation, self-regulation, and prior achievement \\
\cite{abhirami2022student} & Machine learning & Supervised Machine Learning & Online learner behavior & Self-built systematic collection & Confusion matrix, ACC, Precision, Recall, Sensitivity, F1 & A supervised machine learning-based intelligent model is proposed for e-learning systems \\
\cite{cantabella2019analysis} & Integrated Study & Apriori, Hadoop MapReduce  & Number of visits to the LMS, tools used by students and their related incidents & Self-built data & Case Studies & Conducting case studies on student behavior \\
\cite{liu2017mining} & Integrated Study & Sequence Analysis & Cloud classroom learning behavior: Video watching, Submitting assignments, Viewing announcements & Self-built data & Verification in practical experiments & Analyzing the main reasons behind the different behavioral patterns of students in the cloud classroom \\
\cite{kamberovic2023personalized} & Integrated Study & SVD, GB, NN & Errors made while learning an introductory C programming course & Self-built data & MAE, PCC & Exploring the use of historical student data to predict future compiler errors \\
\cite{yu2024raw} & Integrated Study & LLM & Classroom behavior video & Self-built data & Verification in practical experiments & Analyzing classroom student behavior using temporal action detection and advanced large-scale language models \\
\midrule
\label{stu_beha_ana_sum}
\end{longtable}
\footnotetext{LDA: Latent Dirichlet Allocation, PCC: Pearson Correlation Coefficient, SVD: Singular Value Decomposition, GB: Gradient Boosting, IPW: Inverse Probability Weighting, SA: Structured Analysis}
\end{landscape}%
\twocolumn

\onecolumn
\begin{landscape}
\footnotesize 
\begin{longtable}{@{\extracolsep\fill}p{1cm}p{2cm}p{2cm}p{3.5cm}p{3cm}p{2cm}p{4.5cm}}
\caption{Summary of studies on student performance prediction (non-cognitive)}\label{tab3} \\
\toprule
References & Algorithm Category & Technique & \multicolumn{2}{c}{Dataset} & Indicator & Research Outcomes \\
\cmidrule(lr){4-5} 
 &  &   & Name & Link &  &  \\ 
\midrule
\endfirsthead

\caption[]{(continue)} \\ 
\toprule
References & Algorithm Category & Technique & \multicolumn{2}{c}{Dataset} & Indicator & Research Outcomes \\
\cmidrule(lr){4-5}
 &  &   & Name & Link &  &  \\ 
\midrule
\endhead

\midrule
\endfoot

\endlastfoot
\cite{feng2022analysis} & Machine learning & K-means & Self-built data & NA & ACC & Improvement of the traditional K-means algorithm by using objective quantitative analysis to determine the number of clusters \\ 
\cite{yaugci2022educational} & Machine learning & NB, RF, SVM, KNN, NN, LR & Self-built data & Publicly available in the form of an annex & ACC, Precision, Recall, F1 & A simple comparison of existing algorithms\\ 
\cite{verma2022prediction} & Machine learning & LR, LDA, KNN, DT, NB, SVM & Self-built data & NA & ACC, Precision & A simple comparison of existing algorithms \\ 
\cite{zhang2018grade} & Machine learning & NB, DT, MLP, SVM & Self-built data & NA & ACC, Precision, Recall, F1 & A simple comparison of existing algorithms \\ 
\cite{khairy2024prediction} & Machine learning & RF, DT, NB, NN, KNN & Self-built data & NA & confusion matrix, ACC, Precision, Recall, F1 & A simple comparison of existing algorithms \\
\cite{ni2023leverage} & Machine learning & PCA, RF & Self-built data & NA & ACC, Precision, Recall, F1 & An analysis of learning behavior data identified course duration, document study time, average test scores, and video completion rates as key factors in student performance prediction \\ 
\cite{xu2019prediction} & Machine learning & DT, NN, SVM & Self-built data & NA & Spearman’s nonparametric correlation analysis & Examining the correlation between Internet use (surfing) and student academic achievement \\ 
\cite{priyambada2023two} & Machine learning & KNN, SVM, RF & Self-built data & NA & ACC, Precision, Recall, F1 & Integration of existing machine learning algorithms \\ 
\cite{asselman2023enhancing} & Ensemble learning & RF, AdaBoost, XGBoost & ASSISTments, Andes, Synthetic & \url{https: //sites.google.com/site/assistmentsdata/home/2009-2010-assistment-data/skill-builder-data-2009-2010.}, \cite{feng2009addressing}, NA & Confusion matrix & Student performance prediction based on different ensemble models \\ 
\cite{joshi2021catboost} & Ensemble learning & K-Means, RF, DT, CatBoost, XGBoost, AdaBoost, LGBM  & Kaggle & \url{https://www.kaggle.com/aljarah/xAPI-Edu-Data} & ACC, Precision, Recall, F1 & A simple comparison of existing algorithms \\ 
\cite{kukkar2023prediction} & Ensemble learning & LSTM, RF, GB & Self-built data, OULAD & NA, \url{https://analyse.kmi.open.ac.uk/open_dataset\#description} & ACC, Precision, Recall, F1 & A system combining a four-layer stacked LSTM network, RF and GB is designed \\ 
\cite{xu2017progressive} & Ensemble learning & Ensemble Learning, Domain-Specific Knowledge & Self-built data & NA & ACC & Built a two-tier architecture based on integrated learning \\ 
\cite{fan2025complementary} & Ensemble learning & CatBoost, Residual Error & UCI Machine learning repository, Kaggle & \url{https://archive.ics.uci.edu/ml/dat asets/student+performance} & RMSE, MAE, MAC, SD & A residual error model was designed to complement the performance of CatBoost \\ 
\cite{fan2024feature} & Ensemble learning & RF, CatBoost & UCI Machine learning repository & \url{https://archive.ics.uci.edu/ml/dat asets/student+performance} & RMSE, MAE, MAC, SD & Multilayer CatBoost  \\ 
\cite{giannakas2021deep} & Deep learning & DNN & Self-built data & NA & Confusion matrix  & Prediction of team performance and analysis of the impact of positive and negative traits \\ 
\cite{kusumawardani2023transformer} & Deep learning & Transformer & OULAD & \url{https://analyse.kmi.open.ac.uk/open_dataset\#description} & ACC, F1 & Proposed a transform-based approach to grade prediction \\ 
\cite{yang2024framelet} & Deep learning & Hypergraph Neural Networks, Attention Mechanisms & OULA, UCI Machine learning repository & \url{https://analyse.kmi.open.ac.uk/open_dataset\#description}, \url{https://archive.ics.uci.edu/ml/dat asets/student+performance} & ACC & Propose framework based dual hypergraph neural networks \\ 
\cite{oh2024language} & Other & LLM, MF & ASSISTments, i-ScreamEdu & \url{https: //sites.google.com/site/assistmentsdata/home/2009-2010-assistment-data/skill-builder-data-2009-2010.}, NA & ACC, Precision, Recall, F1 & The integration of multimodal auxiliary information through a large language model showcases a novel approach to applying language modeling and multimodal learning in deep learning \\ 
\cite{ni2024enhancing} & Other & Signed Graph Neural Networks, LLM, Contrastive Learning & PeerWise platform's fve real data & NA & Binary-F1, Micro-F1, Macro-F1, AUC & The problem of noise and sparsity in educational data is addressed by integrating symbolic graph neural networks and large language model embeddings \\ 
\midrule
\label{stu_per_pre_sum}
\end{longtable}
\footnotetext{NB: Na\"{i}ve Bayes,  RF: Random Forest, SVM: Support Vector Machine, KNN: k-Nearest Neighbors, NN: Neural Networks, LR: Logistic Regression, LDA: Linear Discriminant Analysis, DT: Decision Tree, PCA: Principal components analysis,  GB: Gradient Boosting, DNN: Deep Neural Networks, MF: Matrix Factorization}
\end{landscape}%
\twocolumn

\onecolumn
\begin{landscape}
\footnotesize 
\begin{longtable}{@{\extracolsep\fill}p{1cm} p{2cm} p{2cm} p{2cm} p{3cm} p{4.5cm}}

\caption{Summary of studies on personalized learning path recommendation} \\
\toprule
References & \multicolumn{2}{c}{Dataset} & Indicator & Technique & Research Outcomes \\
\cmidrule(lr){2-3}
 &  Name & Link &  &  &  \\ 
\midrule
\endfirsthead

\caption[]{(continue)} \\ 
\toprule
References & \multicolumn{2}{c}{Dataset} & Indicator & Technique & Research Outcomes \\
\cmidrule(lr){2-3}
 &  Name & Link &  &  &  \\ 
\midrule
\endhead

\midrule
\endfoot

\endlastfoot
\cite{supic2018case} & Self-build&NA& Average score increase & Case-Based Reasoning & Proposes a case-based reasoning model for personalized learning path recommendation\\

\cite{KraussSM18} & Self-build&NA&Hit Ratio & Knowledge Graph, Transfer Probability & Proposes a method to generate personalized learning paths using knowledge graphs and transition probabilities\\

\cite{ZhouHHZT18} &Moocs &\url{http://moocdata.cn/data/MOOCCube} & Precision & LSTM Neural Networks, Clustering & Proposes a full-path learning recommendation model using LSTM neural networks\\

\cite{li2019personalized} & Canvas&\url{https://www.canvas.net/} & Average Grade& Network Embedding, Learning Effects & Proposes a learning path generation algorithm based on network embedding and learning effects\\

\cite{mansur2019personalized} & Self-build&NA & AUC, F1, Precision, Recall & Deep Learning Algorithm & Proposes a personalized learning model using deep learning\\

\cite{nabizadeh2019estimating} & Enki, Mooshak& NA,NA &MAE & Depth-First Search, Time and Score Estimation & Recommends successful learning paths under time constraints\\

\cite{vanitha2019collaborative} & Self-build&NA&Student Score & Ant Colony Optimization, Genetic Algorithm & Combines ant colony optimization and genetic algorithms to construct personalized learning paths\\

\cite{xia2019peerlens} & Self-build& NA & User interactions & Peer-Inspired Learning Path Planning & Proposes PeerLens, an interactive system for learning path planning based on peer exercise history\\

\cite{liu2020learning} &Moocs &\url{http://moocdata.cn/data/MOOCCube}& Precision &Clustering, Learning Networks & Proposes a learning path combination recommendation method based on learning networks\\

\cite{LiuDW20} & Self-build & NA & Average grades& Deep Learning & Proposes a personalized learning model using deep learning\\

\cite{shi2020learning} & Self-build & NA & Average grades & Multidimensional Knowledge Graph & Proposes a learning path recommendation model based on a multidimensional knowledge graph\\

\cite{li2021optimal} & Self-build&NA& Reward & Reinforcement Learning, Hierarchical Skill Model & Proposes an optimal learning policy using reinforcement learning and a hierarchical skill model\\

\cite{sun2021personalized} &Self-build& NA & NA & Knowledge Graph, Graph Convolutional Network (GCN) & Proposes a personalized English learning recommendation method using knowledge graphs and GCN\\

\cite{ChenWTZ23} & ASSIST09, ASSIST12 &\url{https://sites.google.com/site/assistmentsdata}& Hit Ratio &Temporal Convolutional Network, Graph Attention Network, Reinforcement Learning & Combines TCN and GAT into the knowledge tracing model and uses it as the environment of the reinforcement learning model\\

\cite{KrahnKC23} & Self-build& NA & Questionnaire & Learning Style Questionnaire, Moodle Behavior Analysis & Designs a Moodle plugin that generates personalized learning paths according to students' learning styles\\

\cite{li2023personalized} & Self-build& NA&Precision  & Graph-based Genetic Algorithm (GBGA) & Uses GBGA to optimize the alignment of features between learners and learning objects (LOs)\\

\cite{LiXYSR0CT023} & Junyi, ASSITments2015 & \url{https://www.kaggle.com/datasets/junyiacademy/},\url{https://sites.google.com/site/assistmentsdata/datasets} & Reward & Hierarchical Reinforcement Learning, Graph-based Candidate Selector, DKT Model & Presents a Graph Enhanced Hierarchical Reinforcement Learning framework for goal-oriented learning path recommendation\\

\cite{ZhangLW23} & Movielens-1M, LastFM, Self-build&NA,NA,NA & AUC,F1& Multidimensional Knowledge Graph, Graph Convolutional Network & Proposes a learning path recommendation model based on a multidimensional knowledge graph framework\\

\cite{frej2024finding} &COCO, Xuetang& NA,\url{https://www.xuetangx.com/}& NCDG, Recall, Hit Ratio, Precision& Reinforcement Learning, Knowledge Graphs & Proposes an explainable recommendation system for MOOCs using graph reasoning\\

\cite{abu2024knowledge} & Self-build & NA & Recall, Precision, F1& Knowledge Graphs, Large Language Models (LLMs) & Utilizes knowledge graphs as factual context for LLM prompts to reduce model hallucinations\\

\cite{yekollu2024ai} & Self-build&NA&NA & Artificial Intelligence, Adaptive Systems & Discusses the impact of AI on education, focusing on AI-driven personalized learning paths\\

\cite{xiao2024highlighting}& Moocs &Hit Ratio, NDCG &\url{http://moocdata.cn/data/MOOCCube}& LLMs & Proposes LPR model leveraging LLMs to extract coherent learning pathways from students' course enrollment histories\\

\cite{raj2024improved} & Self-build & NA&RMSE, R2 & Real-time Learning Analytics, Knowledge Building, Learning Performance Analysis & Proposes an adaptive learning path recommendation model considering static and dynamic learner parameters\\

\midrule
\label{path.}
\end{longtable}
\end{landscape}%
\twocolumn

\onecolumn
\begin{landscape}
\footnotesize 
\begin{longtable}{@{\extracolsep\fill}p{1cm} p{2cm} p{2.5cm} p{2cm} p{2.5cm} p{6cm}}
\caption{Summary of studies on personalized course recommendation} \\ \\
\toprule
References & \multicolumn{2}{c}{Dataset} & Indicator & Technique & Research Outcomes \\
\cmidrule(lr){2-3}
 &  Name & Link &  &  &  \\ 
\midrule
\endfirsthead

\caption[]{(continue)} \\ 
\toprule
References & \multicolumn{2}{c}{Dataset} & Indicator & Technique & Research Outcomes \\
\cmidrule(lr){2-3}
 &  Name & Link &  &  &  \\ 
\midrule
\endhead

\midrule
\endfoot

\endlastfoot
\cite{HouZXW18} & MOOCCube& \url{http://moocdata.cn/data/MOOCCube}& Regret Comparison & Hierarchical Bandits, Reinforcement Learning & Proposes a methodology for personalized course recommendation using hierarchical bandits to optimize decision-making at multiple levels \\

\cite{symeonidis2019multi} &MERLOT & \url{https://merlot.org/merlot/}& RMSE& Multi-modal Matrix Factorization, Collaborative Filtering & Proposes xSVD++, a model combining multi-dimensional matrix factorization and collaborative filtering to improve recommendation accuracy \\

\cite{YangJ19} & Xuetangx 2013–2014& \url{https://github.com/THU-KEG/MOOCCubeX} &MRR & HITS Algorithm, Social Learning Network & Proposes a dynamic online course recommendation method using social learning networks and the HITS algorithm \\

\cite{zhang2018mcrs} & T10I4D100K, T25I10D10K &  \url{http://fimi.ua.ac.be/data/},\url{http://www.philippe-fournier-viger.com/spmf/index.php?link=datasets.php} & execution time & Distributed Computing, Association Rule Mining & Introduces MCRS, a course recommendation system for MOOCs based on distributed computing and an improved Apriori algorithm \\

\cite{ZhangHCL0S19} &MOOCCube& \url{http://moocdata.cn/data/MOOCCube}& HR,NDCG &Hierarchical Reinforcement Learning & Proposes a hierarchical reinforcement learning algorithm to revise user profiles and tune the recommendation model \\

\cite{Tan0LZ20} & MOOCCube& \url{http://moocdata.cn/data/MOOCCube}& precision, recall, and F1-score & AMSLSTM Network, Autoencoder & Proposes an AMSLSTM network and autoencoder to construct course relevance and improve recommendation accuracy \\

\cite{ZhuLQSCN20} & Self-built&NA&RMSE, MAE, Precision, Recall, F1 & Graph Neural Network, Tensor Factorization & Proposes a hybrid recommendation model combining graph neural networks and tensor factorization \\

\cite{MaWCS21} & MOOCCube& \url{http://moocdata.cn/data/MOOCCube} & Regret Comparison  & Knowledge Graph, LDA, Contextual Multi-Armed Bandit & Proposes a semantic and relationship-aware online course recommendation scheme using LDA, knowledge graph embedding, and a contextual multi-armed bandit algorithm \\

\cite{ZhaoYLN21} & CSM, Eco & \url{https://www.xuetangx.com/},NA& Precision, Recall, F1-measure & Neural Attention Network, Prerequisite Relation Embeddings & Proposes a method to recommend courses using neural attention networks and prerequisite relation embeddings \\

\cite{TianL21} &MOOCCube& \url{http://moocdata.cn/data/MOOCCube}&Precision,Hit Ratio& Cognitive Diagnosis, MIRT, Collaborative Filtering & Integrates MIRT into recommendation models to dynamically update learners' capacities and improve recommendation effectiveness \\

\cite{XuJSZ21} & Xuetangx& \url{https://github.com/THU-KEG/MOOCCubeX} & Precision, Recall& Knowledge Graph, Collaborative Filtering & Combines knowledge graph representation learning with collaborative filtering to enhance recommendation performance \\

\cite{BanWHLZH22} &POJ, Liulishuo & NA,NA& NDCG,HR & Knowledge-enhanced Multi-task Learning & Proposes a knowledge-enhanced multi-task learning model for course recommendation integrating an improved knowledge tracing task \\

\cite{ChenMJF22} & Xuetangx& \url{https://github.com/THU-KEG/MOOCCubeX} & Hit Ratio, NDCG, MRR& GCN-based Attentive Decay Network & Introduces a GCN-based Attentive Decay Network for course recommendation \\

\cite{HaoLB23} & XuetangX, MOOCCube& \url{https://github.com/THU-KEG/MOOCCubeX},\url{http://moocdata.cn/data/MOOCCube} &Hit Ratio, NDCG, MRR, AUC & Meta-Relationship & Proposes Meta-Relationship Course Recommendation to enrich relational information using graph embedding and optimized matrix factorization \\

\cite{JungJKK22} & ESOF,XuetangX  & \url{https://www.ebssw.kr/},\url{https://www.xuetangx.com/}& user satisfaction & Knowledge Graph Enhanced & Proposes KPCR, a knowledge graph-enhanced personalized course recommendation framework \\

\cite{LinLYZLW22} &  MOOCCube & \url{http://moocdata.cn/data/MOOCCube}& Hit Ratio, NDCG & Context-aware Reinforcement Learning & Introduces Hierarchical and Recurrent Reinforcement Learning for course recommendation \\

\cite{LinLZXLWLM22} &XuetangX & \url{https://www.xuetangx.com/}& Hit Ratio, NDCG & Hierarchical Reinforcement Learning with Dynamic Recurrent Mechanism & Proposes HELAR to address the exploration-exploitation trade-off in user profile construction \\

\cite{ZhangSYWF23} &  MOOCCube & \url{http://moocdata.cn/data/MOOCCube} & Recall, NDCG, AUC & Knowledge Grouping Aggregation Network (KGAN) & Proposes a knowledge graph-enhanced course recommendation model using intra-group and inter-group attention operators \\

\cite{Zhang23a} & Self-build &NA &Recall, Precision, F1 & Deep Learning, BERT, LSTM, Multi-Head Attention & Proposes a personalized course resource recommendation method using BERT, LSTM, and multi-head attention \\

\cite{ZhangYSZY23} & Self-build& NA & RMSE, AUC, NDCG, Recall & Factor Memory Network and Graph Neural Network (MG-CR) & Constructs a heterogeneous information network and uses factor memory network and graph neural network for high recommendation accuracy \\

\cite{amin2024adaptable} &  MOOCCube & \url{http://moocdata.cn/data/MOOCCube} & Precision, Hit Ratio, Recall, NDCG& Deep Reinforcement Learning (DRL) & Integrates DRL and multi-agent approach for personalized course recommendation \\
\cite{li2024quantification} & XuetangX, KDDCUP & \url{https://www.xuetangx.com}, NA & MAE, RMSE, Hit Ratio, Recall, Precision & Quantified Engagement and Engagement Neural Network & Proposes a method to quantify learner engagement and apply it to personalized course recommendations \\
\cite{gharahighehi2025enhancing} &  XuetangX, Canvas& \url{https://www.xuetangx.com/}, \url{https://www.canvas.net/} &NDCG & Survival Analysis & Enhances collaborative filtering-based course recommendations using survival analysis \\

\midrule
\label{course.}
\end{longtable}

\end{landscape}
\twocolumn

\onecolumn
\begin{landscape}
\footnotesize 
\begin{longtable}{@{\extracolsep\fill}p{1cm} p{2cm} p{3cm} p{2cm} p{3cm} p{5cm}}
\caption{Summary of studies on personalized exercise recommendation} \\
\toprule
References & \multicolumn{2}{c}{Dataset} & Indicator & Technique & Research Outcomes \\
\cmidrule(lr){2-3}
 &  Name & Link &  &  &  \\ 
\midrule
\endfirsthead

\caption[]{(continue)} \\ 
\toprule
References & \multicolumn{2}{c}{Dataset} & Indicator & Technique & Research Outcomes \\
\cmidrule(lr){2-3}
 &  Name & Link &  &  &  \\ 

\midrule
\endhead

\midrule
\endfoot

\endlastfoot
\cite{diao2018personalized} &Self-build&NA&Precision & Collaborative Filtering, Knowledge Diagnosis & A novel exercise recommendation algorithm integrating learning objectives and assignment feedback, enhancing precision and recall through real-world dataset experiments \\

\cite{lv2018utilizing} &Self-build&NA&Precision& Knowledge Graph & Pioneering approach leveraging student learning status and prerequisite dependencies for personalized exercise recommendation, improving precision and diversity \\

\cite{XiaLC18} &Self-build&NA&Precision, Recall, F1 & Knowledge Tree, Tree Exploration & An automatic exercise recommendation approach based on learning objectives, utilizing knowledge tree models to enhance personalization \\

\cite{huang2019exploring} & MATH, PROGRRAM &\url{https://zhixue.com},\url{https://poj.org} & NDCG, MAP, F1& Reinforcement Learning & A Deep Reinforcement Learning framework for multi-objective exercise recommendations, optimizing review, difficulty, and engagement through novel reward functions \\

\cite{wu2020exercise} & ASSISTments 2009–2010, Algebra 2005–2006, OLIES 2011& \url{https://sites.google.com/site/assistmentsdata/},\url{https://github.com/hcnoh/knowledge-tracing-collection-pytorch},\url{https://edudata.readthedocs.io/en/latest/build/blitz/OLI_Fall2011/OLI_2011F_transaction.html}& Acc, Novelty, Diversity& RNN, DKT & A hybrid method using RNNs and DKT for personalized exercise recommendation, optimizing difficulty, diversity, and novelty \\

\cite{zhu2020study} & Self-build&NA&ACC, Recall, F1, Diversity & Knowledge Graph, RWMD & A personalized exercise recommendation method for computer network courses using knowledge graphs for efficient recommendations \\

\cite{li2021exercise} & Self-build&NA&Hit Ratio & Cognitive Diagnosis Method, LSTM & An exercise recommendation method improving student performance through enhanced cognitive diagnosis and LSTM \\

\cite{ZhouW21a} & Self-build&NA&Recall, Precision & DINA Model, Clustering, Collaborative Filtering & A personalized exercise recommendation method for English learning, achieving higher precision, recall, and efficiency \\

\cite{chuang2021moocers} & MOOCERS &NA & NDCG& Reinforcement Learning & An exercise recommender system for MOOCs using reinforcement learning, integrating review, difficulty, and learning objectives \\

\cite{he2022exercise} & ASSISTments, Algebra, OLI Statics & \url{https://sites.google.com/site/assistmentsdata/},\url{https://github.com/hcnoh/knowledge-tracing-collection-pytorch},NA,\url{https://edudata.readthedocs.io} & ACC, Novelty, Diversity& Knowledge Graph, Knowledge Tracing & ER-KTCP, a method capturing student knowledge state changes for exercise recommendation, introducing a new metric for performance evaluation \\

\cite{li2022personalized} & Self-build & NA & ACC & Collaborative Filtering, KNN & A personalized exercise recommendation method for teaching objectives, achieving a high prediction success rate \\

\cite{guan2023kg4ex} & Assistments 2009, Algebra 2005, Statics 2012 &\url{https://sites.google.com/site/assistmentsdata/},\url{https://github.com/hcnoh/knowledge-tracing-collection-pytorch},\url{https://edudata.readthedocs.io}   & ACC, Novelty& Knowledge Graph, LSTM & KG4Ex, a knowledge graph-based exercise recommendation method with superior performance and strong explainability \\

\cite{li2023knowledge} & Assistment, Eedi & \url{https://edudata.readthedocs.io/} &AUC,ACC & Knowledge Graph, Cognitive Diagnosis Model & A framework enhancing knowledge graphs for exercise recommendation, improving student performance through neural attentive cognitive diagnosis \\

\cite{ren2023muloer} & Assistments 2009, Algebra 2005, Statics 2011 &\url{https://sites.google.com/site/assistmentsdata/},\url{https://github.com/hcnoh/knowledge-tracing-collection-pytorch},\url{https://edudata.readthedocs.io} & ACC, Novelty, Diversity & Self-Attention Networks, Knowledge Tracing Model & MulOER-SAN, a 2-layer multi-objective framework for exercise recommendation, outperforming state-of-the-art methods \\

\cite{wu2023contrastive} &Assistments 2009, Assist12-13, Ednet, BePKT &\url{https://edudata.readthedocs.io} & Hit Ratio, NDCG & Contrastive Learning, Reinforcement Learning & RCL4ER, a framework combining contrastive learning and reinforcement learning for effective exercise recommendation, promoting student learning ability \\
\midrule
\label{exercise}
\end{longtable}

\end{landscape}

{\footnotesize
\begin{longtable}{%
  m{2.8cm}    
  >{\raggedright\arraybackslash}m{3.2cm}  
  c           
  m{2.8cm}    
  c           
  c           
  c           
  c           
  c           
  c           
  c           
  c           
}

\caption{The existing adaptive learning systems} \label{tab:ada_lea_sys} \\

\toprule
\textbf{System} & \textbf{Web-site} & \textbf{Type} & \textbf{Developers} &
\rotatebox[origin=c]{-90}{\textbf{Year}} &
\rotatebox[origin=c]{-90}{\textbf{Charges}} &
\rotatebox[origin=c]{-90}{\textbf{Cloud}} &
\rotatebox[origin=c]{-90}{\textbf{Mobile}} &
\rotatebox[origin=c]{-90}{\textbf{Local}} &
\rotatebox[origin=c]{-90}{\parbox{1.8cm}{\centering \textbf{Cognitive\\modeling}}} &
\rotatebox[origin=c]{-90}{\parbox{1.8cm}{\centering \textbf{Non-cognitive\\modeling}}} &
\rotatebox[origin=c]{-90}{\parbox{2.4cm}{\centering \textbf{Personalized\\recommendations}}} \\
\midrule
\endfirsthead

\toprule
\textbf{System} & \textbf{Web-site} & \textbf{Type} & \textbf{Developers} &
\rotatebox[origin=c]{-90}{\textbf{Year}} &
\rotatebox[origin=c]{-90}{\textbf{Charges}} &
\rotatebox[origin=c]{-90}{\textbf{Cloud}} &
\rotatebox[origin=c]{-90}{\textbf{Mobile}} &
\rotatebox[origin=c]{-90}{\textbf{Local}} &
\rotatebox[origin=c]{-90}{\parbox{1.8cm}{\centering \textbf{Cognitive\\modeling}}} &
\rotatebox[origin=c]{-90}{\parbox{1.8cm}{\centering \textbf{Non-cognitive\\modeling}}} &
\rotatebox[origin=c]{-90}{\parbox{2.4cm}{\centering \textbf{Personalized\\recommendations}}} \\
\midrule
\endhead

\bottomrule
\endfoot

Knewton & knewton.com & ALS & Jose Ferreira & 2008 & Pay & + & + & - & \checkmark &  & \checkmark \\
Smart Sparrow & smartsparrow.com & ALS & University of New South Wales & 2010 & Trial/Pay & + & + & - & \checkmark &  & \checkmark \\
Codecademy & codecademy.com & ALS & Zach Sims & 2011 & Free/Pay & + & + & - &  &  & \checkmark \\
DreamBox & dreambox.com & ALS & Lou Gray, Ben Slivka & 2006 & Trial/Pay & + & + & - & \checkmark &  & \checkmark \\
ALEKS & aleks.com & ALS & Jean-Claude Falmagne & 1996 & Pay & + & + & - & \checkmark &  & \checkmark \\
Tandem & tandem.net & VEA & Arnd Aschentrup, Tobias Dickmeis, Matthias Kleimann & 2015 & Free/Pay & + & + & - &  &  & \checkmark \\
Duolingo & duolingo.com & VEA & Luis von Ahn, Severin Hacker & 2011 & Free/Pay & + & + & + & \checkmark & \checkmark & \checkmark \\
Jill Watson & - & VEA & Ashok Goel & 2016 & - & + & - & - &  &  &  \\
Woebot & www.woebot.\allowbreak io & VEA & Alison Darcy & 2017 & Pay & + & + & - &  & \checkmark & \checkmark \\
IBM Watson Assistant for Education & ibm.com/watson/\allowbreak education/\allowbreak pearson & VEA & - & 2016 & Pay & + & + & - & \checkmark &  & \checkmark \\
AskAway & askaway.org & VEA & - & - & Free & + & + & - &  &  &  \\
Coursera & coursera.org & OLP & Andrew Ng, Daphne Koller & 2012 & Free/Pay & + & + & + &  &  & \checkmark \\
EdX & edx.org & OLP & Harvard University, MIT & 2012 & Free/Pay & + & + & + &  &  & \checkmark \\
Udemy & udemy.com & OLP & Eren Bali, Gagan Biyani, Oktay Caglar & 2010 & Pay & + & + & + &  &  & \checkmark \\
Udacity & udacity.com & OLP & Sebastian Thrun, David Stavens, Mike Sokolsky & 2011 & Pay & + & + & - &  &  & \checkmark \\
Khan Academy & khanacademy.org & OLP & Salman Khan & 2008 & Free & + & + & + & \checkmark &  & \checkmark \\
Carnegie Learning & carnegielearning.com & OLP & Carnegie Mellon University & 1998 & Pay & + & + & - & \checkmark &  & \checkmark \\
FutureLearn & futurelearn.com & OLP & The Open University & 2012 & Free/Pay & + & + & - &  &  & \checkmark \\
CodeCademy & code.org & OLP & Hadi, Ali Partovi & 2013 & Free & + & + & - &  &  & \checkmark \\
Rosetta Stone & rosettastone.com & OLP & Allen Stoltzfus & 1992 & Pay & + & + & + & \checkmark & \checkmark & \checkmark \\
Learning Locker & www.learning\allowbreak locker.net & LAT & HT2 Labs & 2014 & Pay & + & - & - &  &  &  \\
Blackboard Analytics & blackboard.com & LAT & Michael Chasen, Matthew Pittingsky & 1997 & Pay & + & + & - &  &  &  \\
Cognos Analytics for Education & - & LAT & - & - & Pay & + & + & - &  &  &  \\
Brightspace Insights & community.d2l.com/\allowbreak brightspace & LAT & John Baker & - & Pay & + & + & - &  &  &  \\
IBM Watson Education & - & LAT & - & - & Pay & + & + & - & \checkmark &  & \checkmark \\
Moodle & moodle.org & LMS & Martin Dougiamas & 2002 & Free/Pay & + & + & + &  &  &  \\
Blackboard Learn & blackboard.com & LMS & Michael Chasen, Matthew Pittingsky & 1997 & Pay & + & + & + &  &  &  \\
Canvas & instructure.com/\allowbreak canvas & LMS & Brian Whitmer, Devlin Daley & 2008 & Pay & + & + & + &  &  &  \\
Schoology & schoology.com & LMS & Jeremy Friedman, Ryan Hwang, Tim Trinidad, Bill Kindler & 2009 & Free/Pay & + & + & - &  &  &  \\
Minecraft: Education Edition & education.minecraft.\allowbreak net & EG & Markus Persson, Mojang & 2011 & Free & + & + & + &  & \checkmark & \checkmark \\
BrainPOP & brainpop.com & EG & Avraham Kadar & 1999 & Pay & + & + & - &  &  &  \\
Scratch & scratch.mit.edu & EG & Mitchel Resnick & 2007 & Free & + & + & + &  &  &  \\
Prodigy & prodigygame.com & EG & Alex Peters, Rohan Mahimker & 2011 & Free/Pay & + & + & - & \checkmark &  & \checkmark \\
Tinkercad & tinkercad.com & EG & Kai Backman, Mikko Mononen & 2011 & Free & + & - & - &  &  &  \\
Autodesk Education & autodesk.com/\allowbreak education & EG & John Walker & 1982 & Free & + & + & + &  &  &  \\

\end{longtable}
}

\twocolumn

\bibliographystyle{fcs}
\bibliography{reference}   

\end{document}